\documentclass{article}

\PassOptionsToPackage{numbers, compress}{natbib}
\usepackage[preprint]{neurips_2026}
\usepackage{hyperref}
\usepackage{url}
\usepackage{amsfonts}
\usepackage{amsthm}
\usepackage{amsmath}
\usepackage[capitalize]{cleveref}
\usepackage{xspace}
\usepackage{titletoc}
\usepackage{wrapfig}
\usepackage{float}
\usepackage{graphicx}
\usepackage{caption}
\usepackage{booktabs}
\usepackage{array}
\usepackage{tabularx}
\usepackage{subcaption}
\usepackage{multirow}
\usepackage{multicol}
\usepackage{algorithm}
\usepackage{algpseudocode}
\usepackage{mathtools}
\usepackage{nicefrac}
\usepackage{colortbl}
\usepackage[table]{xcolor}
\usepackage{adjustbox}
\usepackage{pifont}
\usepackage{colortbl}
\usepackage{stfloats}
\usepackage{xspace}
\usepackage{tcolorbox}
\usepackage[dvipsnames]{xcolor}
\usepackage{amssymb}
\usepackage{fontawesome5}
\usepackage[utf8]{inputenc} 
\usepackage[T1]{fontenc}    
\usepackage{microtype}      
\usepackage{csquotes}
\usepackage[normalem]{ulem}
\usepackage{bbm}

\newcommand{\ldm}{AudioLDM2\@\xspace}
\newcommand{\stableaudio}{Stable Audio Open\@\xspace}
\newcommand{\ace}{Ace-Step\@\xspace}

\definecolor{lslavender}{RGB}{130, 92, 194}
\definecolor{darkpink}{RGB}{194, 30, 86} 
\definecolor{mmcyan}{RGB}{0, 139, 139}

\definecolor{linkblue}{HTML}{1a5276}
\definecolor{teaserblue}{HTML}{38789D}

\title{TADA! Tuning Audio Diffusion Models through Activation Steering}

\author{Łukasz Staniszewski$^{1,2}$\thanks{Corresponding author: \url{luks.staniszewski@gmail.com}}\space\space\space\space\space\space Katarzyna Zaleska$^{1}$\space\space\space\space Mateusz Modrzejewski$^{1}$\space\space\space\space Kamil Deja$^{1,2}$\\
    $^1$Warsaw University of Technology \space \space $^2$IDEAS Research Institute \\
}

\newcommand{\hf}{%
  \raisebox{-0.7ex}{\includegraphics[height=1.3\baselineskip]{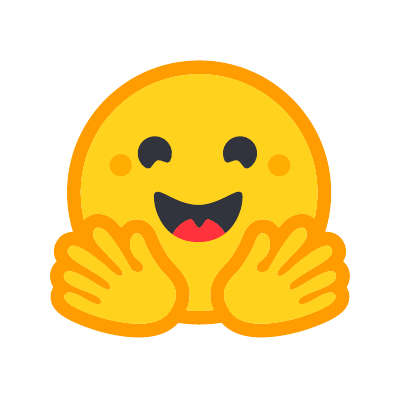}}%
}

\begin{document}

\maketitle

\begin{abstract}
  Audio diffusion models can synthesize high-fidelity music from text, yet achieving fine-grained control over specific musical attributes remains challenging, as their internal mechanisms for representing high-level concepts are poorly understood. In this work, we use activation patching to demonstrate that recent audio diffusion architectures exhibit a semantic bottleneck, where a small, shared subset of consecutive attention layers controls distinct musical concepts, such as the presence of specific instruments, vocals, or genres. Building on this, we systematically evaluate a broad spectrum of steering paradigms, comparing activation steering against prompt-level, score-space, and weight-space interventions, analyzing the interaction between the steering mechanism and the intervention site. Our new benchmark, supported by an extensive user study, demonstrates that localized activation steering establishes a new state-of-the-art in audio concept modulation. 
\end{abstract}
\begin{center}
    \vspace{-1.5em}
    \makebox[\dimexpr\textwidth-2\leftmargini-32pt\relax][s]{%
        \href{https://github.com/luk-st/steer-audio}{\textcolor{black!75!white}{\faGithub\ \space \texttt{\textbf{Code}}}}%
        \hfil
        \href{https://huggingface.co/collections/lukasz-staniszewski/tada-steering-collection}{\textcolor{orange!75!white}{\hf\ \space \texttt{\textbf{Steering Vectors}}}}%
        \hfil
        \href{https://audio-steering.github.io}{\textcolor{black!75!white}{\faHeadphones\ \space \texttt{\textbf{Audio Samples}}}}%
    }%
    \vspace{0.5em}
\end{center}

\begin{figure}[h]
    \centering
    \vspace{-1em}
    \begin{minipage}{\linewidth}
        \centering
        \includegraphics[width=0.82\linewidth]{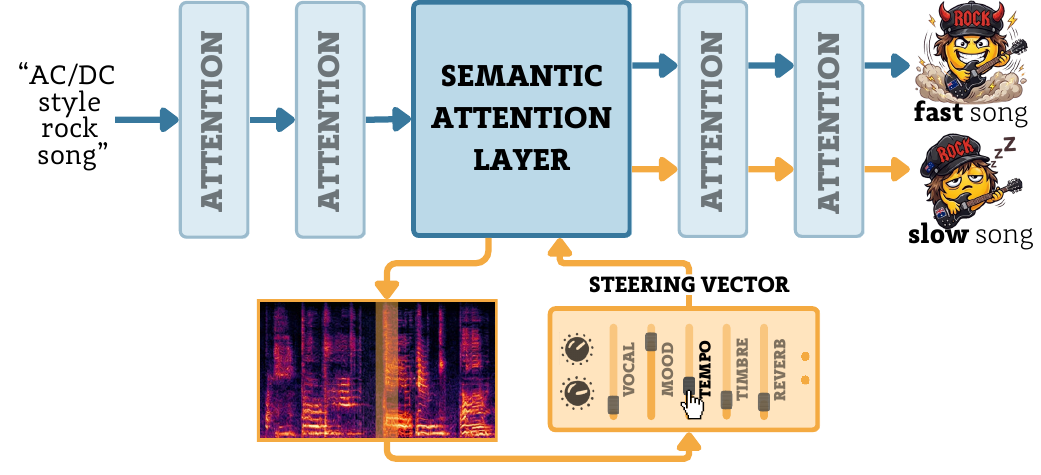}
        \caption{\label{fig:teaser}\textbf{We study localized activation steering in Audio Diffusion Models.} By localizing functional layers, we enable precise modulation of audio concepts through activation steering.}
        
    \end{minipage}
\end{figure}


\section{Introduction}

Diffusion-based text-to-music models such as \ace~\citep{gong2025ace0step0,gong2026ace} now synthesize coherent, full-length musical pieces from a single text prompt. Analogous technologies~\citep{suno2026} establish consumer products in routine use among hobbyists, advertisers, podcasters, and game-audio designers. Practical creative use of these systems, however, exposes a fundamental limitation of the prompt-based interface: it does not support fine-grained control.

A user can request ``a fast, lively rock song'' and obtain a plausible result, but cannot request a ``fast but not too fast tempo'', a ``slightly lower vocal pitch'', or the same arrangement with the more feminine vocal. Re-issuing a perturbed prompt regenerates an entirely different piece. Music is a continuous medium, and creative work with it depends on small adjustments along specific perceptual axes that prompt-based interfaces do not expose.

In language and image generation, this kind of fine-grained control has been achieved by intervening directly inside the model through: manipulations of prompts and their embeddings~\citep{gorgun2026temporal, baumann2025continuous}, weight adaptations~\citep{gandikota2024concept}, or activation-space steering~\citep{rimsky-caa, huben2024sparse, gao2025scaling, feng2026finegrained}. We argue that music's continuous perceptual structure makes it a more natural target for such techniques than in those other domains. Yet, audio adaptations have so far been narrow, focusing on autoregressive music models~\citep{koo2025smitin, facchiano2025activation, singh2025discovering}.

In this work, we address these limitations by posing two core questions: \emph{where} and \emph{how} should we intervene in the audio diffusion models to achieve smooth, precise control over semantic concepts?

To answer \emph{where}, we apply activation patching \citep{NEURIPS2022_6f1d43d5} across a curated set of counterfactual prompt pairs covering vocals, tempo, mooļd, instruments, and genres. We evaluate three models spanning the dominant architectural families: \ldm \citep{liu2024audioldm} (Diffusion U-Net), \stableaudio \citep{evans2024stable} (Diffusion Transformer), and \ace \citep{gong2025ace0step0} (Flow-Matching Transformer).
The empirical picture is consistent across all architectures: every model exhibits a 'semantic bottleneck.' The control for high-level musical concepts is densely concentrated in a small, shared subset of cross-attention layers: two out of twenty-four in the transformer-based models, and a comparably narrow band in the U-Net. 

To answer \emph{how}, we systematically evaluate a wide spectrum of steering paradigms. We primarily focus on activation steering techniques and compare them with prompt-level interventions, score-space modifications, and weight-space techniques that we adapt to the music domain. To understand the dynamics of the discovered bottleneck, we investigate the role of localization by comparing each method's standard global application with a localized variant restricted to the functional layers.

Through a rigorous evaluation protocol grounded in the alignment-preservation trade-off across musical concepts, we uncover a crucial interaction between the steering mechanism and the intervention site. We show that restricting interventions to the semantic bottleneck greatly improves the effectiveness of activation-based techniques. Our localized steering avoids collateral acoustic damage and successfully isolates the conceptual change. Consequently, it establishes a new state-of-the-art in audio concept modulation, significantly outperforming both global activation steering and strong baselines across all other paradigms. These objective metrics are corroborated by an extensive listening study, confirming that localized activation steering yields the most seamless and natural integration of desired musical attributes. Our contributions can be summarized as follows:
\begin{enumerate}
\vspace{-0.5em}
    \item We identify a \emph{semantic bottleneck} in distinct text-to-music diffusion models: a small, shared subset of cross-attention layers governing the majority of interpretable musical concepts.
\vspace{-0.2em}
    \item We introduce a comprehensive suite of steering techniques for text-to-music diffusion models and show that restricting activation steering to selected functional layers establishes a new state-of-the-art in audio concept modulation.
\vspace{-0.2em}
    \item We propose an evaluation protocol grounded in the alignment--preservation trade-off, instantiate it as a benchmark over musical concepts, and validate it with a listening study whose preferences are consistent with the automatic metrics.
\end{enumerate}

We release the benchmark, trained artifacts, and our adaptations of all evaluated steering methods to support further research on controllable music generation.
\section{Related work} \label{sec:related}

\paragraph{Interpretability and Steering of Generative Models.}
Mechanistic interpretability is emerging as a research direction aimed at understanding the internal computations of large-scale architectures. The objective of Causal Mediation Analysis \citep{pearl2001causalTracing, NEURIPS2022_6f1d43d5} is to understand how the model output changes under interventions in its computational graph. Recently, this technique has been applied to the image domain \citep{basu2024-localizing-knowledge, basu2024mechanistic, staniszewski2025precise, zarei2025localizing}, thereby uncovering the mechanistic role of DMs' Attention layers. Precisely, \citet{basu2024-localizing-knowledge} employ activation patching and \citet{basu2024mechanistic} use prompt injection to localize layers controlling model knowledge in U-Net-based DMs. \citet{staniszewski2025precise} combine these techniques to identify layers that control text generated in images with DiTs. Finally, \citet{zarei2025localizing} show that important attentions in DiTs can be efficiently traced by the magnitudes of their outputs.

On the other hand, the linear representation hypothesis~\citep{park2024the} posits that neural networks encode high-level concepts as linear directions in the activation space. Recently, \citet{rimsky-caa} proposed the Contrastive Activation Addition (CAA) technique, showing that differentiating neural network hidden states from runs with prompts describing opposite concepts yields an activation difference that encodes the semantic change between the prompts. This approach has been widely applied to Large Language Models (LLMs), e.g., for steering model behavior~\citep{chen2025persona, feng2026finegrained} or reducing its toxicity~\citep{rodriguezcontrolling}. In vision, \citet{rodriguezcontrolling,rodriguez2025lineas} steer generations by training affine maps that represent transformations between activation distributions in distilled text-to-image models. Similarly, difference-based methods have been applied in the text encoder space of diffusion models \citep{baumann2025continuous} for concept modulation, or in attention layers \citep{gaintseva2025casteer} for concept removal. Beyond activation contrasting, Sparse Autoencoders (SAEs)~\citep{olshausen1997sparse} have been applied to LLMs~\citep{huben2024sparse, bricken2023monosemanticity}  to decompose activations into sparse, interpretable features in an unsupervised manner by training an autoencoder to reconstruct them with a sparsity constraint. In the image domain, \citet{surkov2025onestepenoughsparseautoencoders} demonstrated that SAEs can capture meaningful features inside DMs, and \citet{cywinski2025saeuron} ablated features within the SAE latent space for concept unlearning. Apart from activations, \citet{gandikota2024concept} steer concepts via trainable low-rank adapters that represent meaningful directions in the weight space.

\paragraph{Interpretability and Control of Audio Models.}

Recent research has begun to adapt interpretability techniques to the audio domain to understand models and control their internal computations. Initial efforts \citep{whisper2024sadov} identify interpretable circuits in the ASR model via feature discovery techniques. Prior works have also studied text-to-music models, focusing on autoregressive architectures (Music LLMs). In particular, \citet{wei2024music} introduced a dataset to probe whether music foundation models encode specific music-theory concepts, such as intervals and chords. Similarly, \citet{vasquez2024exploring} probe for the information about instruments and genres. Moving towards control in Music LLMs, \citet{koo2025smitin} use classifier probes to steer attention heads for specific musical traits, \citet{zhao2026steering} adapt Recursive Feature Machines for steering, while \citet{facchiano2025activation} use activation patching to manipulate binary attributes. Recently, SAEs have been employed in the audio domain to steer the residual stream of MusicGen \citep{singh2025discovering}, or map features to acoustic concepts within audio autoencoders \citep{paek2025learning} and speech transformers \citep{audiosae-2026}. However, despite the recent high performance of diffusion-based music generators \citep {gong2026ace}, interpretability and steering of Audio DMs remain largely underexplored. 
Recent works predominantly focus on audio editing by manipulating attention maps \citep{xu2024promptguided, MelodyEdit, yang2025melodia}, or by combining the inversion process with intermediate latents optimization \citep{liang2024audiomorphix,audioeditor, MusicMagus, audio_ddpm_inv, niu2026steermusic, lee2026diffusion}. 
Notably, \citet{yang2025melodia} probe self-attention maps within AudioLDM 2 \citep{liu2024audioldm}, achieving better preservation of temporal structure. 
These methods perform a single instruction-based edit to an existing recording, whereas steering aims to continuously modulate a target concept during generation, yielding generation variants along a concept axis rather than a single track edit. 
Closest to our work, \citet{ezra2025freesliders0} evaluate diffusion score-based methods for steering acoustic concepts with Stable Audio \citep{evans2024stable}. 
In this work, we focus on musical concepts, systematically benchmark our activation steering against their score-based approach, and further analyze it within our localization framework.

\section{Background and methodology} \label{sec:background}

\paragraph{Audio Diffusion Models.} 
Modern audio DMs operate in a compressed latent space and employ a U-Net \citep{ronneberger2015unet} or a Transformer \citep{vaswani2017attention,dosovitskiy2021an} as their backbone architecture. Unlike text-to-image models processing spatial patches, audio latents are structured as sequences of temporal frames $z_t \in \mathbb{R}^{F \times d}$. Thus, each token $f$ corresponds to a distinct timeframe in the audio. The cross-attention role is to introduce the text information into the hidden states $\mathbf{h}\in \mathbb{R}^{F \times d}$ through the prompt embedding $\mathbf{c} \in \mathbb{R}^{C \times d_c}$. The output of the $l$-th Cross-Attention block at step $t$ is then
    \begin{equation}
        \text{CrossAttn}(\mathbf{h}_{l-1}^{(t)}, \mathbf{c}) = \text{softmax}\left(\frac{\mathbf{Q}\mathbf{K}^T}{\sqrt{d_k}}\right)\mathbf{V}\mathbf{W}_O,
    \end{equation}
    where $\mathbf{Q} = \mathbf{h}_{l-1}^{(t)} \mathbf{W}_Q$, $\mathbf{K} = \mathbf{c} \mathbf{W}_K$, $\mathbf{V} = \mathbf{c} \mathbf{W}_V$, and $\{\mathbf{W}_Q, \mathbf{W}_K, \mathbf{W}_V,\mathbf{W}_O\}$ are the learned weights. This output is further added to the residual stream as $\mathbf{h}_l^{(t)} = \mathbf{h}_{l-1}^{(t)} + \text{CrossAttn}(\mathbf{h}_{l-1}^{(t)}, \mathbf{c})$.

\begin{figure}[t]
        \centering
        \includegraphics[width=0.99\linewidth]{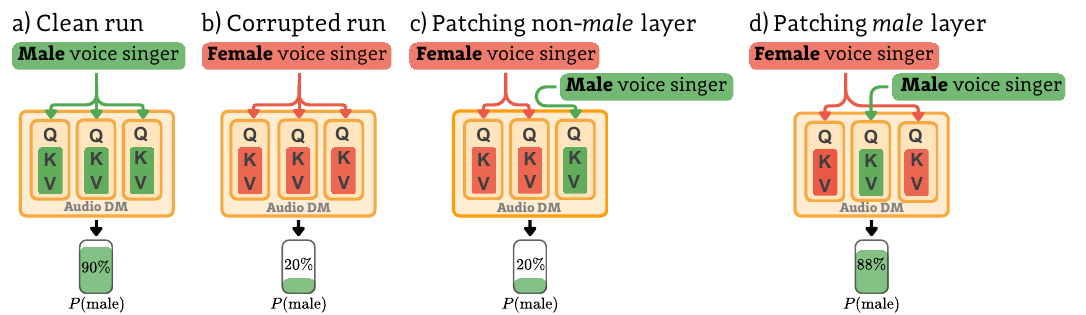}
        \vspace{-0.5em}
        \caption{\textbf{Layer localization via Activation Patching.} For a target concept, we perform a clean run (a) with a prompt referring to the concept and save the attention key (\textbf{K}) and value (\textbf{V}) activations. In a corrupted run (b), we infer with a counterfactual prompt. Next (c,d), we run a corrupted run, substituting the $l$-th layer key and value activations with the saved ones from the clean run. If (d) patching inside $l$ produces audio with the target concept, we identify it as a functional layer.}
        \label{fig:trace_schema}
        \vspace{-1.5em}
    \end{figure}

    \paragraph{Activation Patching.} To identify which layers control specific musical concepts, we employ activation patching~\citep{NEURIPS2022_6f1d43d5} illustrated in \cref{fig:trace_schema}. For a concept $c$ (e.g., ``male voice''), we define a set of counterfactual prompt pairs $(\mathcal{P}_c, \mathcal{P}_{\tilde{c}})$ where $\mathcal{P}_c$ contains the concept and $\mathcal{P}_{\tilde{c}}$ its semantic counterpart.

    \clearpage
    We generate audio with the target prompt $\mathcal{P}_c$, caching the attention keys $\mathbf{K}_l = \mathbf{c}\mathbf{W}_K$ and the values $\mathbf{V}_l = \mathbf{c}\mathbf{W}_V$ at each layer $l$. Then, generating audio with the source prompt $\mathcal{P}_{\tilde{c}}$, we patch layer $l$ by substituting its keys and values with the cached ones. Finally, we compare how the intervention affects the similarity between the generated audio and the prompt that describes the concept $c$. 

    If the output audio for the patched run exhibits concept $c$, we identify $l$ as the diffusion model's \emph{functional layer} for $c$. Precisely, the impact of layer $l$ on concept $c$ is calculated as:
\begin{equation}\label{eq:impact}
    \text{I}(l, c) = \frac{\text{s}(l\leftarrow c,l'\leftarrow \tilde{c})-\text{s}(l\leftarrow \tilde{c},l'\leftarrow \tilde{c})}{\text{s}(l\leftarrow c,l'\leftarrow c)-\text{s}(l\leftarrow \tilde{c},l'\leftarrow \tilde{c})},
\end{equation}
where $\text{s}(l\leftarrow c_1,l'\leftarrow c_2)$ denotes the audio-text similarity between the concept name and a generation where layer $l$ inputs prompt $c_1$ while all other layers $l'=L\setminus\{l\}$ receive $c_2$. The denominator serves as a normalization constant representing the upper bound of similarity gain, ensuring $\text{I}(l,c)\in[0,1]$. Each layer's impact score is calculated separately for each concept, then aggregated by averaging across concepts: $\text{I}(l)=\frac{1}{|C|}\Sigma_{c\in C}\text{I}(l,c)$. The final set of functional layers ($\mathcal{L}$) consists of ones that score above a predefined threshold $\tau$: $\mathcal{L}=\{l\in L:\text{I}(l)\ge\tau\}$.

\paragraph{Activation Steering.} Activation steering intervenes on a model's hidden states (internal representations) at inference time to control its final output. Steering Vector $\mathbf{v}_c$ encoding the concept $c$ is obtained by contrasting model's intermediate activations $\{(h_c^{i}, h_{\tilde{c}}^{i})\}_{i=1}^N$, collected from $N$ paired positive (concept-amplifying, $c$) and negative (concept-suppressing, $\tilde{c}$) prompts runs. Finally, $\mathbf{v}_c$ is added to layer $l$ outputs as $\mathbf{h}'_l = \mathbf{h}_l + \alpha \cdot \mathbf{v}_c$. The coefficient $\alpha \in \mathbb{R}$ controls both the direction of the modulation (positive values amplify the concept, negative values suppress it) and its strength.

During experiments, we employ diverse steering methods, which differ in how $\mathbf{v}_c$ is constructed. For \emph{Contrastive Activation Addition} (CAA \citep{rimsky-caa}), we follow \citet{gaintseva2025casteer} and compute $\mathbf{v}_c$ as the mean difference between positive and negative activations, precisely $\mathbf{v_c^{\text{CAA}}} = \frac{1}{N}\Sigma_{i=1}^N (\bar{h}_{c}^{i} - \bar{h}_{\tilde{c}}^{i})$, with $\bar{h}$ denoting cross-attention layer outputs averaged across temporal frames.

In \emph{AUSteer}~\citep{feng2026finegrained}, $\mathbf{v}_c$ is forced to be sparse by deriving the ranking activations' dimensions with a signed contrastive score. For each layer $\ell$ and activation dimension $d$, AUSteer measures the fraction of contrastive pairs for which the activation difference $h_{c}^{i,\ell}[d] - h_{\tilde{c}}^{i,\ell}[d]$ agrees in sign. This yields scores $\beta_{\ell,d} \in [-1, 1]$ whose magnitude reflects the directional reliability of the given dimension. A global top-$s$ selection over scores $|\beta_{\ell,d}|$ produces a sparse steering vector at layer $\ell$ through dimensions $d\in1\dots D$ as $\mathbf{v_c^{\text{AUS}}}[d]=\mathbbm{1}_{|\beta_{\ell,d}|\in\text{top-}s}\cdot\beta_{\ell,d}$. Notably, AUSteer, through its ranking-and-selection mechanism, inherently performs localization. 

Finally, to discover interpretable directions within activations, we train TopK \emph{Sparse Autoencoders} (SAEs \citep{Olshausen1997SparseCW,gao2025scaling,bussmann2024batchtopksparseautoencoders}). Here, an autoencoder architecture is trained to reconstruct cross-attention activations with a TopK sparsity constraint. To identify concept-specific features, we use a set of positive and negative activations and perform TF-IDF-like scoring within the autoencoder latent space. The final steering vector for concept $c$ is built by summing SAE's decoder $W_{\text{dec}}$ columns corresponding to the top-$k_c$ scoring features $\mathcal{F}_c$ within SAE latent space, as $\mathbf{v_c^\text{SAE}}=\sum_{j \in \mathcal{F}_c} W_{\text{dec}}[:, j]$. In Appendix~\ref{app:activation_steering_methods}, we describe these methods in detail.

\section{Layer localization} \label{sec:layer_localization}

To understand \emph{where} an intervention enables precise concept steering, we measure the causal importance of individual cross-attention layers in text-to-audio models. We construct a dataset of counterfactual prompt pairs that considers the following musical concepts: \textit{vocal gender} (female vs.\ male), \textit{tempo} (slow vs.\ fast), \textit{mood} (happy vs.\ sad), instruments (\textit{drums}, \textit{flute}, \textit{guitar}, \textit{maracas}, \textit{trumpet}, \textit{violin}), and genres (\textit{jazz}, \textit{techno}, \textit{reggae}). For each contrasting pair $(c, \tilde{c})$, we select captions from the MusicCaps~\citep{agostinelli2023musiclm} containing keywords associated with concept $c$ (e.g., `female voice') but do not contain any terms related to the alternative variant $\tilde{c}$ (e.g., `male voice'). For genres and instruments, we replace concepts (e.g., `violin') with alternatives (e.g., `trumpet'). We select up to 256 such prompts as target ones $\mathcal{P}_c$ and construct their alternatives $\mathcal{P}_{\tilde{c}}$ via rule-based replacement of concept-associated terms with their counterparts. We employ GPT-4 \citep{achiam2023gpt} to find alternative forms for concepts and propose substitution rules (see examples in App.~\ref{app:prompt_examples}), which we manually verify.

\begin{wrapfigure}{r}{.4\linewidth}
    \vspace{-1.5em}
    \centering
    \includegraphics[width=\linewidth]{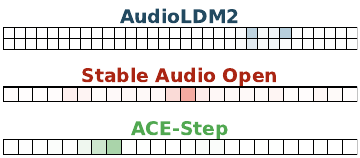}
    \caption{\textbf{Average impact scores} $\text{I}(l)$ \textbf{of cross-attention layers (denoted with color intensity in $\square$) within text-to-music DMs.} Singular layers control different concepts: vocal, tempo, mood, instruments, and genres across diverse audio diffusion architectures. Plots for all concepts are provided in Appendix~\ref{app:localization}.}
    \vspace{-1.5em}
    \label{fig:layers_all_compact}
\end{wrapfigure}
Using a constructed dataset, we apply the activation patching procedure described in \cref{sec:background} to three state-of-the-art audio diffusion models: \ldm~\citep{liu2024audioldm}, \stableaudio~\citep{evans2024stable}, and \ace~\citep{gong2025ace0step0}. We generate 10-second (\ldm, \stableaudio) and 30-second (\ace) waveforms, using 8 different random seeds per prompt, resulting in 2048 generations per concept. We use MuQ~\citep{zhu2025muq} for assessing mood, tempo, instruments, and genres, and CLAP~\citep{wu2022largescale} for distinguishing vocal gender.

\cref{fig:layers_all_compact} presents the layer-wise impact scores (aggregated across concepts) for each model. For all three architectures, we observe that a small subset of layers concentrates the control over musical concepts. Using a selection threshold of $\tau=0.10$, in \ldm (U-Net), we localize key components in the decoder, specifically the layers 45--51 (7 out of 64). In transformer-based architectures, we observe a high degree of control in the middle layers. Namely, in cross-attentions \{7,8\} for \ace, and \{12,13,14\} in \stableaudio. 

These findings show that audio diffusion models develop a \textbf{semantic bottleneck} -- a set of interpretable, functionally specialized layers shared across various semantic concepts. Additionally, such specialized control is not limited to a single model but appears to be a general property of music DMs. These observations contrast with studies on image generators \cite{basu2024mechanistic, staniszewski2025precise, zarei2025localizing}, where localized components remain largely disjoint across concept categories. Such surprising regularity opens the door to future studies (see App.~\ref{app:future_work}) on layer transfer and interchangeability \citep{model_stitching} across text-to-audio models.

\section{Continuous Audio Concept Modulation} \label{sec:audio_concept_steering}
The high degree of semantic control concentrated in single layers suggests their activations as a natural intervention site: restricting modifications to the semantic bottleneck should isolate the desired change, leaving the rest of the audio intact. To answer the question of \emph{how} to precisely modulate the concept $c$ in the generated track, we adapt a broad set of steering paradigms from other domains to music generation and benchmark them against activation steering localized to the bottleneck.

\subsection{Baselines}

In addition to the activation steering methods (Sec.~\ref{sec:background}), we evaluate techniques operating in other levels: prompts, weights, and diffusion prediction. Detailed descriptions are provided in App.~\ref{app:steering_methods}.

\textbf{Prompt-level interventions.} The most direct way to influence a text-to-music model is through its conditioning. For fine-grained control, \emph{PCI}~\citep{gorgun2026temporal} inputs the model with a prompt describing the modulated concept starting from a chosen diffusion timestep, ensuring smooth, natural manipulation in the conditioning space. Alternative approaches alter conditioning embedding itself: \emph{Text Embeddings}~\citep{ezra2025freesliders0} add to prompt embeddings a difference between concept-amplifying and concept-suppressing embeddings, while \emph{Token Embeddings}~\citep{baumann2025continuous} restrict this approach to tokens referring to the modulated subject (e.g., \emph{'instrument'} when adding piano), leaving the rest of the prompt untouched.

\textbf{Weights-level interventions.} \emph{Concept Sliders}~\citep{gandikota2024concept} encode concepts directions as low-rank weight updates~\citep{hu2022lora} learned during model fine-tuning. During inference, learned weights are further merged with the base model at varying scales, which serve as the steering-strength coefficients $\alpha$.
\vspace{-0.3em}

\textbf{Score-level interventions.} \emph{FreeSliders}~\citep{ezra2025freesliders0} offer a training-free alternative to Concept Sliders, where steering direction is encoded as the difference between the diffusion model's noise prediction on contrastive (amplifying and suppressing) prompts and further added to standard DM prediction.
\vspace{-0.3em}
\subsection{Evaluation protocol}\label{sec:metrics}
\vspace{-0.3em}
For robust evaluation, we construct a new benchmark that assesses the modulation of $9$ musical concepts. Methods are compared over a range of steering strengths $\alpha \in \{-\alpha_{\text{max}} \ldots0\dots \alpha_{\text{max}}\}$, with positive (concept-amplifying, $\alpha>0$) and negative (concept-suppressing, $\alpha<0$) directions. 
\vspace{-0.3em}

\paragraph{Steering effectiveness.}
No single metric adequately captures the performance of a steering method.
Modulation involves an inherent trade-off between conserving the characteristics of the original audio (\textit{preservation}) and increasing the presence of the target concept (\textit{alignment}) -- e.g., raising the \textit{tempo} inevitably alters the song structure.
An effective intervention is therefore one that, at a matched level of preservation, achieves better alignment. It introduces the desired change, leaving the remaining musical content intact.
Inspired by the perception--distortion trade-off~\citep{blau2018perception}, we quantify \textit{steering effectiveness} as the area under the alignment--preservation curve, as in \cref{fig:loc_steering}.
\vspace{-0.2em}

For audio preservation, we use LPAPS~\citep{iashin2021taming}. We measure perceptual distance from the unsteered baseline (when $\alpha = 0$), and define the preservation score for audio generated with strength $\alpha$ as $p(\alpha) = \text{LPAPS}(\alpha_{\text{max}}) - \text{LPAPS}(\alpha)$, where higher values indicate better conservation of the original audio. To compare methods fairly, we calibrate their steering range $(-\alpha_{\text{max}}, \alpha_{\text{max}})$ so that their maximum distortion matches the distance between two audios generated by the diffusion model from different textual prompts. Such a reference represents the natural level of audio change expected from a prompt-level intervention, provides an interpretable reference, and, technically, is obtained by the \emph{PCI} method when steered at maximum strength. In App.~\ref{app:preservation_metrics_section}, we present an analysis of additional decomposable audio preservation metrics, including harmony, rhythm, melody, and musical structure. 
\vspace{-1.3em}

For alignment, we use the sign-corrected delta $\Delta a(\alpha) = \text{sign}(\alpha) \cdot (a(\alpha) - a(0))$, where $a(\cdot)$ is the similarity between audio steering with given strength and the textual description of the target direction. Similarity is measured with CLAP~\citep{wu2022largescale} and MuQ~\citep{zhu2025muq}. Sign correction ensures that both steering directions are on a common axis, with higher values indicating stronger alignment. The final \textbf{AUC ($\uparrow$)} is then defined as $\int_0^{P_{\text{max}}} \Delta a(p)\, dp$ and computed over the $\alpha$-swept pairs: $(p(\alpha), \Delta a(\alpha))$.
\vspace{-0.3em}

\paragraph{Alternative steering axes.} To complete the benchmark, we measure \textbf{Smoothness ($\downarrow$)} \citep{ezra2025freesliders0}, which captures how uniformly a method increases alignment along the steering scale. We normalize the alignment deltas to $[0,1]$, compute consecutive gaps $g_i = \Delta a_{i+1} - \Delta a_i$ ordered by increasing $|\alpha|$, and report $\text{S} = \text{std}(\{g_i\})$. We further assess \textbf{Audio Quality ($\uparrow$)} with Audiobox Aesthetics~\citep{tjandra2025aes}, averaging the Content Enjoyment (CE), Content Usefulness (CU), Production Complexity (PC), and Production Quality (PQ) scores at 30 uniformly spaced preservation values.
\vspace{-0.3em}

\paragraph{Technical details.} We adapt the methods to the \ace \citep{gong2025ace0step0}, a state-of-the-art text-to-music model which significantly outperforms \ldm and \stableaudio in the generative capabilities. We generate $N=100$ thirty-second tracks across $15$ positive and $15$ negative steering strengths. For activation steering, concept-suppressing vectors are constructed from their respective concept-amplifying vectors through negation. Working with diffusion models, we construct steering vectors for each diffusion step separately. More technical details, including method ablations and hyperparameter analyses, are provided in Appendix~\ref{app:benchmark_details}. 
\vspace{-0.3em}

\subsection{Results}
\vspace{-0.3em}

\cref{tab:steering_auc_avg} compares all evaluated steering paradigms, averaged across nine musical concepts, while individual results are provided in App.~\ref{app:loc_steering_tables}.  Activation steering restricted to the functional layers resolves the alignment--preservation trade-off most effectively: $\text{SAE}$ tops the AUC ranking under both alignment metrics ($0.118$ MuQ, $0.070$ CLAP), with localized $\text{CAA}$ and $\text{AUSteer}$ completing the top three under MuQ. Under CLAP, Concept Sliders ($0.067$) is a close runner-up, with localized activation steering variants following closely. The non-localized activation steering, PCI, and FreeSliders form a middle tier, while remaining prompt-level interventions trail. In particular, Text Embeddings achieves a near-zero AUC because its embedding-level distortions quickly fall outside the preservation window.

In smoothness, our methods distribute concept changes uniformly along the intervention scale rather than producing abrupt jumps. The localized activation steering closely matches Concept Sliders and FreeSliders at $0.05$--$0.07$, three to twelve times lower than the prompt-level methods.  Audio Quality is stable across the table, with a majority of aesthetic scores remaining within a narrow range, confirming that none of the methods, except Text Embeddings, cause severe perceptual degradation. 

\begin{table}[t]
    \centering
    \caption{\textbf{Audio concept modulation with \ace.} Localized activation steering outperfroms prompt-, score-, and weight-level methods. Results are averaged over 9 musical concepts, steered in both positive and negative directions (see App.\ref{app:steering_benchmark} for details). \textbf{Best} and \uline{second} results are highlighted.}
    \label{tab:steering_auc_avg}
    \resizebox{1.0\linewidth}{!}{
        \begin{tabular}{lccccc}
            \toprule
            \multirow{2}{*}{\textbf{Method}}
            & \multicolumn{2}{c}{\textbf{AUC} ($\uparrow$)}
            & \multicolumn{2}{c}{\textbf{Smoothness} ($\downarrow$)}
            & \multirow{2}{*}{\shortstack{\textbf{ Audio}\\\textbf{Quality} ($\uparrow$)}} \\
            \cmidrule(lr){2-3}\cmidrule(lr){4-5}
            & MuQ & CLAP & MuQ & CLAP & \\
            \midrule
            \multicolumn{6}{l}{\textcolor{gray}{\textsc{\textbf{Prompt-level interventions}}}} \\
            PCI~\textcolor{gray}{(\citet{gorgun2026temporal})} & $0.084_{\pm.019}$ & $0.049_{\pm.010}$ & $0.185_{\pm.064}$ & $0.223_{\pm.066}$ & $\mathbf{6.803}_{\pm.016}$ \\
            Text Embeddings~\textcolor{gray}{(\citet{ezra2025freesliders0})} & $0.014_{\pm.007}$ & $0.011_{\pm.004}$ & $0.439_{\pm.224}$ & $0.681_{\pm.277}$ & $6.690_{\pm.027}$ \\
            Token Embeddings~\textcolor{gray}{(\citet{baumann2025continuous})} & $0.035_{\pm.011}$ & $0.024_{\pm.006}$ & $0.267_{\pm.063}$ & $0.211_{\pm.058}$ & $6.776_{\pm.023}$ \\
            \multicolumn{6}{l}{\textcolor{gray}{\textsc{\textbf{Score-space steering}}}} \\
            FreeSliders~\textcolor{gray}{(\citet{ezra2025freesliders0})} & $0.079_{\pm.012}$ & $0.049_{\pm.009}$ & \uline{$0.055$}$_{\pm.007}$ & $0.066_{\pm.012}$ & $6.790_{\pm.017}$ \\
            \multicolumn{6}{l}{\textcolor{gray}{\textsc{\textbf{Weight-space steering}}}} \\
            Concept Sliders~\textcolor{gray}{(\citet{gandikota2024concept})} & $0.086_{\pm.010}$ & \uline{$0.067$}$_{\pm.011}$ & $0.056_{\pm.006}$ & $\mathbf{0.045}_{\pm.006}$ & $6.762_{\pm.019}$ \\
            \multicolumn{6}{l}{\textcolor{gray}{\textsc{\textbf{Activation steering}}}} \\
            AUSteer~\textcolor{gray}{(\citet{feng2026finegrained})} & $0.081_{\pm.020}$ & $0.058_{\pm.013}$ & $0.212_{\pm.069}$ & $0.065_{\pm.011}$ & $6.758_{\pm.014}$ \\
            CAA~\textcolor{gray}{(\citet{rimsky-caa})} & $0.075_{\pm.016}$ & $0.046_{\pm.010}$ & $0.092_{\pm.034}$ & $0.071_{\pm.015}$ & $6.787_{\pm.016}$ \\
            \multicolumn{6}{l}{\textcolor{gray}{\textsc{\textbf{Localized steering (ours)}}}} \\
            AUSteer \textbf{(loc.)} & $0.096_{\pm.016}$ & $0.063_{\pm.012}$ & $\mathbf{0.054}_{\pm.006}$ & \uline{$0.050$}$_{\pm.005}$ & $6.775_{\pm.018}$ \\
            CAA \textbf{(loc.)} & \uline{$0.104$}$_{\pm.019}$ & $0.065_{\pm.011}$ & $0.063_{\pm.018}$ & $0.052_{\pm.010}$ & \uline{$6.792$}$_{\pm.017}$ \\
            SAE \textbf{(loc.)} & $\mathbf{0.118}_{\pm.019}$ & $\mathbf{0.070}_{\pm.013}$ & $0.057_{\pm.007}$ & $0.055_{\pm.004}$ & $6.770_{\pm.020}$ \\
            \bottomrule
        \end{tabular}
    }
\end{table}

\begin{figure}[t]
    \vspace{-0.65em}
    \centering
    \begin{subfigure}{0.3\linewidth}
        \centering
        \includegraphics[width=\linewidth]{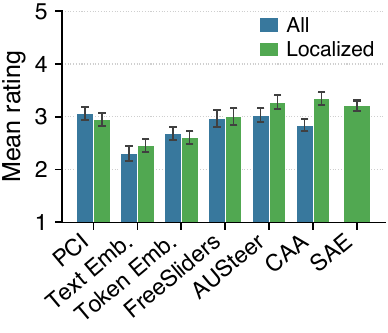}
        \vspace{-1.5em}
        \caption{\textbf{Seamless Edit}}
    \end{subfigure}
    \hfill
    \begin{subfigure}{0.3\linewidth}
        \centering
        \includegraphics[width=\linewidth]{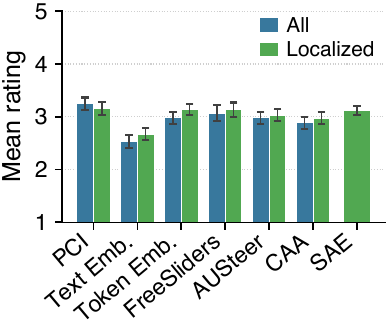}
        \vspace{-1.5em}
        \caption{\textbf{Audio Quality}}
    \end{subfigure}
    \hfill
    \begin{subfigure}{0.3\linewidth}
        \centering
        \includegraphics[width=\linewidth]{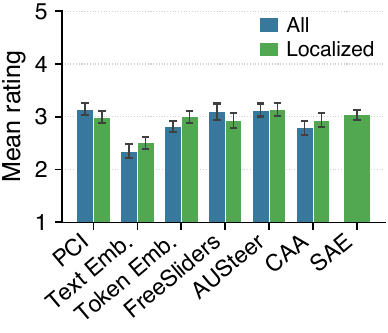}
        \vspace{-1.5em}
        \caption{\textbf{Smoothness}}
    \end{subfigure}
    \vspace{-0.6em} 
    \caption{\textbf{Human evaluation of audio steering methods.} We report mean Likert rating ($1$--$5$) for each method in the \textit{standard} (all layers) and \textit{localized} configurations across the three questions.}
    \label{fig:user_study_methods}
    \vspace{-1.6em}
\end{figure}

\vspace{-0.7em}
\paragraph{Human Evaluation.} As the objective metrics only loosely correlate with perceived musical quality, we additionally ran an extensive listening study. We collected $1279$ ratings from $32$ listeners over $15$ trials covering the \textit{piano}, \textit{female vocals}, and \textit{tempo} concepts (see App.~\ref{app:user_study}).
For each trial, we presented the seven steering methods (PCI, Text Emb., Token Emb., FreeSliders, AUSteer, CAA, SAE) in both the \textit{standard} (all layers) and the \textit{localized} (functional layers) configuration. We asked participants to rate clips on a $1$--$5$ Likert scale along: \emph{Seamless Edit} (was the concept introduced, and without breaking the original audio?), \emph{Audio Quality}, and \emph{Smoothness} (how gradual is the transition with $\alpha$?).
\vspace{-1.35em}

\cref{fig:user_study_methods} reports the mean rating per method. Methods that act on the cross-attention activations (CAA, SAE, AUSteer) consistently lead on \emph{Seamless Edit}, with the localized CAA reaching the highest rating ($3.32$) followed closely by SAE ($3.22$) and localized AUSteer ($3.24$). Embedding-level methods, particularly Text Emb., trail across all three dimensions, mirroring their lower Audio Quality scores in \cref{tab:steering_auc_avg}. Differences between methods are smaller on \emph{Audio Quality} and \emph{Smoothness}, suggesting that listeners perceive most interventions as comparably clean once the concept is properly introduced.
\vspace{-0.2em}

\subsection{Does localization improve all steering methods?}
\vspace{-0.3em}

\paragraph{Steering inside vs.\ outside the bottleneck.}
We test our bottleneck hypothesis, using three architecturally distinct backbones: \ldm (U-Net), \stableaudio (diffusion transformer), and ACE-Step (flow transformer), with CAA as a representative activation-steering method. We compare three intervention sites: \emph{all layers}, the \emph{functional layers only}, and an \emph{ablation} setup that steers every layer except the functional ones ($57/64$ in AudioLDM2, $21/24$ in Stable Audio, and $22/24$ in Ace).

\clearpage

\begin{figure}[ht]
    \centering
    \includegraphics[width=1.0\linewidth]{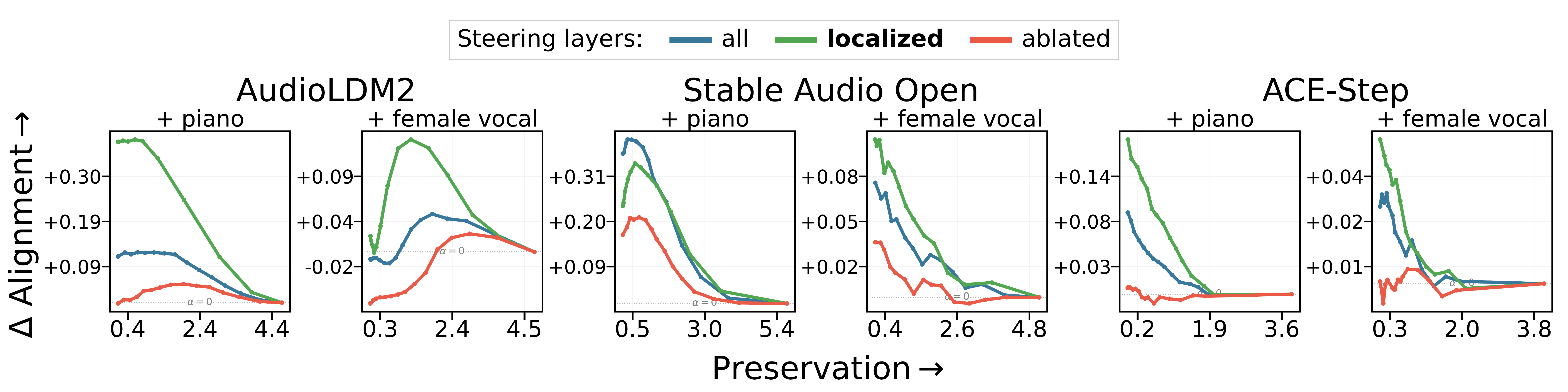}
    \vspace{-1.7em}
    \caption{\textbf{Localized activation steering outperforms the global approach.} For the same audio preservation level, intervening in functional layers offers higher gain in the alignment. Additionally, omitting functional layers for steering often yields no conceptual gain.}
    \label{fig:loc_steering}
    \vspace{-1.2em}
\end{figure}

As visible in \cref{fig:loc_steering}, across all concepts, localized steering achieves the highest alignment at matched preservation level, while the ablation curve flattens close to the unsteered baseline or leads to negative results (female vocal). This indicates that our identified subset acts as a true semantic bottleneck. The residual capacity outside it is insufficient to introduce the target concept on its own. We provide quantitative results for this experiment in App.~\ref{app:loc_steering_tables}.
\vspace{-1em}

\paragraph{Localization across steering approaches.} So far, we have established that localization is the right intervention site for activation steering. \cref{tab:steering_auc_localization_compact} extends this analysis to other steering paradigms. We report the relative gain (\%) achieved by methods when their intervention is restricted to localized layers, with green and red shading marking improvements and regressions. 
\vspace{-0.3em}

Localization is not a universal improvement but interacts strongly with the steering paradigm. Activation-space methods benefit substantially without degrading quality: CAA gains (MuQ/CLAP) $39\%/40\%$ in AUC and $32\%/26\%$ in Smoothness. Notably, constraining AUSteer to our layers significantly improves its Effectiveness ($19\%/8\%$) and Smoothness ($74\%/22\%$), indicating that its original discriminative localization mechanism prioritizes different layers than those identified by activation patching, potentially leading to suboptimal results.  Conversely, weight-space steering degrades sharply with localization: Concept Sliders lose $21\%$/$25\%$ of AUC and $77\%$/$34\%$ on Smoothness. We attribute this behavior to Concept Sliders' methodology of adding new mechanisms to the existing model rather than leveraging its intrinsic capabilities. 

\providecolor{gainpos}{HTML}{51A851}   
\providecolor{gainneg}{HTML}{EA5A47}   
\providecommand{\dpos}[1]{\textcolor{gainpos}{\textbf{$#1$}}}
\providecommand{\dneg}[1]{\textcolor{gainneg}{\textbf{$#1$}}}
\providecommand{\dposf}[1]{\textcolor{gainpos!35}{\textbf{$#1$}}}
\providecommand{\dnegf}[1]{\textcolor{gainneg!35}{\textbf{$#1$}}}

\begin{table}[h]
    \vspace{-1.2em}
    \centering
    \caption{\textbf{Effect of layer localization on each steering method.} We compare the methods' global variant (all layers steering) against the localized one. Gain from localization is expressed as a \% change of the mean: \textcolor{gainpos}{\textbf{improvement}} or \textcolor{gainneg}{\textbf{degradation}} of steering quality, faded when the absolute gain is below the standard error of the difference. We provide full results in App.~\ref{app:subsection_methods_comparison}.}
    \label{tab:steering_auc_localization_compact}
    \resizebox{1.0\linewidth}{!}{
        \begin{tabular}{lccccc}
            \toprule
            \multirow{2}{*}{\textbf{Method}}
            & \multicolumn{2}{c}{\textbf{AUC} ($\uparrow$)}
            & \multicolumn{2}{c}{\textbf{Smoothness} ($\downarrow$)}
            & \multirow{2}{*}{\shortstack{\textbf{ Audio}\\\textbf{Quality} ($\uparrow$)}} \\
            \cmidrule(lr){2-3}\cmidrule(lr){4-5}
            & MuQ & CLAP & MuQ & CLAP & \\
            \midrule
            PCI~\textcolor{gray}{(\citet{gorgun2026temporal})} & $0.084_{\pm.019}$ & $0.049_{\pm.010}$ & $0.185_{\pm.064}$ & $0.223_{\pm.066}$ & $6.803_{\pm.016}$ \\
            \quad $\Delta$  $\text{(localization)}$ & \dneg{-19\%} & \dneg{-22\%} & \dneg{-50\%} & \dpos{+21\%} & \dneg{-0.2\%} \\
            \midrule
            Text Embeddings~\textcolor{gray}{(\citet{ezra2025freesliders0})} & $0.014_{\pm.007}$ & $0.011_{\pm.004}$ & $0.439_{\pm.224}$ & $0.681_{\pm.277}$ & $6.690_{\pm.027}$ \\
            \quad $\Delta$  $\text{(localization)}$ & \dpos{+62\%} & \dpos{+31\%} & \dpos{+25\%} & \dposf{+48\%} & \dpos{+1.4\%} \\
            \midrule
            Token Embeddings~\textcolor{gray}{(\citet{baumann2025continuous})} & $0.035_{\pm.011}$ & $0.024_{\pm.006}$ & $0.267_{\pm.063}$ & $0.211_{\pm.058}$ & $6.776_{\pm.023}$ \\
            \quad $\Delta$  $\text{(localization)}$ & \dnegf{-8\%} & \dneg{-20\%} & \dposf{+18\%} & \dneg{-33\%} & \dpos{+0.3\%} \\
            \midrule
            FreeSliders~\textcolor{gray}{(\citet{ezra2025freesliders0})} & $0.079_{\pm.012}$ & $0.049_{\pm.009}$ & $0.055_{\pm.007}$ & $0.066_{\pm.012}$ & $6.790_{\pm.017}$ \\
            \quad $\Delta$  $\text{(localization)}$ & \dposf{+1\%} & \dnegf{-1\%} & \dneg{-18\%} & \dneg{-13\%} & \dnegf{-0.0\%} \\
            \midrule
            Concept Sliders~\textcolor{gray}{(\citet{gandikota2024concept})} & $0.086_{\pm.010}$ & $0.067_{\pm.011}$ & $0.056_{\pm.006}$ & $0.045_{\pm.006}$ & $6.762_{\pm.019}$ \\
            \quad $\Delta$  $\text{(localization)}$ & \dneg{-21\%} & \dneg{-25\%} & \dneg{-77\%} & \dneg{-34\%} & \dpos{+0.2\%} \\
            \midrule
            AUSteer~\textcolor{gray}{(\citet{feng2026finegrained})} & $0.081_{\pm.020}$ & $0.058_{\pm.013}$ & $0.212_{\pm.069}$ & $0.065_{\pm.011}$ & $6.758_{\pm.014}$ \\
            \quad $\Delta$  $\text{(localization)}$ & \dpos{+19\%} & \dposf{+8\%} & \dpos{+74\%} & \dpos{+22\%} & \dpos{+0.3\%} \\
            \midrule
            CAA~\textcolor{gray}{(\citet{rimsky-caa})} & $0.075_{\pm.016}$ & $0.046_{\pm.010}$ & $0.092_{\pm.034}$ & $0.071_{\pm.015}$ & $6.787_{\pm.016}$ \\
            \quad $\Delta$  $\text{(localization)}$ & \dpos{+39\%} & \dpos{+40\%} & \dpos{+32\%} & \dpos{+26\%} & \dpos{+0.1\%} \\
            \bottomrule
        \end{tabular}
    }
    \vspace{-0.5em}
\end{table}

\begin{figure}[h]
    \centering
    \begin{subfigure}{0.32\linewidth}
        \centering
        \includegraphics[width=\linewidth]{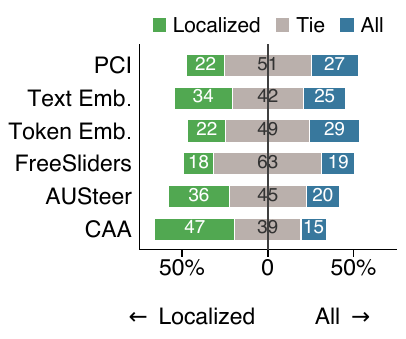}
        \vspace{-1.5em}
        \caption{\textbf{Seamless Edit}}
    \end{subfigure}
    \hfill
    \begin{subfigure}{0.32\linewidth}
        \centering
        \includegraphics[width=\linewidth]{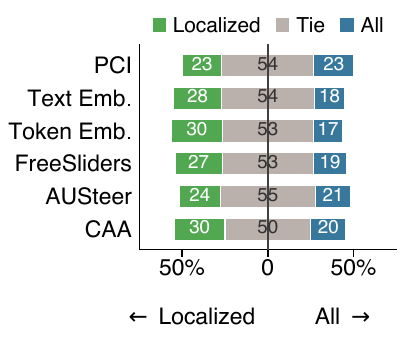}
        \vspace{-1.5em}
        \caption{\textbf{Audio Quality}}
    \end{subfigure}
    \hfill
    \begin{subfigure}{0.32\linewidth}
        \centering
        \includegraphics[width=\linewidth]{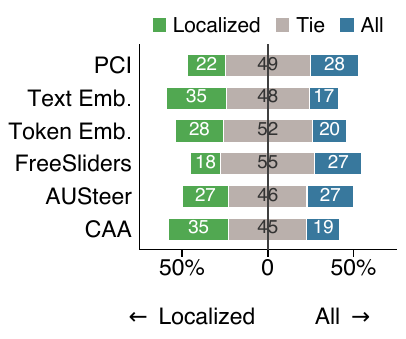}
        \vspace{-1.5em}
        \caption{\textbf{Smoothness}}
    \end{subfigure}
        \vspace{-0.6em}
    \caption{\textbf{Human preference for localized vs.\ global setups of steering methods.} We compare user study ratings of both configurations for the same method. Bars present pairs in which the localized variant is preferred (green), the two are tied (grey), or the global steering is preferred (blue).}
    \label{fig:user_study_winrate}
    \vspace{-1em}
\end{figure}

Our listening study (\cref{fig:user_study_winrate}) corroborates these trends in the pairwise perceptual preference test. For \emph{Seamless Edit}, restricting steering to the functional layers yields a clear advantage for the activation-space methods: localized CAA is preferred over the global one ($47\%$ versus $15\%$), as well as for AUSteer and TextEmb. In \emph{Audio Quality}, the localized variant is preferred or tied for every method. On \emph{Smoothness}, the two configurations are statistically close for most methods, with CAA again favoring the localized version. 

\subsection{Multi-concept steering}\label{sec:multi-concept-steering}

In our final experiment, we investigate whether narrowing down the intervention scope to the semantic bottleneck can improve audio modulation when steering the model in multiple directions simultaneously. For example, such an intervention can add two new instruments to a song or manipulate vocal gender while decreasing tempo. In this setting, we compare three activation steering methods: CAA and AUSteer in \textit{localized} and \textit{global} configurations, as well as SAE. In all three cases, the combined vector is obtained by uniformly weighted summation of the single-direction steering vectors, optionally negated for concept-suppressing directions.

Results in \cref{tab:multi-concept-auc-muq} show that our localization extends to the multi-concept steering setting. Across $9$ concept combinations (5 pairs and 4 triples), the three localized methods outperform global steering. This suggests that applying combined steering vectors at every layer accumulates collateral changes that mask the intended ones. Within the localized group, SAE is the most consistent performer (with $5/9$ wins), ahead of localized CAA (with $3/9$ wins). Even when including negated directions in vector summation, restricting the intervention to the semantic bottleneck preserves a clear steering semantics, whereas spreading it across the entire stack substantially weakens it.

\begin{table}[h]
\centering
\small
\caption{\textbf{Multi-concept steering performance measured with average Area Under LPAPS-MUQ Curve.} Best result per combination of concepts is \textbf{bolded}. Single column denotes steering on combinations of concepts, where \textbf{P} = Piano, \textbf{V} = Violin, \textbf{FV} = Female Vocal, \textbf{A} = Acoustic Guitar, \textbf{J} = Jazz Music, \textbf{M} = Brighter Mood, $\overline{\text{\textbf{FV}}}$ = Male Vocal, $\overline{\text{\textbf{T}}}$ = Slower Tempo, and $\overline{\text{\textbf{M}}}$ = Darker Mood. In App.~\ref{app:multi_concept_steering}, we provide more details regarding the experiment.}
\label{tab:multi-concept-auc-muq}
\resizebox{\linewidth}{!}{%
\begin{tabular}{lccccccccc}
\toprule
\multirow{2}{*}{\textbf{Method}} & \multicolumn{5}{c}{Steering Two Concepts} & \multicolumn{4}{c}{Steering Three Concepts} \\
\cmidrule(lr){2-6}\cmidrule(lr){7-10}
& \textbf{P\,+\,V} & \textbf{P\,+\,FV} & \textbf{A\,+\,FV} & \textbf{P\,+\,$\overline{\text{FV}}$} & \textbf{J\,+\,$\overline{\text{T}}$} & \textbf{P\,+\,V\,+\,J} & \textbf{A\,+\,FV\,+\,M} & \textbf{P\,+\,V\,+\,$\overline{\text{T}}$} & \textbf{A\,+\,$\overline{\text{FV}}$\,+\,$\overline{\text{M}}$} \\
\midrule
AUSteer & +0.057 & +0.001 & +0.022 & +0.013 & +0.090 & +0.062 & +0.005 & +0.048 & +0.045 \\
CAA & +0.048 & +0.011 & +0.028 & +0.008 & +0.068 & +0.047 & +0.026 & +0.029 & +0.032 \\
\midrule
\textbf{AUSteer (loc.)} & +0.099 & +0.018 & +0.029 & \textbf{+0.039} & +0.086 & +0.088 & +0.021 & +0.069 & +0.051 \\
\textbf{CAA (loc.)} & +0.094 & \textbf{+0.034} & \textbf{+0.052} & +0.027 & +0.103 & +0.081 & \textbf{+0.035} & +0.061 & +0.039 \\
\textbf{SAE (loc.)} & \textbf{+0.114} & +0.027 & +0.041 & +0.034 & \textbf{+0.110} & \textbf{+0.100} & +0.018 & \textbf{+0.081} & \textbf{+0.063} \\
\bottomrule
\end{tabular}
}
\end{table}
\vspace{-0.3em}
\section{Conclusions}
In this work, we demonstrated that audio diffusion models exhibit a semantic bottleneck -- a small subset of attention layers that govern high-level musical concepts. By identifying this bottleneck via activation patching, we systematically benchmarked activation steering against prompt-, score-, and weight-based baselines, evaluating all paradigms within our localized framework. Supported by both objective metrics and a comprehensive listening study, we showed that restricting activation steering (using AUSteer, Contrastive Activation Addition, and Sparse Autoencoders) to these functional layers outperforms both global baselines and other paradigms in terms of precision and fidelity. This layer-specific approach offers a robust method for fine-grained musical control, overcoming the limitations of standard text prompting.

\bibliographystyle{abbrvnat}
\bibliography{bibliography}


\clearpage
\appendix
\startcontents[app]
\begin{center}
\LARGE\bfseries Appendix
\end{center}
 {
  \renewcommand{\addvspace}[1]{} 
  \printcontents[app]{l}{1}
  {
      \setcounter{tocdepth}{2}
      \section*{\Large Table of contents}\vspace{-0.1em}
  }
 }

\newpage

\section{Limitations}\label{app:limitations}

\paragraph{Architectural scope.} Our localization study and findings about the existence of "semantic bottleneck" are limited to audio diffusion architectures: \ldm~\citep{liu2024audioldm}, \stableaudio~\citep{evans2024stable}, and \ace~\citep{gong2025ace0step0}. We believe a similar study could be conducted on Music-LLMs, such as MusicGen. The evaluation protocol we propose can be easily applied to benchmark audio modulation with autoregressive text-to-music models. We further note that our comparison is conducted on a single model, \ace. However, in \cref{fig:loc_steering} and App.~\ref{app:loc_steering_tables}, we show that our central finding -- restricting the intervention to the semantic bottleneck drives effective steering -- generalizes to all three models. 

\paragraph{Intervention site.} We localize functional components at the level of cross-attention layers. The rationale is that those modules are the only ones that incorporate text conditioning into generated audio. We do not study activations in residual streams, self-attention, or feed-forward layers and leave this for follow-up work. We believe that the proposed methodology for localization and steering can be easily adapted to other layer types.

\paragraph{Concept coverage.} Our steering benchmark considers nine directions: piano, violin, acoustic/electric guitar, mood, tempo, vocal gender, vocal style, jazz/rock, and classical/electronic genres. The localization analysis covers additional instruments (drums, flute, maracas, trumpet), and genres (jazz, techno, reggae). Together, they span the dimensions commonly used in music-generation evaluation, but remain a finite slice of musical structure. We emphasize that our choice of concept depends on the generative model's intrinsic capabilities, as it is impossible to perform text-based localization of concepts that the model cannot generate when prompted.

\paragraph{Evaluation metrics.} Across steering paradigms, we calibrate the maximum steering strength using an LPAPS preservation cutoff. Hence, the reported AUC values inherit the choice of LPAPS as a perceptual distance measurement. In App.~\ref{app:preservation_metrics_section}, we show that our findings generalize to alternative preservation metrics. Additionally, to measure the presence of target concepts in generated audio, we employ CLAP~\citep{wu2022largescale} and MuQ~\citep{zhu2025muq}. We emphasize that such models may exhibit their own biases and rate certain features more confidently than others. To ensure robust evaluation, during experiments, we report both (CLAP, MuQ) metrics values, include a user study, and, in the App.~\ref{sec:external_eval}, show that our evaluation protocol is aligned with external music evaluators, such as tempo detectors or vocal gender classifiers. We note, however, that inference-time steering applied with excessive strength can push activations off the training manifold, degrading generation quality. In App.~\ref{app:quality_over_strength} we analyze this effect and show that, within the calibrated steering range, quality remains stable, with localized steering avoiding the monotone degradation exhibited by global interventions.

\section{Broader impact}\label{app:broader_impact}

Our work targets a long-standing limitation of audio generators that rely on prompting. Localized activation steering provides a continuous control surface without changing the underlying composition, which is the kind of fine-grained adjustment musicians or composers may need in practice.

We demonstrate that semantic musical attributes are concentrated in a small, shared subset of layers. This advances scientific understanding of audio diffusion models, which until now have largely been treated as black boxes, providing a foundation for downstream interpretability work.

The capability our work adds is fine-grained \emph{control} over existing models, not new generative capacity. The standard concerns associated with text-to-music systems, such as training-data copyright, may apply to the underlying generators we analyze, but are not amplified by our approach.

\section{Future work}\label{app:future_work}
We discover semantic bottlenecks independently in three text-to-music models, showing that a narrow, contiguous band of functional layers recurs across diverse architectures. While we observe this structural regularity, we do not study whether layer indices can be transferred across architectures. An open question is whether the bottlenecks of different models are representationally compatible. A potential future direction is to study the stitching \citep{model_stitching} of the semantic bottlenecks across audio DMs to test the transfer of steering vectors or layers.

\newpage

\section{Compute resources}\label{app:compute_resources}

For the experiments, we used a scientific cluster consisting of 110 nodes with CrayOS operating system. Each node is powered by 288 CPU cores, stemming from 4 NVIDIA Grace processors, each with 72 cores and a clock speed of 3.1 GHz. The nodes are equipped with substantial memory, featuring 480 GB of RAM per node. For GPU acceleration, each node in the cluster consists of 4 NVIDIA GH200 96GB GPUs with 120 GB of RAM and 72 CPUs per GPU.

\section{Dataset construction for layer localization}\label{app:prompt_examples}

\cref{tab:prompt_examples} presents examples of counterfactual prompt pairs used in our localization experiments. Each pair consists of an original prompt $\mathcal{P}_c$ from the MusicCaps dataset and a modified prompt $\mathcal{P}_{\tilde{c}}$ where concept-associated terms are replaced with their counterparts. \cref{tab:concept_keywords} lists the keywords used to filter MusicCaps captions for each concept and the corresponding replacement terms used to generate counterfactual prompts.

\begin{table}[h]
    \centering
    \small
    \caption{\textbf{Examples of counterfactual prompt pairs.} For each concept category, we show an original prompt from MusicCaps and its counterfactual version with the target concept replaced. Modified terms are highlighted in \textbf{bold}.}
    \vspace{0.5em}
    \label{tab:prompt_examples}
    \begin{tabular}{p{1.3cm}p{5.5cm}p{5.5cm}}
        \toprule
        \textbf{Concept} & \textbf{Original Prompt} $\mathcal{P}_c$                                                                                                                                                                                                                          & \textbf{Counterfactual Prompt} $\mathcal{P}_{\tilde{c}}$                                                                                                                                                                                                             \\
        \hline
        Vocal Gender                  & \textit{"The low quality recording features a ballad song that contains sustained strings, mellow piano melody, and soft \textbf{female} vocal singing over it. It sounds sad and soulful."}                                                                       & \textit{"The low quality recording features a ballad song that contains sustained strings, mellow piano melody, and soft \textbf{male} vocal singing over it. It sounds sad and soulful."}                                                                            \\
        \hline
        Tempo  & \textit{"This is a country music piece. There is a fiddle playing the main melody. The acoustic guitar and electric guitar are playing gently. The song has a \textbf{slow} tempo. The atmosphere is sentimental."}                                               & \textit{"This is a country music piece. There is a fiddle playing the main melody. The acoustic guitar and electric guitar are playing gently. The song has a \textbf{fast} tempo. The atmosphere is sentimental."}                                                  \\
        \hline
        Mood & \textit{"This is a Hindustani classical music piece. There is a harmonium playing the main tune. A bansuri joins in to play, supporting a melody. The rhythmic background consists of tabla percussion and electronic drums. The atmosphere is \textbf{joyful}."} & \textit{"This is a Hindustani classical music piece. There is a harmonium playing the main tune. A bansuri joins in to play, supporting a melody. The rhythmic background consists of tabla percussion and electronic drums. The atmosphere is \textbf{sorrowful}."} \\
        \hline
        Instrument & \textit{"This folk song features a male voice singing the main melody in an emotional mood. This is accompanied by an accordion playing fills in the background. A \textbf{violin} plays a droning melody."}                                                      & \textit{"This folk song features a male voice singing the main melody in an emotional mood. This is accompanied by an accordion playing fills in the background. A \textbf{trumpet} plays a droning melody."}                                                        \\
        \hline
        Genre & \textit{"The low quality recording features a \textbf{reggae}/dub song that consists of a flat male vocal singing over punchy 808 bass, punchy snare, shimmering hi hats and groovy piano chords. It sounds energetic, groovy and the recording is noisy and in mono."}    & \textit{"The low quality recording features a \textbf{metal}/dub song that consists of a flat male vocal singing over punchy 808 bass, punchy snare, shimmering hi hats and groovy piano chords. It sounds energetic, groovy and the recording is noisy and in mono."}         \\
        \bottomrule
    \end{tabular}
\end{table}

\begin{table}[h]
    \centering
    \small
    \caption{\textbf{Keywords for dataset construction.} For each concept, we list the examples of keywords proposed by an LLM (GPT-4o) to filter captions from MusicCaps (for a given concept, we select prompts containing these terms) and the replacement keywords used to generate counterfactual prompts.}
    \label{tab:concept_keywords}
    \vspace{0.5em}
    \begin{tabular}{m{1.6cm}m{6.4cm}m{4cm}}
        \toprule
        \textbf{Concept} & \textbf{Example Filter Keywords} & \textbf{Example Replacements} \\
        \midrule
        \textbf{female}& female, woman, girlish                                 & female $\rightarrow$ male \newline woman $\rightarrow$ man \newline girlish $\rightarrow$ boyish \\
        \midrule
        \textbf{male}& male, man                                              & man $\rightarrow$ woman \newline his $\rightarrow$ her \newline himself $\rightarrow$ herself \\
        \midrule
        \textbf{slow}& slow, slowed, slower, slowly, slow-paced, leisurely    & slow $\rightarrow$ fast \newline slower $\rightarrow$ faster \newline slow-paced $\rightarrow$ fast-paced \\
        \midrule
        \textbf{fast}& fast, rapid, dynamic, quick, energetic, groovy, lively & fast $\rightarrow$ slow \newline quickly $\rightarrow$ slowly \newline lively $\rightarrow$ slowly \\
        \midrule
        \textbf{happy}& happy, joyful, upbeat, uplifting, cheerful             & joyful $\rightarrow$ sorrowful \newline cheerful $\rightarrow$ gloomy \newline warm $\rightarrow$ cold \\
        \midrule
        \textbf{sad}& sad, depressing, dejected                              & sad $\rightarrow$ happy \newline lonely $\rightarrow$ connected \newline emotional $\rightarrow$ lighthearted \\
        \midrule
        \textbf{drums}& drums, drum, percussion                                & drums $\rightarrow$ saxophone \newline percussionist $\rightarrow$ saxophonist \\
        \midrule
        \textbf{flute}& flute, flutes                                          & flute $\rightarrow$ maracas \newline flutist $\rightarrow$ maracas player \\
        \midrule
        \textbf{maracas}& maracas, maraca                                        & maracas $\rightarrow$ flute \newline maraca $\rightarrow$ flute \\
        \midrule
        \textbf{trumpet}& trumpet, trumpets                                      & trumpet $\rightarrow$ violin \newline trumpeter $\rightarrow$ violinist \\
        \midrule
        \textbf{violin}& violin, fiddle                                         & violin $\rightarrow$ trumpet \newline string $\rightarrow$ $\emptyset$ \\
        \midrule
        \textbf{jazz}& jazz, blues                                            & jazz $\rightarrow$ opera \newline band $\rightarrow$ choir \\
        \midrule
        \textbf{reggae}& reggae, jamaican, tropical, beach                      & reggae $\rightarrow$ metal \newline chill $\rightarrow$ heavy \newline beach $\rightarrow$ concert \\
        \bottomrule
    \end{tabular}
\end{table}

\clearpage
\section{Layer localization}
\label{app:localization}

In this section, we describe our localization experiment and its results in detail. In \ref{app:sub_loc_heatmaps}, we provide full results for all three architectures. In \ref{app:loc_param_analysis}, we analyze how the choice of prompts and the selection threshold $\tau$ affect the results. Finally, in \ref{app:loc_cost}, we quantify the computational cost of the localization experiment.

\subsection{Per-concept impact maps}
\label{app:sub_loc_heatmaps}

\begin{figure}[ht]
    \centering
    \includegraphics[width=\linewidth]{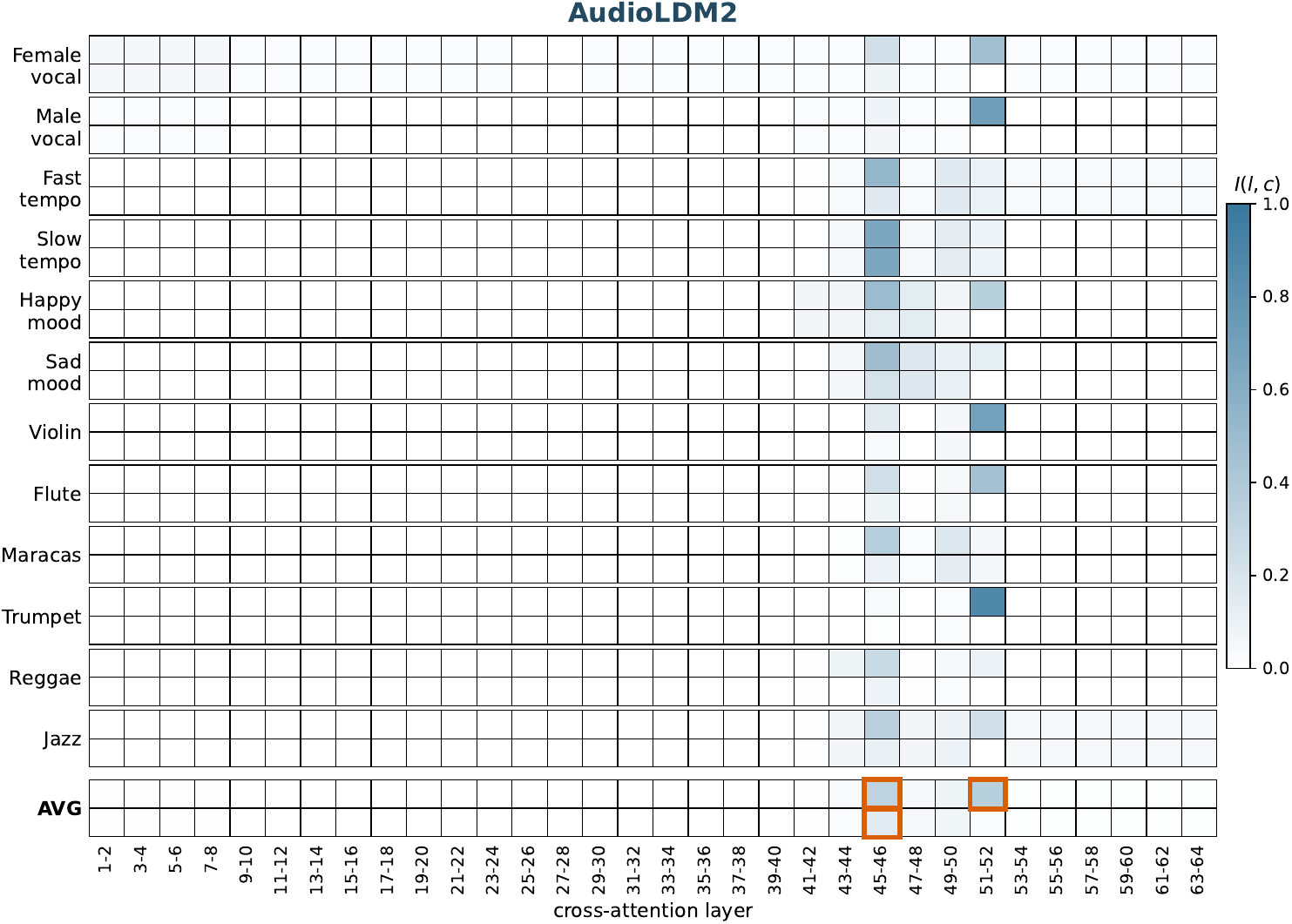}
    \caption{\textbf{Impact of every cross-attention layer on every concept in \ldm
    \citep{liu2024audioldm}.} Each cell color intensity denotes layer $l$ impact on concept $c$ as $\text{I}(l,c)$ (\cref{eq:impact}). Last row denotes the aggregated impact $\text{I}(l)$ of layer $l$ averaged over all concepts. The final selected layers (where $\text{I}(l) \geq \tau=0.10$) are outlined in orange.}
    \label{fig:layers_audioldm2}
\end{figure}

\begin{figure}[ht]
    \centering
    \includegraphics[width=\linewidth]{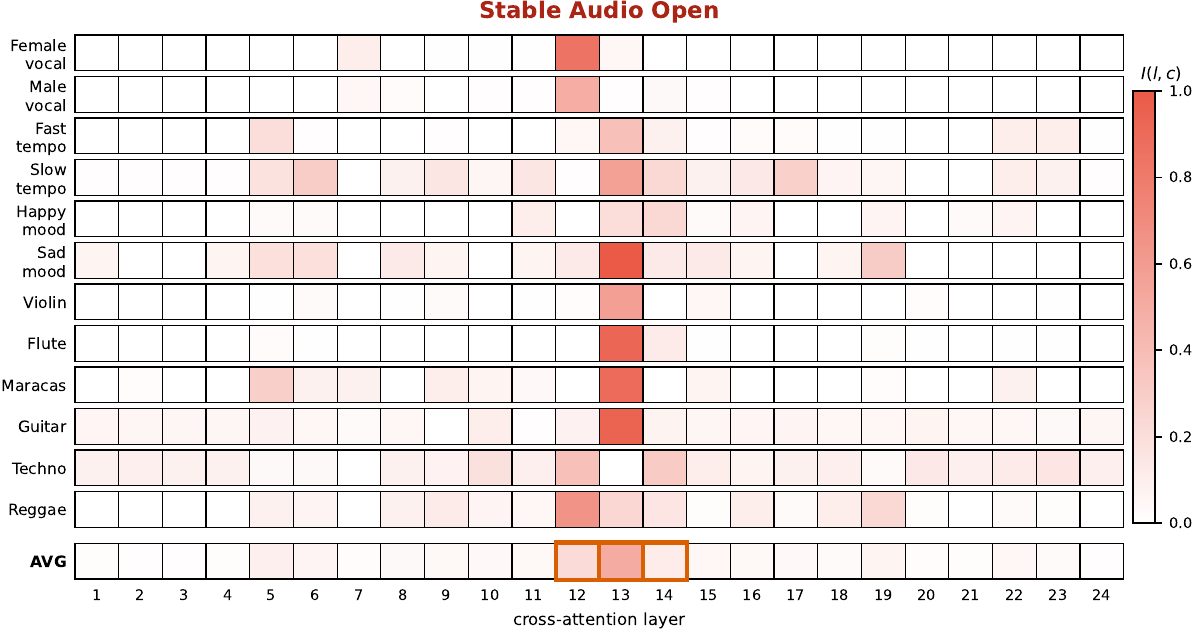}
    \caption{\textbf{Impact of every cross-attention layer on every concept in
    \stableaudio \citep{evans2024stable}.} Each cell color intensity denotes layer $l$ impact on concept $c$ as $\text{I}(l,c)$ (\cref{eq:impact}). Last row denotes the aggregated impact $\text{I}(l)$ of layer $l$ averaged over all concepts. The final selected layers (where $\text{I}(l) \geq \tau=0.10$) are outlined in orange.}
    \label{fig:layers_stableaudio}
\end{figure}

\begin{figure}[ht]
    \centering
    \includegraphics[width=\linewidth]{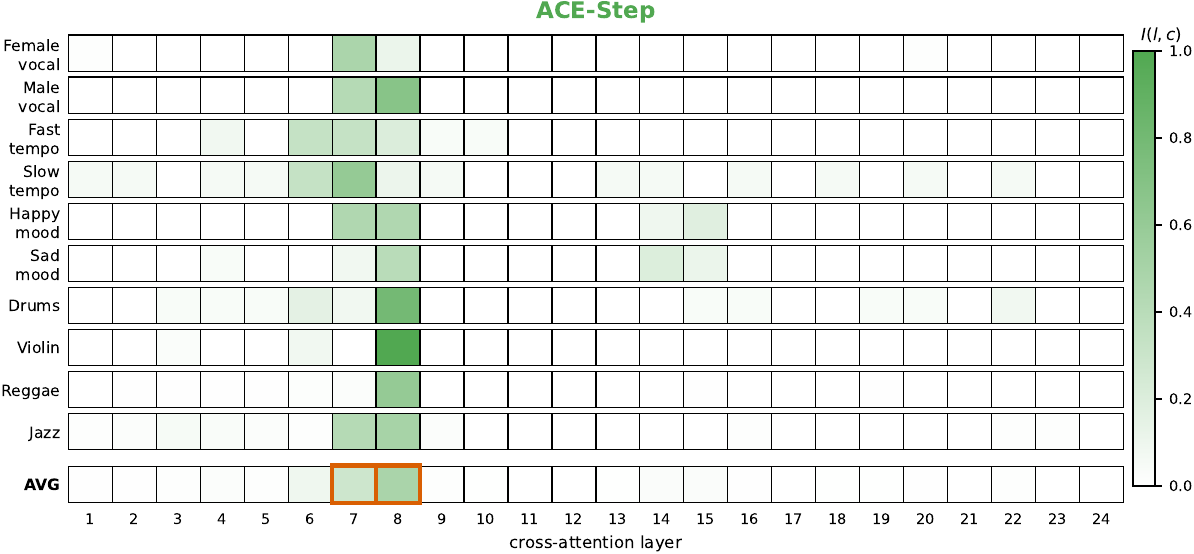}
    \caption{\textbf{Impact of every cross-attention layer on every concept in \ace
    \citep{gong2025ace0step0}.} Each cell color intensity denotes layer $l$ impact on concept $c$ as $\text{I}(l,c)$ (\cref{eq:impact}). Last row denotes the aggregated impact $\text{I}(l)$ of layer $l$ averaged over all concepts. The final selected layers (where $\text{I}(l) \geq \tau=0.10$) are outlined in orange.}
    \label{fig:layers_ace}
\end{figure}

\cref{fig:layers_audioldm2,fig:layers_stableaudio,fig:layers_ace} present the impact scores $\text{I}(l,c)$ (see \cref{eq:impact}) of every cross-attention layer on probed concepts across all three architectures: \ldm, \stableaudio, and \ace. In the figures, we provide the aggregated (across concepts) impact scores $\text{I}(l)=\frac{1}{|C|}\sum_{c}$, and highlight (orange) the layers that are selected to carry the control, i.e., the layers that score above the selection threshold $\tau$.

Score maps make the \textit{semantic bottleneck} visible per concept. In \ldm (\cref{fig:layers_audioldm2}), control over vocal gender, tempo, mood, instruments and genres concentrates in layers 45--51 (out of 64), in \stableaudio (\cref{fig:layers_stableaudio}) in layers 12--14 (out of 24), and in \ace (\cref{fig:layers_ace}) in layers 7--8 (out of 24). The aggregate row is strongly peaked across all architectures. In \ldm: $\text{I}(\texttt{attn51})=0.36$, $\text{I}(\texttt{attn45})=0.33$, $\text{I}(\texttt{attn46})=0.15$, while other score below $0.09$. In \stableaudio: $\text{I}(\texttt{attn13})=0.51$, $\text{I}(\texttt{attn12})=0.22$, and $\text{I}(\texttt{attn14})=0.12$, while other score below $0.10$. In \ace: $\text{I}(\texttt{attn8})=0.49$, $\text{I}(\texttt{attn7})=0.29$, and $\text{I}(\texttt{attn6})=0.09$, while other score below $0.04$.

\clearpage

\subsection{Parameter analysis} \label{app:loc_param_analysis}

\paragraph{Prompt count.}

In our experiments, to ensure robust results, we run the localization experiment on up to 256 counterfactual prompt pairs per concept (230 on average).

In \cref{tab:loc_prompts}, we show that exact localization for ACE-Step is achievable with far fewer prompts. We report the Jaccard similarity between the sets of layers localized using the entire (full-budget) tracing dataset and a random subset of it. As shown, ACE's semantic bottleneck $\{7,8\}$ can be recovered exactly using only $34$ prompt pairs per concept.

\begin{table}[h]
\centering
\caption{\textbf{Impact of number of prompts on ACE-Step localization results.} $J$(top-j) denotes Jaccard similarity between sets of $j$ highest scoring layers selected by full-budget localization and a random subset of prompts. Results are averaged across 200 random subsamples.}
\label{tab:loc_prompts}
\begin{tabular}{llccccc}
\toprule
\textbf{Prompts / concept} & \textbf{Budget \%} &
$J$\textbf{(top-1)} & $J$\textbf{(top-2)} & $J$\textbf{(top-3)} & $J$\textbf{(top-5)} \\
\midrule
4   & 2\%    & 0.54 & 0.82 & 0.66 & 0.45 \\
12  & 5\%    & 0.65 & 0.98 & 0.81 & 0.56 \\
17  & 8\%    & 0.67 & 0.99 & 0.90 & 0.63 \\
23  & 10\%   & 0.60 & 0.99 & 0.92 & 0.67 \\
34  & 15\%   & 0.65 & \textbf{1.00} & 0.97 & 0.74 \\
46  & 20\%   & 0.69 & \textbf{1.00} & 0.99 & 0.80 \\
69  & 30\%   & 0.72 & \textbf{1.00} & \textbf{1.00} & 0.87 \\
115 & 50\%   & 0.89 & \textbf{1.00} & \textbf{1.00} & 0.98 \\
230 & 100\%  & \textbf{1.00} & \textbf{1.00} & \textbf{1.00} & \textbf{1.00} \\
\bottomrule
\end{tabular}
\end{table}

\paragraph{Threshold sensitivity.}

In \cref{tab:loc_tau}, we analyze the impact of the selection threshold $\tau$ on localization results. We show that \ace selection $\{7,8\}$ is unchanged over $\tau\in[0.10,0.20]$, and \ldm keeps layers
45 and 51 over the whole range. In our experiments, we set $\tau=0.10$.

\begin{table}[h]
\centering
\small
\caption{\textbf{Selected semantic bottleneck in Audio Diffusion Models as a function of selection threshold $\tau$}. Setting used in our experiments is \textbf{bolded}.}
\label{tab:loc_tau}
\begin{tabular}{lccc}
\toprule
$\tau$ & \textbf{\ldm} & \textbf{\stableaudio} & \textbf{\ace} \\
\midrule
0.05  & 45, 46, 49, 50, 51 & 5, 6, 12, 13, 14, 19 & 6, 7, 8 \\
0.075 & 45, 46, 49, 50, 51 & 5, 12, 13, 14        & 6, 7, 8 \\
\textbf{0.10} & \textbf{45, 46, 51} & \textbf{12, 13, 14} & \textbf{7, 8} \\
0.15  & 45, 51             & 12, 13               & 7, 8 \\
0.20  & 45, 51             & 12, 13               & 7, 8 \\
0.30  & 45, 51             & 13                   & 8 \\
\bottomrule
\end{tabular}
\end{table}

\subsection{Computational cost}
\label{app:loc_cost}

Localizing $C$ concepts in a model with $L$ cross-attention layers costs
\begin{equation}\label{eq:loc_cost}
    N_{\text{gen}} = C \times (L + 2) \times P \times S
\end{equation}
generations, where $P$ is the number of counterfactual prompt pairs per concept, and $S$ is the number of noise seeds per prompt (in our experiments, we set $S=8$). For \ace, where $L=24$, $C=10$, and $P=34$, localization involves $70.720$ generations. On the NVIDIA GH200 GPU, where, on average (with batch size $32$), a single generation takes $1.68$ GPU seconds, the entire localization process takes roughly $33$ GPU hours.

\clearpage
\section{Prompts: steering vectors construction and evaluation}\label{app:steering_prompts}

This section details the prompts used in the steering experiments.

\subsection{Contrastive prompts for steering vectors and SAE features}

To compute steering vectors and identify concept-specific SAE features, we generate audio from contrastive prompt pairs. For each concept, we construct positive prompts $\mathcal{P}_c$ containing the target attribute and negative prompts $\mathcal{P}_{\tilde{c}}$ with the contrasting attribute. \cref{tab:steering_prompts} shows the prompt templates used for each concept.

\begin{table}[h]
\centering
\small
\caption{\textbf{Contrastive prompt templates for steering.} The \{base\} placeholder is filled with diverse musical descriptions (e.g., ``a song'', ``a jazz piece'', ``electronic music'').}
\label{tab:steering_prompts}
\begin{tabular}{lp{5.5cm}p{5.5cm}}
\toprule
\textbf{Concept} & \textbf{Positive Prompt} $\mathcal{P}_c$ & \textbf{Negative Prompt} $\mathcal{P}_{\tilde{c}}$ \\
\midrule
Piano & ``\{base\}, with piano'' & ``\{base\}'' \\
\midrule
Violin & ``\{base\}, with violin'' & ``\{base\}'' \\
\midrule
Guitar Type & ``\{base\}, with acoustic guitar'' & ``\{base\}, with electric guitar'' \\
\midrule
Mood & ``happy song, \{base\}'' & ``sad song, \{base\}'' \\
\midrule
Tempo & ``fast song, \{base\}'' & ``slow song, \{base\}'' \\
\midrule
Vocal Gender & ``\{base\}, with female vocal'' & ``\{base\}, with male vocal'' \\
\midrule
Vocal Style & ``\{base\}, with rap vocal'' & ``\{base\}, with sing vocal'' \\
\midrule
Classical Genre & ``classical song, \{base\}'' & ``electronic song, \{base\}'' \\
\midrule
Jazz Genre & ``jazz song, \{base\}'' & ``rock song, \{base\}'' \\
\bottomrule
\end{tabular}
\end{table}

The base prompts span 50 diverse musical styles and genres, i.a.: ``a song'', ``a melody'', ``music'', ``a tune'', ``a track'', ``instrumental music'', ``a pop song'', ``a rock song'', ``a jazz piece'', ``a classical piece'', ``electronic music'', ``acoustic music'', ``orchestral music'', ``hip hop music'', ``country music'', ``blues music'', ``folk music'', ``reggae music'', ``ambient music'', ``lofi music'', ``a ballad'', ``a love song'', ``energetic music'', ``calm music'', ``dramatic music'', among others.

\subsection{Benchmark prompts}

For steering evaluation, we use 100 diverse prompts (separate from those used to calculate steering vectors) that span a wide range of musical styles. Here, we provide examples:
\begin{itemize}
    \item \textbf{Pop \& Rock}: `Upbeat indie pop track with jangly guitars and handclaps', `90s grunge with fuzzy guitars, angst-filled dynamics'
    \item \textbf{Electronic}: `Dark synthwave anthem with pulsing bass, retro analog synths', `Techno industrial with distorted kicks, metallic textures'
    \item \textbf{Jazz \& Blues}: `Melancholic jazz ballad with smooth saxophone, walking bassline', `Delta blues with slide guitar, stomping rhythm'
    \item \textbf{World Music}: `Traditional Irish jig with fiddle, tin whistle, bodhran drums', `Afrobeat groove with polyrhythmic percussion, brass stabs'
    \item \textbf{Classical}: `Romantic piano nocturne with expressive dynamics', `Epic orchestral score with full brass, timpani rolls'
    \item \textbf{Hip Hop \& R\&B}: `Boom bap with punchy drums, scratched samples', `Neo-soul with warm keys, silky bassline'
    \item \textbf{Country \& Folk}: `Acoustic folk song with fingerpicked guitar, gentle harmonies', `Bluegrass breakdown with banjo rolls, fiddle solo'
\end{itemize}

\clearpage

\section{Evaluation metrics}\label{sec:external_eval}

\subsection{Audio-text alignment prompts}

To measure how the generated audio exhibits target concepts, we compute audio-text similarity between steered generations and concept-specific text queries using CLAP \citep{wu2022largescale} and MuQ \citep{zhu2025muq} models. In \cref{tab:alignment_prompts}, we present the evaluation prompts used for computing the similarity scores.

\vspace{-1.2em}

\begin{table}[h]
\centering
\caption{\textbf{Audio-text alignment queries for audio-text alignment measurement.} These prompts are used to compute similarity scores between generated audio and target concepts.}
\label{tab:alignment_prompts}
\begin{tabular}{ll}
\toprule
\textbf{Concept} & \textbf{Alignment query} \\
\midrule
\textbf{Piano} & ``a piano song'' \\
\textbf{Violin} & ``a song with violin'' \\
\textbf{Guitar Type} & ``a song with acoustic guitar'' \\
\textbf{Mood} & ``a cheerful track'' \\
\textbf{Tempo} & ``a fast track'' \\
\textbf{Vocal Gender} & ``female vocal singing'' \\
\textbf{Vocal Style} & ``rap vocal'' \\
\textbf{Classical Genre} & ``a classical song'' \\
\textbf{Jazz Genre} & ``a jazz song'' \\
\bottomrule
\end{tabular}
\end{table}

\subsection{Alternative concept alignment measures}

To assess whether text-audio alignment models properly score concepts in generated audio, we compare them with external evaluators that do not rely on contrastive text--audio alignment. We re-evaluate the SAE-steered generations across the three concepts (\textit{mood}, \textit{tempo}, \textit{vocal gender}) using measurements from prior works.

For \textit{mood}, inspired by \citet{facchiano2025activation}, we compute the \textbf{Spectral Centroid}, the magnitude-weighted mean frequency of the spectrum, which serves as a standard proxy for spectral brightness. Brighter timbres tend to correlate with more cheerful content and darker timbres with sadder content. For \textit{tempo}, we follow \citet{zhao2026steering} and compute the \textbf{onset event rate} (events/s), defined as the number of \texttt{librosa} onset peaks divided by clip duration, which captures rhythmic density. Finally, for \textit{vocal gender}, we first \textbf{extract a vocal stem} from each song with \texttt{SAM Audio} \citep{shi2025sam}, conditioned on the text prompt ``singing''. We then run an off-the-shelf \textbf{gender classifier}, a version of \texttt{Wav2Vec2-XLS-R} \citep{babu2022xls} fine-tuned for speaker gender classification, and report $P(\text{female}\mid\text{stem})$ averaged over the generated clips at each $\alpha$.

\begin{figure}[h]
  \centering
  \begin{subfigure}{0.32\linewidth}
      \centering
      \includegraphics[width=\linewidth]{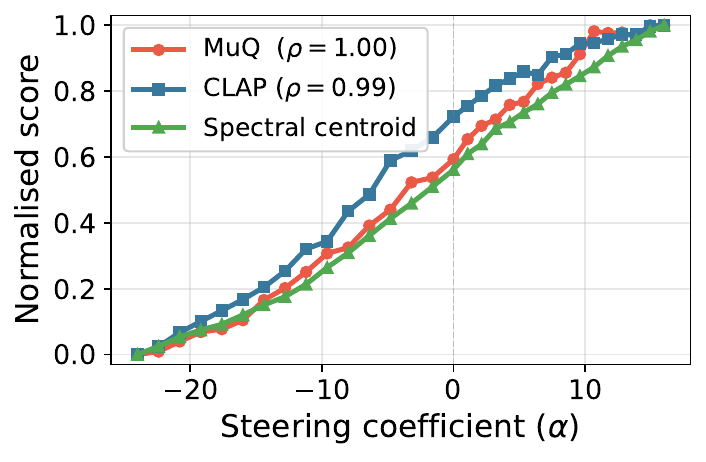}
      \vspace{-1.6em}
      \caption{\textbf{Mood} (darker$\rightarrow$brigher)}
      \label{fig:ext_mood}
  \end{subfigure}
  \hfill
  \begin{subfigure}{0.32\linewidth}
      \centering
      \includegraphics[width=\linewidth]{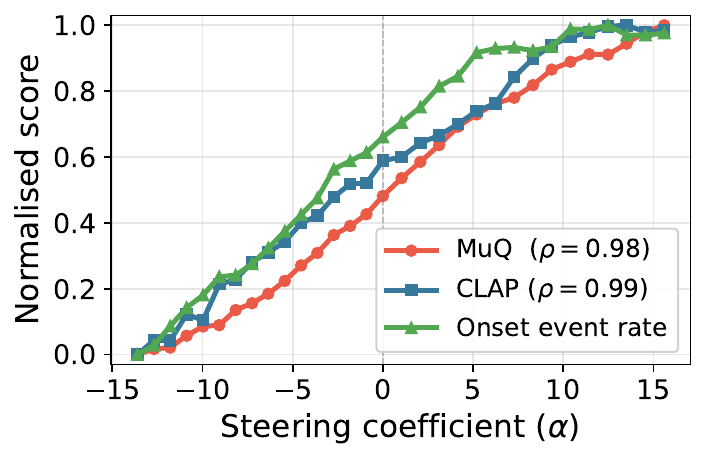}
      \vspace{-1.6em}
      \caption{\textbf{Tempo} (slow$\rightarrow$fast)}
      \label{fig:ext_tempo}
  \end{subfigure}
  \hfill
  \begin{subfigure}{0.32\linewidth}
      \centering
      \includegraphics[width=\linewidth]{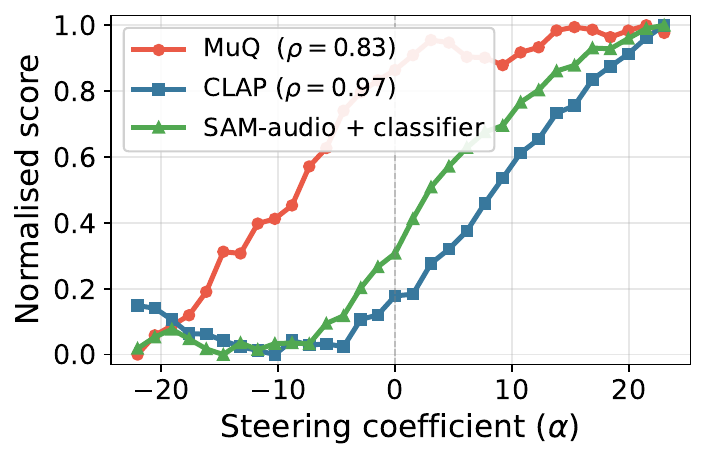}
      \vspace{-1.6em}
      \caption{\textbf{Vocal gender} (male$\rightarrow$female)}
      \label{fig:ext_vocals}
  \end{subfigure}
  \vspace{-0.6em}
  \caption{\textbf{Text-audio alignment models track the presence of concepts in generated audios.} We compare alignment metrics from our evaluation benchmark with external acoustic evaluators on \textit{mood}, \textit{tempo}, and \textit{vocal gender} concepts. We plot the per-$\alpha$ averages of MuQ, CLAP, and the external evaluator, each normalized to $[0,1]$. As shown, CLAP and MuQ track real audio properties. In legend, we report the Pearson correlation $\rho$ between each text-based metric and the evaluator.}
  \label{fig:external_eval}
  \vspace{-0.8em}
\end{figure}

As presented in \cref{fig:external_eval}, across all three concepts, the text-alignment models score similarly to the independent acoustic descriptors. Quantitatively, in legend, the Pearson correlation ($\rho$) is very high for \textit{mood} ($\rho_{\text{MuQ}}=0.998$, $\rho_{\text{CLAP}}=0.987$ vs.\ spectral centroid) and for \textit{tempo} ($\rho_{\text{MuQ}}=0.982$, $\rho_{\text{CLAP}}=0.986$ vs.\ onset event rate). For \textit{vocal gender}, CLAP tracks the SAM-Audio + classifier signal almost as tightly as it does for the other two concepts ($\rho_{\text{CLAP}}=0.965$), while MuQ shows a weaker but still strong correlation ($\rho_{\text{MuQ}}=0.830$). Overall, these correlations support the view that CLAP and MuQ are faithful evaluators of the target concepts across the full $\alpha$ range.

\subsection{Alternative audio preservation metrics}\label{app:preservation_metrics_section}

Our evaluation methodology heavily depends on the choice of preservation metric, which, in our main experiments, is LPAPS. However, as shown in \cref{tab:mir_preservation}, our localized activation steering achieves the best steering performance across other preservation metrics.

\begin{figure}[h]
    \centering
    \includegraphics[width=\linewidth]{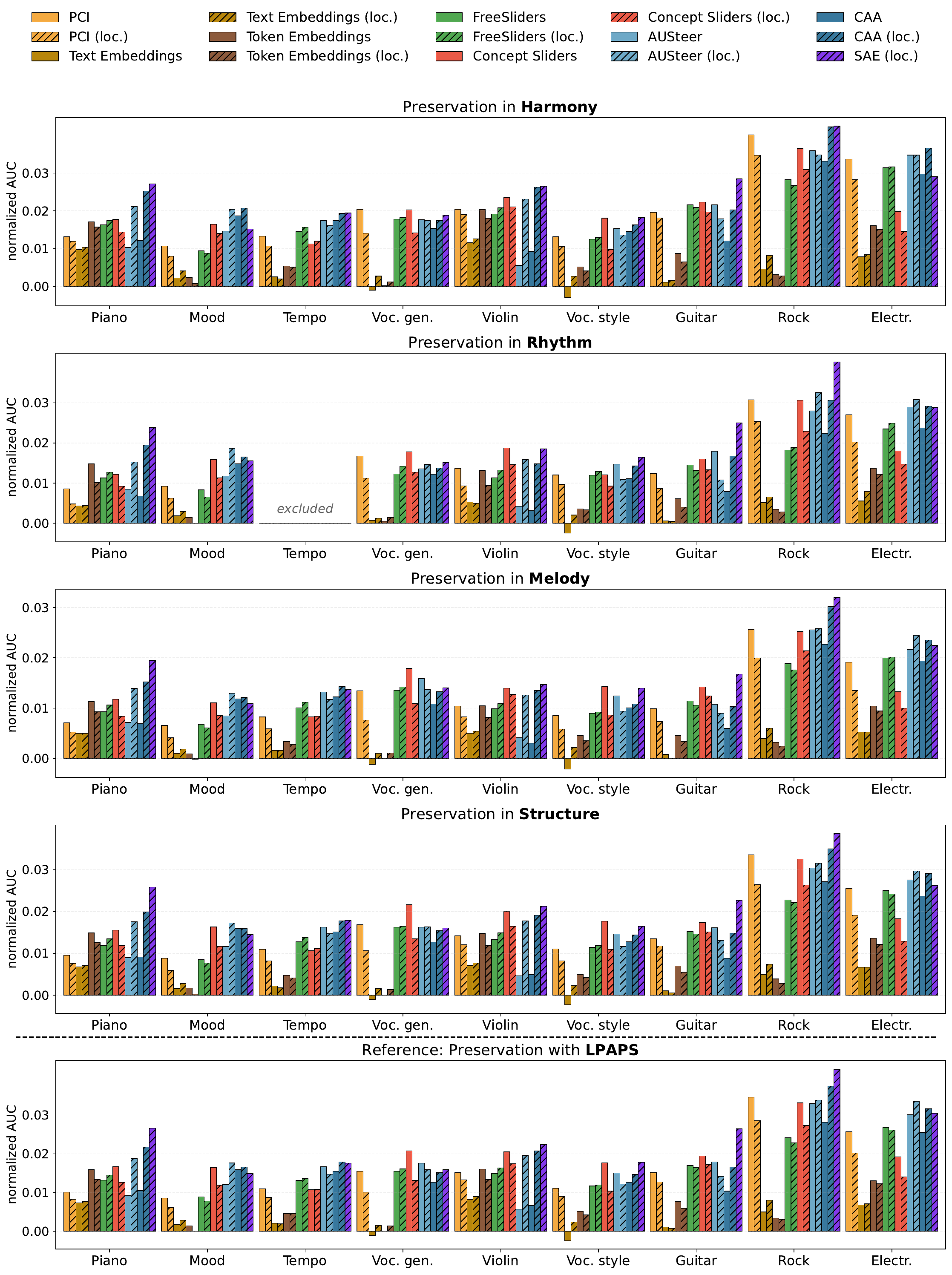}
    \caption{\textbf{Per-concept steering methods comparison across different preservation metrics: Harmony, Rhythm, Melody, Structure.} We report preservation--alignment AUC per concept, averaged over both alignment metrics: MuQ and CLAP. Hatched bars represent localized variants of the methods. Localized activation steering (CAA, AUSteer, SAE) leads on nearly every concept and axis.}
    \label{fig:mir_per_concept}
\end{figure}

We employ measurements from the Musical Information Retrieval (MIR) domain. We alter our evaluation protocol (\cref{sec:metrics}), replacing only the preservation axis with four decomposed measures, each expressed as a distance metric ($f_{\text{dist}} = 1-f_{\text{sim}}$). We assess similarity ($f_{\text{sim}}$) between two audio tracks in terms of their:
\begin{itemize}
    \item \textbf{Harmony} -- the mean cosine similarity between corresponding frames of their \texttt{librosa}-computed CQT chromagrams.
    \item \textbf{Rhythm} -- the beat F-measure calculated with \texttt{mir\_eval} \citep{raffel2014mir_eval} between their \texttt{librosa}-tracked beat sequences. Here, we exclude the \textit{tempo} concept.
    \item \textbf{Melody} -- the raw pitch accuracy calculated with \texttt{mir\_eval} \citep{raffel2014mir_eval} between their pYIN-estimated fundamental frequencies ($f_0$ contours).
    \item \textbf{Structure} -- the Pearson correlation, clamped to $[0,1]$, between self-similarity matrices of their \texttt{librosa}-computed CQT chromagrams.
\end{itemize}
We emphasize that these metrics do not rely on trained models and therefore share neither training data nor perceptual priors with LPAPS. As presented, localized activation steering offers the best steering performance under $7$ out of $8$ additional metrics. In \cref{fig:mir_per_concept} we compare methods under the new preservation metrics across concepts.

\begin{table}[h]
    \centering
    \caption{\textbf{Localized steering offers best steering performance across diverse preservation metrics.} We report AUC-MuQ and AUC-CLAP under LPAPS and four MIR preservation axes, measuring harmony, rhythm, melody, and structure. Results are averaged over the nine benchmark concepts (except \textit{tempo} on the Rhythm axis) and across both steering directions. \textbf{Best} and \uline{second} results are highlighted.}
    \label{tab:mir_preservation}
    \vspace{0.5em}
    \resizebox{1.0\linewidth}{!}{
        \begin{tabular}{lcccccccccc}
            \toprule
            \multirow{2}{*}{\textbf{Method}}
                & \multicolumn{2}{c}{\textbf{LPAPS} ($\uparrow$)}
                & \multicolumn{2}{c}{\textbf{Harmony} ($\uparrow$)}
                & \multicolumn{2}{c}{\textbf{Rhythm} ($\uparrow$)}
                & \multicolumn{2}{c}{\textbf{Melody} ($\uparrow$)}
                & \multicolumn{2}{c}{\textbf{Structure} ($\uparrow$)}
            \\
            \cmidrule(lr){2-3}\cmidrule(lr){4-5}\cmidrule(lr){6-7}\cmidrule(lr){8-9}\cmidrule(lr){10-11}
                & MuQ & CLAP & MuQ & CLAP & MuQ & CLAP & MuQ & CLAP & MuQ & CLAP \\
            \midrule
            PCI & $0.021$ & $0.012$ & $0.026$ & $0.015$ & $0.020$ & $0.013$ & $0.015$ & $0.009$ & $0.020$ & $0.012$ \\
            PCI (loc.) & $0.017$ & $0.009$ & $0.022$ & $0.012$ & $0.015$ & $0.009$ & $0.011$ & $0.006$ & $0.015$ & $0.009$ \\
            Text Embeddings & $0.004$ & $0.003$ & $0.005$ & $0.003$ & $0.003$ & $0.003$ & $0.002$ & $0.002$ & $0.003$ & $0.003$ \\
            Text Embeddings (loc.) & $0.006$ & $0.004$ & $0.007$ & $0.004$ & $0.004$ & $0.003$ & $0.004$ & $0.002$ & $0.005$ & $0.003$ \\
            Token Embeddings & $0.009$ & $0.006$ & $0.010$ & $0.007$ & $0.008$ & $0.006$ & $0.006$ & $0.005$ & $0.008$ & $0.006$ \\
            Token Embeddings (loc.) & $0.008$ & $0.005$ & $0.010$ & $0.006$ & $0.007$ & $0.004$ & $0.005$ & $0.004$ & $0.007$ & $0.005$ \\
            FreeSliders & $0.020$ & $0.012$ & $0.023$ & $0.015$ & $0.017$ & $0.011$ & $0.015$ & $0.010$ & $0.019$ & $0.012$ \\
            FreeSliders (loc.) & $0.020$ & $0.012$ & $0.024$ & $0.015$ & $0.018$ & $0.012$ & $0.015$ & $0.010$ & $0.019$ & $0.012$ \\
            Concept Sliders & $0.022$ & \uline{0.017} & $0.023$ & $0.018$ & $0.019$ & \uline{0.016} & $0.016$ & \uline{0.013} & $0.021$ & \textbf{0.017} \\
            Concept Sliders (loc.) & $0.017$ & $0.013$ & $0.019$ & $0.014$ & $0.015$ & $0.012$ & $0.013$ & $0.010$ & $0.017$ & $0.012$ \\
            AUSteer & $0.020$ & $0.015$ & $0.022$ & $0.016$ & $0.018$ & $0.014$ & $0.015$ & $0.011$ & $0.019$ & $0.014$ \\
            CAA & $0.019$ & $0.012$ & $0.022$ & $0.014$ & $0.015$ & $0.011$ & $0.014$ & $0.009$ & $0.018$ & $0.011$ \\
            \midrule
            AUSteer (loc.) & $0.024$ & $0.016$ & $0.027$ & $0.017$ & $0.022$ & $0.016$ & $0.018$ & $0.012$ & $0.023$ & $0.015$ \\
            CAA (loc.) & \uline{0.026} & $0.016$ & \uline{0.031} & \textbf{0.019} & \uline{0.023} & $0.016$ & \uline{0.019} & $0.013$ & \uline{0.025} & $0.016$ \\
            SAE (loc.) & \textbf{0.030} & \textbf{0.018} & \textbf{0.032} & \uline{0.018} & \textbf{0.028} & \textbf{0.018} & \textbf{0.022} & \textbf{0.013} & \textbf{0.028} & \uline{0.016} \\
            \bottomrule
        \end{tabular}
    }
\end{table}

\clearpage

\subsection{Generation quality over steering range}
\label{app:quality_over_strength}

Activation steering intervenes in the model's hidden states, propagating to subsequent layers a signal that was never encountered during training. A natural concern is that such interventions push activations off the training manifold, producing distortions and disrupting the quality of generated audio, even when the target concept is successfully modulated. In this section, we discuss this problem both theoretically and experimentally. 

First, the common architectural design of transformers softens the train--inference activation mismatch. We intervene at the output level of the attention block, which, in all evaluated backbones, is followed by a normalization layer: RMSNorm in \ace and LayerNorm in both \ldm and \stableaudio. One beneficial aspect of these layers is that they renormalize the perturbed activations back toward the statistics observed during training, before the next block consumes them.

Second, during our evaluation, the steering ranges are bounded by the distortion induced by prompt replacement, which keeps all methods within the model's natural operating regime. 
Thus, any off-distribution damage would surface directly as an audio quality drop in \cref{tab:steering_auc_avg}.
Under calibrated conditions, activation steering does not produce tracks of worse quality than other baselines. To further quantify this phenomenon, we explicitly analyze the degradation of quality as a function of steering strength.

We measure the Production Quality (PQ) score from Audiobox Aesthetics~\citep{tjandra2025aes}, which targets acoustic ``clarity and fidelity'' rather than content and therefore serves as a good probe of off-manifold degradation. The unsteered baseline scores $\text{PQ} = 7.40$ on the $1$--$10$ scale. \Cref{tab:pq_vs_alpha} reports the change in PQ relative to this baseline for the activation-steering methods in their global and localized configurations, as a function of the methods' steering strength ($\alpha / \alpha_{\max}$). We report the average across the benchmark concepts in the concept-adding direction. We exclude the \textit{mood} concept from this experiment, as amplifying it directly biases the quality score assigned by the external model.

The overall effect of steering on quality is small throughout the calibrated range. Even at maximum strength, the PQ change stays within $\pm 0.16$, roughly $2\%$ of the baseline. However, we expect PQ to degrade more noticeably at strengths well beyond this range. Although the magnitudes are small, the trends differ qualitatively by intervention site. The localized variants show almost no degradation at any strength (at full $\alpha$: CAA $+0.10$, AUSteer $-0.01$, SAE $+0.03$), whereas both global variants decline with $\alpha$. This suggests that injecting the change only into the layers that carry semantic control achieves the intended shift with a far smaller perturbation of the network's internal state, and less of the representation is pushed out of distribution.

\begin{table}[h]
    \centering
    \caption{\textbf{Change in Production Quality (PQ) under increasing strength in activation steering methods.} We report $\Delta$PQ relative to the unsteered baseline ($\text{PQ}=7.40$) at increasing steering strengths. Localized variants show little-to-no degradation at any strength, whereas global methods decline as steering strength ($\alpha$) increases.}
    \label{tab:pq_vs_alpha}
    \begin{tabular}{lccccc}
        \toprule
        $\alpha / \alpha_{\max}$ & CAA & CAA (loc.) & AUSteer & AUSteer (loc.) & SAE (loc.) \\
        \midrule
        0.0 & $0.000$ & $0.000$ & $0.000$ & $0.000$ & $0.000$ \\
        0.2 & $-0.028$ & $+0.024$ & $+0.008$ & $+0.030$ & $+0.046$ \\
        0.4 & $-0.045$ & $+0.041$ & $-0.043$ & $+0.043$ & $+0.067$ \\
        0.6 & $-0.051$ & $+0.049$ & $-0.090$ & $+0.015$ & $+0.045$ \\
        0.8 & $-0.055$ & $+0.059$ & $-0.137$ & $-0.001$ & $+0.034$ \\
        1.0 & $-0.047$ & $+0.096$ & $-0.154$ & $-0.010$ & $+0.030$ \\
        \bottomrule

    \end{tabular}
\end{table}

\clearpage

\section{Steering methods}

\subsection{Descriptions}\label{app:steering_methods}

We adapt seven steering paradigms from prior work to \ace and group them by the space in which they intervene: prompt-level (PCI \cite{gorgun2026temporal}, Text Embeddings \cite{ezra2025freesliders0}, Token Embeddings \cite{baumann2025continuous}), weight-space (Concept Sliders \cite{gandikota2024concept}), score-space (FreeSliders \cite{ezra2025freesliders0}), and activation-space (CAA \cite{rimsky-caa}, AUSteer \cite{feng2026finegrained}, SAEs \cite{olshausen1997sparse}).

\subsubsection{Prompt-level interventions}\label{app:prompt_methods}

These methods modify the model's text conditioning, leaving its weights and internal activations untouched.

\textbf{Prompt-Conditioned Intervention (PCI)}~\citep{gorgun2026temporal} modifies the conditioning prompt during diffusion process. For each test prompt, a triple $(\mathcal{P}_n, \mathcal{P}_c, \mathcal{P}_{\tilde{c}})$ of neutral, positive (concept-promoting), and negative (concept-suppressing) prompts is used. The denoising trajectory uses $\mathcal{P}_n$ for the first $T - |\alpha|$ inference steps and the directional prompt ($\mathcal{P}_c$ or $\mathcal{P}_{\tilde{c}}$) for the remaining $|\alpha|$ steps. Thus, the integer $\alpha \in [-T,\dots,0,\dots,T]$ encodes both \emph{direction} (sign) and \emph{strength} (number of switched steps) for modulation. PCI's maximum perceptual distortion, when applying directional prompt in all diffusion steps, serves as the reference point used to calibrate $\alpha_{\max}$ for all other methods (see  \cref{sec:metrics}).

\textbf{Text Embeddings}~\citep{ezra2025freesliders0} method interpolates in the text-encoder hidden-state space. For each test (neutral) prompt, we build the positive and negative prompts, which are constructed by adding prefixes and suffixes from App.~\ref{app:steering_prompts}.
Next, all three prompts are encoded with the frozen T5 encoder to obtain embeddings $\mathbf{e}_n, \mathbf{e}_c, \mathbf{e}_{\tilde{c}} \in \mathbb{R}^{L\times d}$ (with $L$ denoting prompt length in tokens) and form the conditioning used by the model during inference $\mathbf{e}'(\alpha) = \mathbf{e}_n + \alpha \cdot \big(\mathbf{e}_c - \mathbf{e}_{\tilde{c}}\big)$,
with steering strength $\alpha$. Within each triple, the prompts are constructed so that all three have the same token count and the same padding after being converted to embeddings.

\textbf{Token Embeddings}~\citep{baumann2025continuous} restrict the perturbation to the embeddings of the \emph{target-concept token} in the prompt rather than the full embedding sequence. For each concept, a target token (e.g., \texttt{instrument} for the piano, \texttt{vocal} for vocal-gender, \texttt{song} for tempo/mood/genre) is selected. A direction vector $\mathbf{d}_c \in \mathbb{R}^{d}$ is precomputed by averaging $(\mathbf{e}_c^{(i)} - \mathbf{e}_{\tilde{c}}^{(i)})$ over the $50$ steering base prompts of Sec.~\ref{app:steering_prompts}, where $\mathbf{e}_c^{(i)}$ is the encoder hidden state at the concept-token position $i$. At inference time, the test prompt is unchanged, and $\alpha \cdot \mathbf{d}_c^{(i)}$ is added only to the hidden state at the concept-token position. Compared to Text Embeddings, this leaves the rest of the prompt structurally identical and isolates the edit to the encoding of the named subject.

\subsubsection{Weight-based steering}
\textbf{Concept Sliders} \citep{gandikota2024concept} encode a steering direction as a low-rank weight delta. LoRA adapters~\citep{hu2022lora} ($\Delta \theta$) of rank $r$ are attached to the attention projections ($W^Q$,$W^K$,$W^O$,$W^V$) of every transformer block in the model and trained with the enhanced-direction objective
\begin{equation}
    \mathcal{L}_{\text{CS}} = \big\| \boldsymbol{\epsilon}_{\theta+\Delta\theta}(x_t, \mathcal{P}_n) - \big[ \boldsymbol{\epsilon}_{\theta}(x_t, \mathcal{P}_{n}) + \eta \cdot \big(\boldsymbol{\epsilon}_{\theta}(x_t, \mathcal{P}_c) - \boldsymbol{\epsilon}_{\theta}(x_t, \mathcal{P}_{\tilde{c}})\big) \big] \big\|_2^2,
\end{equation}
where $\eta$ is a guidance hyperparameter at training time. At inference time, trained LoRAs are merged into the base weights with a scalar $\alpha$, i.e. $\theta' = \theta + \alpha \cdot \Delta\theta$. Negative $\alpha$ inverts the trained direction (suppressing the concept).

\subsubsection{Score-based steering}
\textbf{FreeSliders}~\citep{ezra2025freesliders0} method, a training-free version of Concept Sliders, edits the noise prediction directly using a contrastive direction in score (diffusion prediction) space. Given the neutral, positive, and negative prompts triple, three conditional forward passes per denoising step are run and combined as
\begin{equation}
    \boldsymbol{\epsilon}'(x_t, \alpha) = \boldsymbol{\epsilon}_{\text{cfg}}(x_t, \mathcal{P}_n) + \alpha \cdot \big(\boldsymbol{\epsilon}_{\theta}(x_t, \mathcal{P}_c) - \boldsymbol{\epsilon}_{\theta}(x_t, \mathcal{P}_{\tilde{c}})\big),
\end{equation}
where $\boldsymbol{\epsilon}_{\text{cfg}}(x_t, \mathcal{P}_n)$ is the standard CFG-guided velocity for the neutral prompt and $\boldsymbol{\epsilon}_{\theta}(x_t, \mathcal{P}_c)$ denotes a single conditional pass. 

\subsubsection{Activation steering}\label{app:activation_steering_methods}

Activation-steering methods leave the prompt and the weights unchanged and intervene in the outputs of the cross-attention layer $l$, $\mathbf{h}_l \in \mathbb{R}^{T \times d}$, during the forward pass. With classifier-free guidance, we always apply the intervention only to the conditional pass, leaving the unconditional pass untouched. In our experiments, we calculate steering vectors for each diffusion step separately.

\paragraph{Contrastive Activation Addition (CAA).} We compute steering vectors in Cross-Attention (C-A) layers' outputs following CASteer~\citep{gaintseva2025casteer}. Given $N$ contrastive prompt pairs $\{(\mathcal{P}_c^{(i)}, \mathcal{P}_{\tilde{c}}^{(i)})\}_{i=1}^N$ for concept $c$, we collect activations and compute the steering vector $\mathbf{v}_c^{\text{CAA}}$ as:
    \begin{equation}
        \mathbf{v}_c^{\text{CAA}} = \frac{\mathbf{v}_c}{\|\mathbf{v}_c\|_2} \text{, where} \quad \mathbf{v}_c = \frac{1}{N}\sum_{i=1}^N \left(\bar{\mathbf{h}}_{c}^{(i)} - \bar{\mathbf{h}}_{\tilde{c}}^{(i)}\right),
    \end{equation}
    with $\bar{\mathbf{h}}$ denoting C-A outputs averaged across temporal frames. 
    During inference, we intervene by modifying activations at functional layers as $\mathbf{h}'_l = \mathbf{h}_l + \alpha \cdot \mathbf{v}_c^{\text{CAA}}$, where $\alpha \in \mathbb{R}$ controls steering strength, with positive values adding and negative values removing the modulated concept. Steering vectors are computed separately for each diffusion step.

\paragraph{AUSteer.} AUSteer~\citep{feng2026finegrained} replaces the dense mean-difference of CAA with a sparse, momentum-based selection of discriminative dimensions. Given the same contrastive activation pairs $\{(\mathbf{h}_c^{(i)}, \mathbf{h}_{\tilde{c}}^{(i)})\}_{i=1}^N$ at cross-attention layer $l$ and diffusion step $t$, per-dimension momentum is computed across all $N$ pairs and all temporal frames $f\in {1,\dots,F}$ as $m_j^{(i,f)} = h_{c,j}^{(i,f)} - h_{\tilde{c},j}^{(i,f)}$, 
and each dimension $j$ is scored by the consistency of the sign of $m_j$ over the $S$ pooled samples (pairs $\times$ timeframes):
\begin{equation} \label{eq:scores_austeer}
    r^+_j = \frac{1}{S} \sum_{i,f} [m_j^{(i,f)} > 0], \quad
    r^-_j = \frac{1}{S} \sum_{i,f} [m_j^{(i,f)} < 0], \quad
    \beta_j = \mathrm{sign}(r^+_j - r^-_j) \cdot \max(r^+_j, r^-_j).
\end{equation}
The dimension-wise score $|\beta_j|$ thus lies in $[0, 1]$ and is large only where the sign is stable across the dataset. Next, a \emph{global} top-$s$ selection is performed across all active layers at each diffusion step, zeroing out $\beta$ outside the top-$s$, and obtaining the sparse vector $\mathbf{v}_{c,l,t}^{\text{AUSteer}}$. 

While the original AUSteer formulation~\citep{feng2026finegrained} employs scores to scale single activation dimensions
\begin{equation}\label{eq:austeer_mult}
    \mathbf{h}'_l = \mathbf{h}_l \odot \big(\mathbf{1} + \alpha \cdot \mathbf{v}_{c,l,t}^{\text{AUSteer}}\big),
\end{equation}when adapted to diffusion models, our preliminary experiments resulted in zero gain in concept steering (see \cref{tab:austeer_mode_perconcept}). Thus,  we evaluate a CAA-style \emph{additive} formulation that uses the sparse $\beta$ vector as a steering direction added to the activation:
\begin{equation}\label{eq:austeer_add}
    \mathbf{h}'_l = \mathbf{h}_l + \alpha \cdot \mathbf{v}_{c,l,t}^{\text{AUSteer}}.
\end{equation}
Compared to CAA, AUSteer (a) selects \emph{which} dimensions to edit using a discriminative criterion rather than averaging, (b) allocates a fixed global top-$s$ budget, encouraging sparsity, and (c) implicitly performs a soft (discriminative) layer selection, where layers which dimensions never enter the global top-$s$ are not steered, even when nominally included in the set of layers. 

Interestingly, in \cref{tab:austeer_layer_allocation}, we show that the discriminative localization performed with AUSteer does not necessarily result in the same layer selection as our Activation Patching, often leading to suboptimal results compared with our approach. 

\newpage

    \paragraph{Sparse Autoencoders (SAEs).} We train a TopK SAEs \citep{gao2025scaling, bussmann2024batchtopksparseautoencoders} on the layers localized with activation patching. The SAE with encoder weights $\mathbf{W}_{\text{enc}} \in \mathbb{R}^{md \times d}$, decoder weights $\mathbf{W}_{\text{dec}} \in \mathbb{R}^{d \times md}$, and bias $\mathbf{b}_{\text{pre}} \in \mathbb{R}^{d}$ maps activations $\mathbf{h} \in \mathbb{R}^d$ to a sparse code $\mathbf{f} \in \mathbb{R}^{md}$ (with $m$ being the expansion factor):
    \begin{equation}
        \mathbf{f} = \text{TopK}\left(\mathbf{W}_{\text{enc}}(\mathbf{h} - \mathbf{b}_{\text{pre}})\right),
    \end{equation}
    and further reconstructs them:
    \begin{equation}
        \hat{\mathbf{h}} = \mathbf{W}_{\text{dec}}\mathbf{f} + \mathbf{b}_{\text{pre}}.
    \end{equation}
     The $\text{TopK}(\cdot)$ operation retains only the $k$ largest activations in the autoencoder's latent space, zeroing out the rest. The SAE is trained to minimize reconstruction error $\|\mathbf{h} - \hat{\mathbf{h}}\|_2^2$. To identify concept-specific features, we use two contrastive sets of $N$ prompts $(\mathcal{P}_c, \mathcal{P}_{\tilde{c}})$ for concept $c$ to compute importance scores using a TF-IDF-based criterion:
    \begin{equation} \label{eq:tfidf}
        \text{score}(j, c) = \underbrace{\mu_j(\mathcal{P}_c)}_{\text{TF}} \cdot \underbrace{\log\left(1 + \frac{1}{\mu_j(\mathcal{P}_{\tilde{c}}) + \epsilon}\right)}_{\text{IDF}},
    \end{equation}
    where $\mu_j(\mathcal{P_c}) = \frac{1}{|\mathcal{P_c}|}\sum_{\mathbf{h} \in \mathcal{P_c}} f_j(\mathbf{h})$ is the mean activation of feature $j$ in the SAE's latent space on generated audio with prompts $\mathcal{P_c}$. Features that activate highly for concept $c$ but rarely for other samples receive high scores. We select the top-$k_c$ scoring features $\mathcal{F}_c$ and, by summing their corresponding decoder columns, we construct the steering vector $\mathbf{v}_c^{\text{SAE}}$ as
\begin{equation}
    \mathbf{v}_c^{\text{SAE}} = \sum_{j \in \mathcal{F}_c} \mathbf{W}_{\text{dec}}[:, j],
\end{equation}
and add it directly to the output of the cross-attention layer:
\begin{equation}
    \mathbf{h}'_l = \mathbf{h}_l + \alpha \cdot \mathbf{v}_c^{\text{SAE}}.
\end{equation}

\subsection{Implementation and hyperparameter details}
\label{app:benchmark_details}

\paragraph{Activation Patching.} For Activation Patching \citep{NEURIPS2022_6f1d43d5}, we adapt audio diffusion models to the NNsight \cite{fiotto2025nnsight} API and perform localization using their library. 


\paragraph{Shared setup.} Unless stated otherwise, all steering experiments use a single shared inference configuration for \ace: $T = 30$ flow-matching steps with the Euler-discrete scheduler, CFG scale $5.0$, audio duration $30$s, and the same $100$ test prompts (Sec.~\ref{app:steering_prompts}). For each method, $\alpha$ is swept over $31$ values: $15$ negative, $\alpha = 0$, and $15$ positive. The $\alpha = 0$ run uses the unmodified pipeline and serves as the per-prompt preservation baseline against which LPAPS distortion is measured. Each $\alpha$ produces $100$ steered audio clips, yielding a total of $30 \times 100$ samples per concept per method. To choose method settings, we sweep over key hyperparameters using a held-out dataset of $20$ prompts with $9$ steering strengths.

\paragraph{Concept Sliders.} We attach LoRA adapters to all attention projections of every transformer block. Each concept's LoRA is trained for $500$ iterations using AdamW with a learning rate of $10^{-4}$ and training-time guidance $\eta = 7$. In \cref{tab:cs_hparam_sweep}, we show that choosing different values for these parameters does not yield a significant difference in steering performance. Additionally, in \cref{tab:cs_rank_sweep_perconcept}, we find, for each concept, on the held-out dataset, the optimal LoRA rank $r$. Concept Sliders in our main experiments use LoRAs trained with the optimal rank for each concept.
\begin{table}[h]
    \centering
    \caption{\textbf{Hyperparameter impact on audio concept modulation with Concept Sliders.} We report preservation-alignment AUC (averaged over piano and vocal concepts) on the held-out dataset. The setting used in the experiments is marked $\star$, the best per axis is in \textbf{bold}.}
    \label{tab:cs_hparam_sweep}
    \begin{tabular}{lccccccccc}
    \toprule
    & \multicolumn{3}{c}{\textbf{Learning rate}} & \multicolumn{3}{c}{\textbf{Train guidance}} & \multicolumn{3}{c}{\textbf{Training iters}} \\
    \cmidrule(lr){2-4} \cmidrule(lr){5-7} \cmidrule(lr){8-10}
    Value & 1e-4\,$\star$ & 5e-4 & 5e-3 & 3 & 5 & 7\,$\star$ & 500\,$\star$ & 1000 & 1500 \\
    \midrule
    Avg AUC & \textbf{0.102} & 0.101 & 0.083 & \textbf{0.104} & 0.103 & 0.102 & \textbf{0.102} & 0.101 & 0.101 \\
    \bottomrule
    \end{tabular}
\end{table}

\begin{table}[h]
    \centering
    \caption{\textbf{Per-concept LoRA-rank sweep for Concept Sliders.} We report preservation--alignment AUC. The per-concept highest scoring parameter (used in the benchmark) is \textbf{bolded}.}
    \label{tab:cs_rank_sweep_perconcept}
    \resizebox{1.0\linewidth}{!}{
    \begin{tabular}{l*{9}{c}c}
    \toprule
    Rank & Piano & Mood & Tempo & Voc.\,gen. & Violin & Voc.\,style & Guitar & Rock & Electr. & Avg \\
    \midrule
    4 & $0.061$ & $\mathbf{0.039}$ & $0.038$ & $\mathbf{0.104}$ & $0.080$ & $0.162$ & $0.069$ & $0.076$ & $0.088$ & $0.080$ \\
    8 & $\mathbf{0.061}$ & $0.036$ & $0.038$ & $0.102$ & $\mathbf{0.082}$ & $0.155$ & $0.071$ & $0.082$ & $\mathbf{0.091}$ & $0.080$ \\
    16 & $0.061$ & $0.036$ & $\mathbf{0.043}$ & $0.103$ & $0.080$ & $\mathbf{0.165}$ & $\mathbf{0.075}$ & $\mathbf{0.082}$ & $0.090$ & $\mathbf{0.082}$ \\
    \bottomrule

    \end{tabular}
    }
\end{table}

\newpage
\paragraph{AUSteer.} We rank activation dimensions across all active layers at each diffusion step, and select the top-$s$ of them to construct the steering vector. In the localized variant, only dimensions in $\{7, 8\}$ are considered in ranking and selection. In \cref{tab:austeer_hparam_sweep_tops}, we sweep budget $s$ separately for both method configurations on the held-out dataset: $s \in \{256, 1024, 2048, 4096, 8192, 16384\}$ for all-layer steering, and $s \in \{256, 1024, 2048, 4096, 5120\}$ for the localized variant (two active layers expose a total of $2d = 5120$ dimensions). In our experiments, we use the best-scoring $s$ for each concept and method's variant. Results in \cref{tab:austeer_hparam_sweep_tops} suggest that the best $s$ depends on the steered concept and localized steering is more effective than a global approach, with exceptions for \textit{tempo} and \textit{vocal gender} concepts.

\begin{table}[h]
    \centering
    \caption{\textbf{Per-concept top-$s$ sweep for AUSteer.} We report preservation--alignment AUC (mean of MuQ-T and CLAP over both steering directions) on the held-out dataset. The best value in each column is \textbf{bolded}.}
    \label{tab:austeer_hparam_sweep_tops}

    \resizebox{1.0\linewidth}{!}{
    \begin{tabular}{l*{9}{c}c}
    \toprule
    Top-$s$ & Piano & Mood & Tempo & Voc.\,gen. & Violin & Voc.\,style & Guitar & Rock & Electr. & Avg \\
    \midrule
    \multicolumn{11}{l}{\textit{All-layer steering}} \\
    256 & 0.049 & 0.034 & 0.035 & 0.034 & \textbf{0.038} & 0.024 & 0.052 & 0.087 & 0.097 & 0.050 \\
    1024 & 0.048 & \textbf{0.046} & 0.042 & 0.037 & 0.030 & 0.069 & 0.068 & 0.108 & 0.135 & 0.065 \\
    2048 & \textbf{0.054} & 0.039 & 0.048 & 0.039 & 0.022 & 0.082 & \textbf{0.075} & \textbf{0.118} & 0.143 & 0.069 \\
    4096 & 0.040 & 0.030 & \textbf{0.055} & 0.063 & 0.034 & 0.106 & 0.069 & 0.109 & \textbf{0.151} & \textbf{0.073} \\
    8192 & 0.024 & 0.028 & 0.046 & \textbf{0.065} & 0.029 & \textbf{0.124} & 0.051 & 0.105 & 0.115 & 0.065 \\
    16384 & 0.016 & 0.033 & 0.042 & 0.056 & 0.035 & 0.124 & 0.044 & 0.104 & 0.108 & 0.062 \\
    \midrule
    \multicolumn{11}{l}{\textit{Localized steering (layers} $\{7,8\}$\textit{)}} \\
    256 & 0.029 & 0.030 & 0.030 & 0.032 & 0.029 & 0.103 & 0.056 & 0.082 & 0.110 & 0.056 \\
    1024 & 0.060 & 0.035 & 0.042 & 0.029 & 0.048 & 0.130 & 0.068 & 0.124 & 0.164 & 0.078 \\
    2048 & 0.047 & 0.040 & 0.042 & \textbf{0.045} & \textbf{0.074} & 0.133 & 0.078 & 0.127 & \textbf{0.166} & \textbf{0.083} \\
    4096 & \textbf{0.061} & 0.045 & \textbf{0.046} & 0.034 & 0.066 & 0.139 & 0.079 & 0.126 & 0.145 & 0.082 \\
    5120 & 0.049 & \textbf{0.049} & 0.046 & 0.034 & 0.059 & \textbf{0.148} & \textbf{0.080} & \textbf{0.129} & 0.140 & 0.082 \\
    \bottomrule
    \end{tabular}
    }
\end{table}

In \cref{tab:austeer_mode_perconcept}, we analyze methods for constructing steering vectors from AUSteer's scoring: scaling activation dimensions by scores (\textit{multiplicative}, \cref{eq:austeer_mult}) or concatenating into a single vector and, as in CAA, adding to intermediate activations (\textit{additive}, \cref{eq:austeer_add}). Across all evaluated concepts, the CAA-style approach performs better.

\begin{table}[h]
    \centering
    \caption{\textbf{Impact of steering vector construction on AUSteer performance.} We report preservation--alignment AUC on the held-out set over all concepts and \textbf{highlight} the best method.}
    \label{tab:austeer_mode_perconcept}
    \resizebox{1.0\linewidth}{!}{
    \begin{tabular}{l*{9}{c}c}
    \toprule
    Method & Piano & Mood & Tempo & Voc.\,gen. & Violin & Voc.\,style & Guitar & Rock & Electr. & Avg \\
    \midrule
    add. (eq.~\ref{eq:austeer_add}) & $\mathbf{0.107}$ & $\mathbf{0.031}$ & $\mathbf{0.042}$ & $\mathbf{0.087}$ & $\mathbf{0.102}$ & $\mathbf{0.219}$ & $\mathbf{0.109}$ & $\mathbf{0.191}$ & $\mathbf{0.211}$ & $\mathbf{0.122}$ \\
    mult. (eq.~\ref{eq:austeer_mult}) & $0.028$ & $-0.005$ & $-0.009$ & $0.028$ & $0.003$ & $-0.008$ & $0.009$ & $0.028$ & $-0.013$ & $0.007$ \\
    \bottomrule

    \end{tabular}
    }
\end{table}

In \cref{tab:austeer_layer_allocation}, we analyze how AUSteer's ranking-and-selection mechanism, which can be considered an alternative to our localization, distributes activation dimensions across \ace layers. Interestingly, we find that their top-$s$ selection produces surprisingly distinct layer allocations across concepts. In most cases, our functional layers $\{7, 8\}$ contribute less than $50\%$ to the total budget, with the majority of selected dimensions coming from later layers.

\clearpage

Considering AUSteer's layer allocation results (\cref{tab:austeer_layer_allocation}) and per-concept steering performance (\cref{tab:steering_auc_piano,tab:steering_auc_tempo,tab:steering_auc_mood,tab:steering_auc_vocal_gender,tab:steering_auc_violin,tab:steering_auc_vocal_style,tab:steering_auc_guitar,tab:steering_auc_classical_electronic,tab:steering_auc_popjazz}), we find that AUSteer's discriminative localization often degrades steering performance. Precisely, on \textit{piano} (\cref{tab:steering_auc_piano}), \textit{mood} (\cref{tab:steering_auc_mood}), and \textit{violin} (\cref{tab:steering_auc_violin}) concepts, where our layers contribute, respectively, in $2.4\%$, $12.0\%$, and $16.3\%$ to the budget, all-layer AUSteer scores, respectively, two ($0.049$ vs. $0.102$), two ($0.026$ vs. $0.053$), and three ($0.032$ vs. $0.104$) times worse than localized approach. Usually, steering across all layers outperforms the localized one when our layers contribute to the budget, i.e., for the \textit{tempo} ($47.7\%$) and \textit{guitar} ($63.7\%$) concepts.

This suggests that AUSteer's discriminative-momentum score selects layers where the concept is most reliably encoded (discriminative) rather than where intervention is most effective (causal), thereby limiting the method's performance on audio concept modulation compared to our localized approach.

  \begin{table}[h]
      \centering
      \caption{
\textbf{Ace-Step layers allocated with AUSteer's rank-and-select localization.} We report the percentage of activation dimensions in attention layers, averaged across diffusion steps, for all nine concepts. Layers with at least $1\%$ allocation to at least one concept are listed, and the functional layers ($\{7, 8\}$) we localize are highlighted.}
      \label{tab:austeer_layer_allocation}
      \begin{tabular}{l*{9}{c}}
          \toprule
          Layer & Piano & Mood & Tempo & Voc.\,gen. & Violin & Voc.\,style & Guitar & Rock & Electr. \\
          \midrule
          5 & 9.6 & 0.3 & 1.3 & 0.3 & 17.5 & 11.1 & 1.8 & 0.2 & 17.7 \\
          6 & 5.3 & 0.0 & 0.4 & 0.0 & 8.4 & 2.7 & 2.2 & 0.0 & 0.2 \\
          \rowcolor{gray!15} \textbf{7} & 0.6 & 0.0 & 17.6 & 17.6 & 1.4 & 3.0 & 1.6 & 8.8 & 1.6 \\
          \rowcolor{gray!15} \textbf{8} & 1.8 & 12.0 & 30.1 & 37.6 & 14.9 & 17.3 & 62.1 & 53.5 & 71.8 \\
          9 & 0.0 & 0.7 & 1.1 & 0.0 & 0.2 & 15.2 & 0.2 & 2.5 & 0.4 \\
          12 & 0.4 & 0.0 & 0.0 & 5.8 & 0.6 & 0.1 & 0.1 & 0.0 & 0.1 \\
          13 & 1.5 & 0.1 & 0.0 & 0.0 & 0.7 & 3.3 & 0.0 & 0.0 & 0.0 \\
          14 & 0.1 & 58.7 & 3.4 & 0.0 & 0.4 & 3.9 & 3.2 & 5.4 & 0.6 \\
          15 & 0.0 & 11.6 & 0.0 & 0.0 & 0.1 & 0.0 & 0.1 & 0.0 & 0.0 \\
          16 & 0.2 & 0.0 & 1.1 & 1.2 & 0.0 & 3.4 & 5.4 & 0.4 & 0.1 \\
          17 & 1.6 & 0.0 & 0.0 & 4.4 & 0.5 & 0.0 & 0.1 & 0.0 & 0.0 \\
          18 & 11.1 & 3.1 & 0.8 & 1.0 & 4.1 & 9.0 & 0.5 & 0.4 & 0.0 \\
          19 & 6.7 & 11.6 & 24.1 & 0.0 & 8.1 & 14.2 & 4.7 & 0.1 & 0.0 \\
          20 & 0.2 & 0.0 & 6.8 & 0.4 & 0.8 & 0.8 & 3.4 & 0.8 & 2.2 \\
          22 & 59.2 & 1.8 & 11.8 & 7.5 & 41.3 & 6.6 & 10.6 & 21.2 & 3.1 \\
          23 & 0.1 & 0.0 & 0.1 & 10.1 & 0.2 & 1.9 & 2.4 & 0.1 & 1.9 \\
          24 & 0.2 & 0.0 & 1.3 & 14.1 & 0.2 & 6.9 & 1.2 & 6.5 & 0.0 \\
          \bottomrule
      \end{tabular}
  \end{table}

\paragraph{Contrastive Activation Addition (CAA).} For steering vectors calculation, we first gather activations from the outputs of the cross-attention blocks across all denoising steps. Our methodology is roughly equivalent to CASteer \citep{gaintseva2025casteer}, a method designed for text-to-image diffusion models. Unlike vision, where inputs are split spatially into a grid of patches, each a vector of activations, audio models split signals along the time axis into timeframes. The collected layer outputs from positive and negative prompt runs (see \cref{app:steering_prompts}) are first averaged along the time axis, then contrasted, and finally averaged over all $N$ prompt pairs. During experiments, we steer only the conditional (\textit{cond}) prediction within classifier-free guidance (CFG). Our analysis in \cref{tab:caa_hparam_sweep}, conducted on the held-out dataset, shows that steering both conditional and unconditional (\textit{null}) predictions separately does not yield significantly better results, despite being twice as costly.

\begin{table}[h]
    \centering
    \caption{\textbf{Impact of model's classifier-free guidance (CFG) prediction on concept steering with CAA.} We report preservation-alignment AUC over concepts. Steering separately both conditional and unconditional (\textit{cond + null}) predictions does not yield significantly different results than steering only the conditional prediction (\textit{cond}).}
    \label{tab:caa_hparam_sweep}
    \resizebox{1.0\linewidth}{!}{
    \begin{tabular}{l*{9}{c}c}
    \toprule
    Steered CFG passes & Piano & Mood & Tempo & Voc.\,gen. & Violin & Voc.\,style & Guitar & Rock & Electr. & Avg \\
    \midrule
    cond & $0.104$ & $0.035$ & $\mathbf{0.059}$ & $0.075$ & $0.105$ & $\mathbf{0.239}$ & $\mathbf{0.117}$ & $0.192$ & $\mathbf{0.191}$ & $0.124$ \\
    cond + null & $\mathbf{0.110}$ & $\mathbf{0.040}$ & $0.058$ & $\mathbf{0.089}$ & $\mathbf{0.110}$ & $0.230$ & $0.114$ & $\mathbf{0.203}$ & $0.177$ & $\mathbf{0.126}$ \\
    \bottomrule
    \end{tabular}
    }
\end{table}

\newpage

\begin{figure}[h]
      \centering
      \begin{subfigure}[t]{0.48\linewidth}
          \centering
          \includegraphics[width=0.8\linewidth]{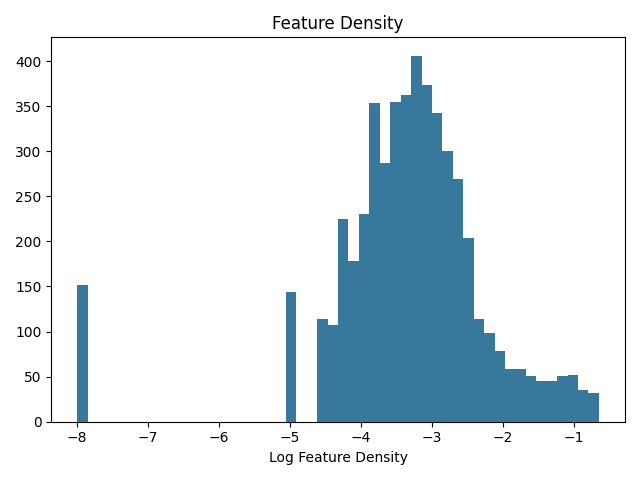}
          \vspace{-0.5em}
          \caption{Layer $7$.}
          \label{fig:sae_feature_density_layer7}
      \end{subfigure}
      \hfill
      \begin{subfigure}[t]{0.48\linewidth}
          \centering
          \includegraphics[width=0.8\linewidth]{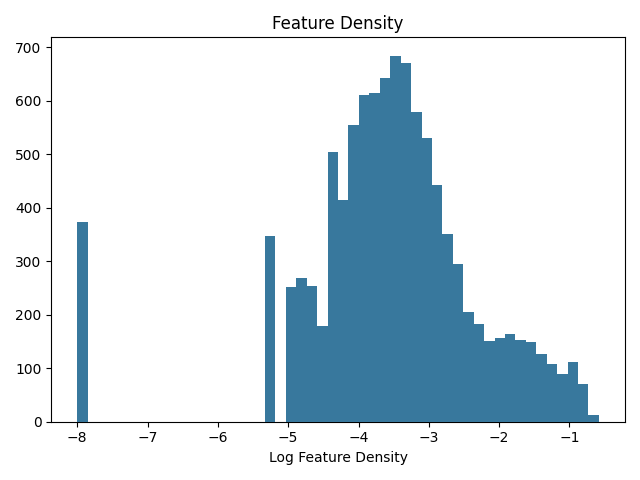}
          \vspace{-0.5em}
          \caption{Layer $8$.}
          \label{fig:sae_feature_density_layer8}
      \end{subfigure}
      \vspace{-0.5em}
      \caption{\textbf{Feature-density histograms for Sparse Autoencoders trained \ace semantic bottleneck.} Features in SAE latent space are considered dead (never used) when their density $<10^{-7}$, while high-frequency features (activating at least once per batch) are the ones with density $>10^{-2}$. We consider features in between to be prospective for steering.}
      \label{fig:sae_feature_density}
  \end{figure}

\paragraph{SAEs.} We train one BatchTopK SAE~\citep{bussmann2024batchtopksparseautoencoders, gao2025scaling} per localized layer, on the outputs of the attention layers $\{7, 8\}$. Activations are collected at different diffusion steps (every $5$ of $30$) for MusicCaps~\citep{agostinelli2023musiclm} prompts. We sweep expansion factors $m \in \{2, 4, 8, 16\}$ and sparsity $k \in \{16, 32, 64\}$, and select the configuration that minimizes reconstruction error while keeping the fraction of dead and high-frequency features low: $(m, k) = (2, 32)$ for layer $\{7\}$, trained for $10$ epochs, and $(m, k) = (4, 64)$ for layer $\{8\}$, trained for $15$ epochs. In \cref{tab:sae_quality}, we report the values: \emph{FVU} (fraction of variance unexplained, measures reconstruction error), \emph{$\mathrm{dead}_{\%}$} (the fraction of features that have not fired for more than $10^{7}$ inputs), and \emph{$\mathrm{fire}_{\%}$} (the fraction of features that fire at least once every batch), while in \cref{fig:sae_feature_density}, we present feature density in trained SAEs.

\begin{table}[h]
      \centering
      \caption{\textbf{Training quality metrics for our BatchTopK SAEs.} Lower FVU indicates better SAE reconstruction, lower $\mathrm{dead}_{\%}$ indicates a smaller fraction of features that never activate; lower $\mathrm{fire}_{\%}$ indicates fewer features that activate every batch (global).}
      \label{tab:sae_quality}
      \begin{tabular}{lccc}
          \toprule
          ACE-Step Layer & FVU ($\downarrow$) & $\mathrm{dead}_{\%}$ ($\downarrow$) & $\mathrm{fire}_{\%}$ ($\downarrow$) \\
          \midrule
          $7$ &  0.216 & 0.37\% & 19.4\%  \\
          $8$ &  0.243 & 0.25\% & 20.1\%  \\
          \bottomrule
      \end{tabular}
  \end{table}
          \vspace{-1em}

\begin{figure}[h]
    \centering
    \includegraphics[width=\linewidth]{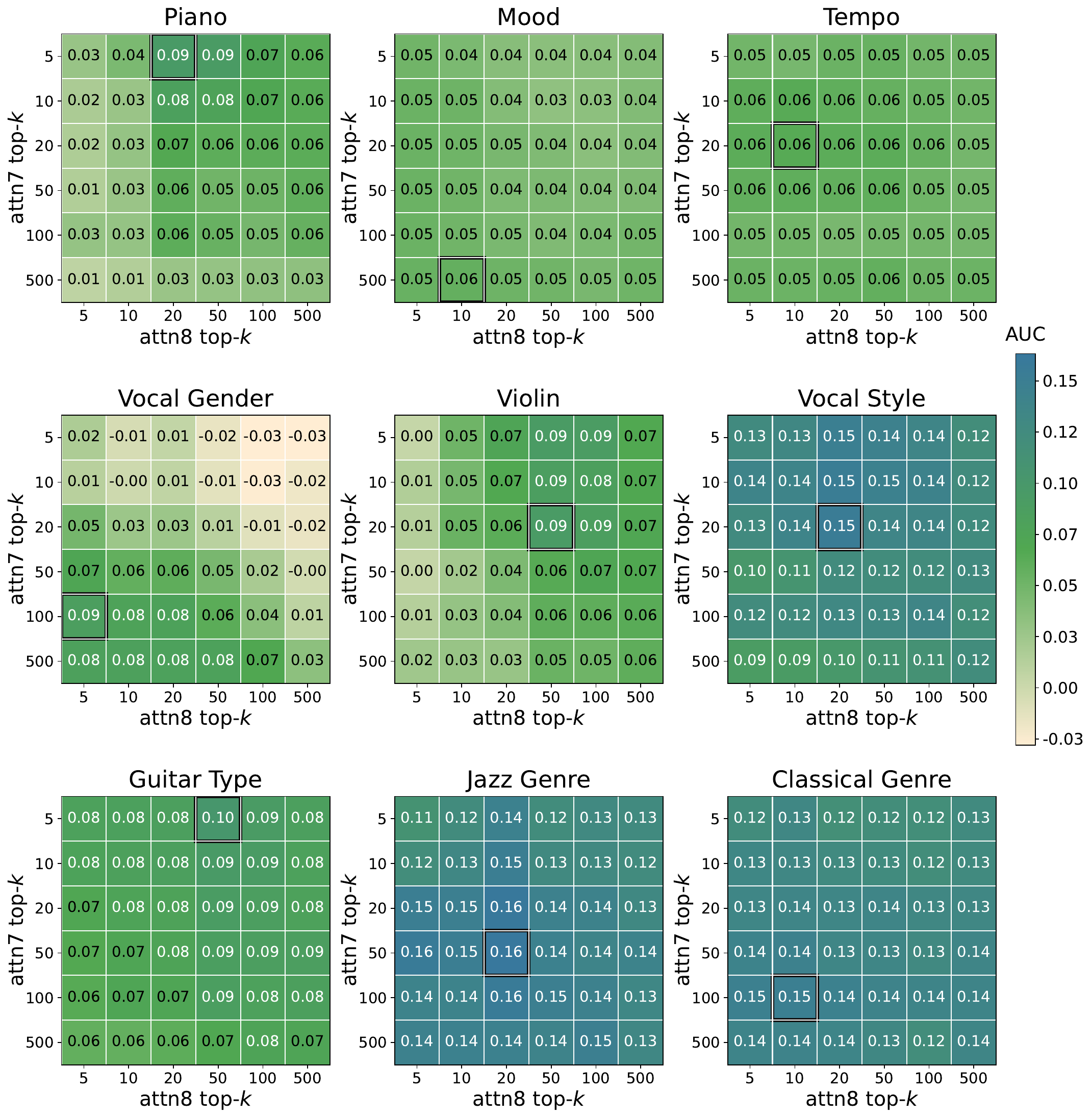}
    \caption{\textbf{Impact of SAE feature count ($k$) on each concept modulation performance.} We find that optimal configuration of $k$ for layers $(attn7,attn8)$ highly depends on given concept. In our experiments, we use the best-scoring pair for each concept, which is highlighted.}
    \label{fig:topk2d_pci_per_concept}
\end{figure}

Concept features set $\mathcal{F}_c$ is selected per layer by ranking SAE features under the TF--IDF criterion (see eq.~\ref{eq:tfidf}), scored on model activations collected from the same contrastive prompt pairs that one in CAA and AUSteer. To construct $\mathcal{F}_c$, we use the $k$ highest-scoring SAE features, where the optimal $k$ is chosen separately for each concept on the held-out dataset. In \cref{fig:topk2d_pci_per_concept}, we sweep $k\in\{5,10,20,50,100,500\}$ over both $\{7,8\}$ functional layers in ACE-Step. Our main experiments with SAE use winning pairs for concepts:
piano $(5,20)$, mood $(500,10)$, tempo $(20,10)$, vocal gender $(100,5)$, violin $(20,50)$, vocal style $(20,20)$, guitar type $(5,50)$, jazz genre $(50,20)$ and classical genre $(100,10)$. Interestingly, steering performance for some concepts (e.g., \textit{piano}, \textit{vocal gender}) depends significantly on $k$, whereas for others (e.g., \textit{mood}, \textit{tempo}) it is uniform across pairs.

The concept's steering vector $\mathbf{v}_c^{\text{SAE}}$ is constructed as a unit-weighted summation of SAE's decoder columns, corresponding to its top-$k$ selected dimensions. Alternatively, during summation, one can weight decoder columns by their corresponding TF--IDF (\cref{eq:tfidf}) scores or by their average (across prompts) value in latent space (TF). We analyze this approach in \cref{tab:sae_weighting_perconcept} and show that it is competitive to our uniform feature weighting, when weights are rescaled in a way that they cumulatively equal $k$. We denote such rescaling as $\Sigma=k$. 

We perform negative-direction (concept-suppressing) steering with SAE by applying a positive-direction steering vector with the strength parameter $\alpha$ negated. Constructed from decoder columns, the steering vector lives in the model's activation space. Negated, under the linear representation hypothesis~\citep{park2024the}, should correspond to reducing the amount of the concept in the generated audio. An alternative approach would be to identify a negative feature set via TF-IDF scoring over the concept-suppressing prompts dataset and create a steering vector from the corresponding decoder columns. In \cref{tab:sae_negdir_perconcept} we compare both approaches to negative-direction steering and show that ours (applying on-direction vector with negative steering strength) offers better concept suppression.

\begin{table}[h]
    \centering
    \caption{\textbf{Impact of SAE's features weighting on steering performance.} We report AUC on the held-out set. While weighting decoder columns by raw scores yields suboptimal results, rescaling scores $(\Sigma=k)$ yields performance competitive with our ($\star$) approach.}
    \label{tab:sae_weighting_perconcept}
    \resizebox{1.0\linewidth}{!}{
    \begin{tabular}{l*{9}{c}c}
    \toprule
    Weighting & Piano & Mood & Tempo & Voc.\,gen. & Violin & Voc.\,style & Guitar & Rock & Electr. & Avg \\
    \midrule
    unit $\star$ & $0.124$ & $0.031$ & $\mathbf{0.065}$ & $0.126$ & $\mathbf{0.117}$ & $0.233$ & $0.123$ & $\mathbf{0.236}$ & $\mathbf{0.174}$ & $\mathbf{0.137}$ \\
    TF & $0.110$ & $\mathbf{0.040}$ & $0.034$ & $0.129$ & $0.074$ & $0.239$ & $0.137$ & $0.173$ & $0.143$ & $0.120$ \\
    TF--IDF & $0.095$ & $0.032$ & $0.041$ & $0.112$ & $0.092$ & $0.220$ & $0.143$ & $0.169$ & $0.160$ & $0.118$ \\
    TF, $\Sigma=k$ & $\mathbf{0.128}$ & $\mathbf{0.040}$ & $0.052$ & $\mathbf{0.138}$ & $0.090$ & $\mathbf{0.247}$ & $\mathbf{0.167}$ & $0.225$ & $0.143$ & $\mathbf{0.137}$ \\
    TF--IDF, $\Sigma=k$ & $0.113$ & $0.033$ & $0.053$ & $0.104$ & $0.108$ & $0.215$ & $0.160$ & $0.225$ & $0.165$ & $0.131$ \\
    \bottomrule

    \end{tabular}
    }
\end{table}

\begin{table}[h]
    \centering
    \caption{\textbf{Constructing negative direction (concept-removal) steering vectors with SAE.} We report AUC on the held-out dataset. Our ($\star$) approach of applying a positive (concept-amplifying) vector with negative steering strength performs better than identifying a negative-feature set.}
    \label{tab:sae_negdir_perconcept}
    \resizebox{1.0\linewidth}{!}{
    \begin{tabular}{l*{9}{c}c}
    \toprule
    Negative SV Construction & Piano & Mood & Tempo & Voc.\,gen. & Violin & Voc.\,style & Guitar & Rock & Electr. & Avg \\
    \midrule
    negated positive dir. vector $\mathbf{\star}$ & $\mathbf{0.065}$ & $\mathbf{0.083}$ & $\mathbf{0.063}$ & $\mathbf{0.046}$ & $0.070$ & $\mathbf{0.075}$ & $\mathbf{0.081}$ & $\mathbf{0.091}$ & $\mathbf{0.123}$ & $\mathbf{0.077}$ \\
    TF--IDF on negative prompts & $0.008$ & $0.033$ & $0.058$ & $0.035$ & $\mathbf{0.075}$ & $0.045$ & $0.016$ & $0.028$ & $0.039$ & $0.037$ \\
    \bottomrule

    \end{tabular}
    }
\end{table}

\clearpage
\section{Details on user study}\label{app:user_study}

To complement the automatic evaluation, we conducted an anonymous in-browser listening study that compares the seven steering paradigms in their \emph{Standard} (all-layer) and \emph{Localized} configurations on three musical concepts: \textit{piano}, \textit{vocal gender}, and \textit{tempo}. 

\paragraph{Stimulus design.}

The study uses $15$ trials, five per concept. Each trial is anchored by a single music prompt drawn from a stylistically diverse pool covering jazz ballads, shoegaze, reggae, Latin music, cinematic soundtracks, trap beats, classic rock, and dreamy meditation. For every trial we pre-render the full $\alpha$-sweep over seven steering strengths $\alpha \in \{-\alpha_{\text{max}},-\frac{2}{3}\alpha_{\text{max}},-\frac{1}{3}\alpha_{\text{max}},0,+\frac{1}{3}\alpha_{\text{max}},+\frac{2}{3}\alpha_{\text{max}},+\alpha_{\text{max}}\}$ (calibrated to match \cref{tab:steering_auc_avg} strengths) for all $13$ steering methods, yielding $13 \times 7 = 91$ short MP3 clips per trial that are served as static assets. Each clip is rendered to loop seamlessly, so listeners can sweep $\alpha$ without hearing playback restart. The negative direction is meaningful: positive $\alpha$ \emph{adds} the concept (e.g., introduces piano) and negative $\alpha$ \emph{suppresses} it (e.g., removes piano-like content); $\alpha = 0$ is the unsteered baseline and serves as a perceptual anchor.

\paragraph{Per-trial method assignment.}

Showing all $13$ methods on every trial would extend the per-trial duration well beyond the $\sim$1\,min target, with predictable drop-off. We instead present \emph{seven} samples per trial, drawn as follows:
\begin{itemize}
    \item $\text{SAE}(\{7,8\})$ is included on every trial as an upper anchor; given the high overall performance and the fact that it has no Standard counterpart, it would otherwise be undersampled relative to the other paradigms.
    \item Three of the six remaining paradigm families (PCI, Text Emb, Token Emb, FreeSliders, AUSteer, CAA) are drawn uniformly at random per (session, trial). For each chosen family, both the \emph{Standard} and \emph{Localized} variants are included, contributing 6 samples.
\end{itemize}
The full $7$-sample list is then shuffled so position cannot leak method identity. The order of the $15$ trials is also independently shuffled per session. Both the trial order and the per-trial sample order are persisted to \texttt{localStorage}, so a refresh or a browser tab close does not reroll either. Concept Sliders is not included in this listening study.

\begin{figure}
    \centering
    \includegraphics[width=0.7\linewidth]{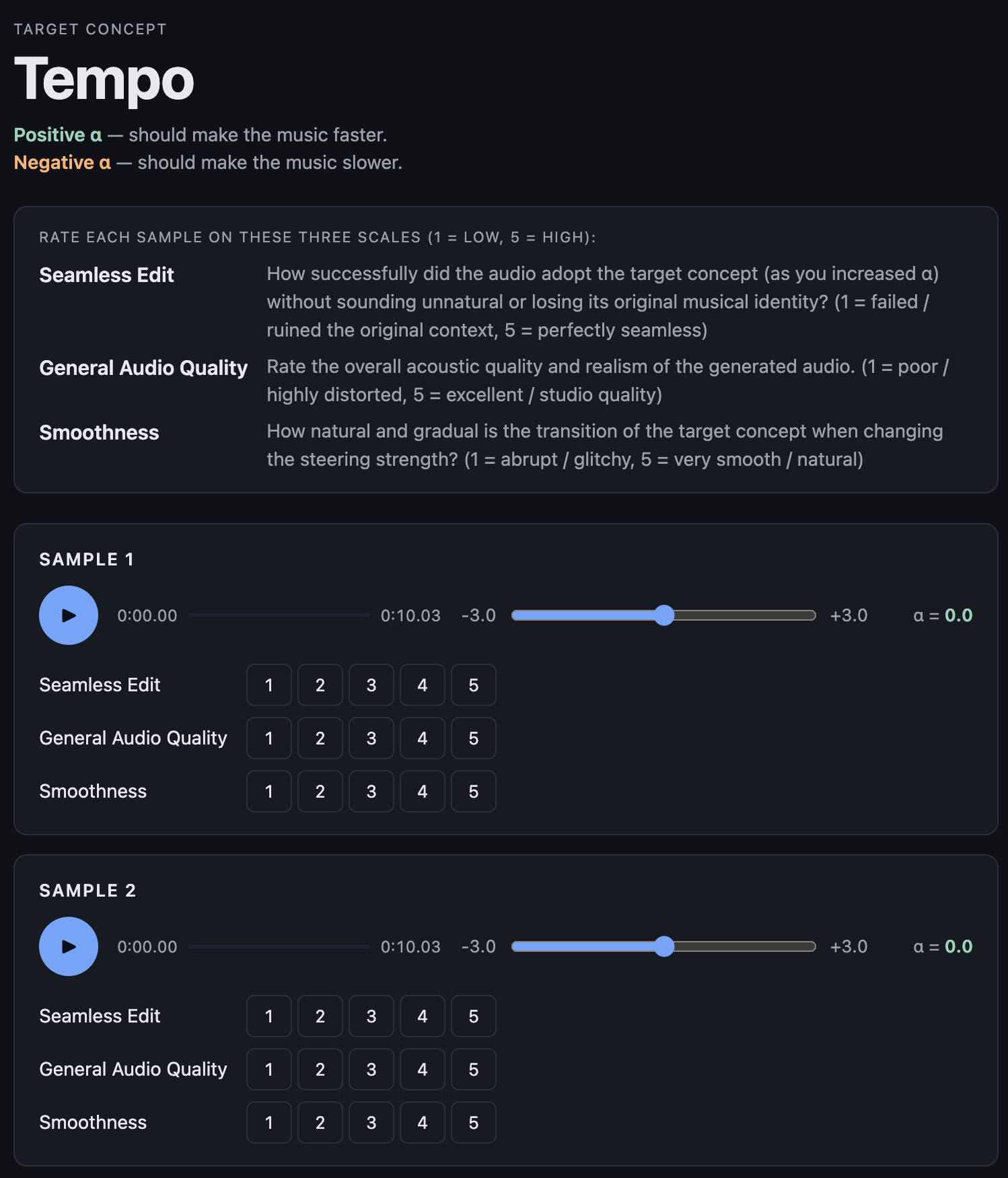}
    \caption{Print screen of the user study interface}
    \label{fig:audio_study_print_screen}
\end{figure}

\paragraph{Listening interface.}
The listening interface is presented in \cref{fig:audio_study_print_screen}. Each trial displays the target concept (e.g., ``Tempo'') and a textual hint of what positive vs.\ negative $\alpha$ should accomplish (e.g., for \textit{tempo}: ``positive $\alpha$ should make the music faster; negative $\alpha$ should make it slower''). The listener sees seven rows, one per sample, anonymized as ``Sample 1''\,\ldots\,``Sample 7''. Each row exposes a single play button and an $\alpha$ slider with seven stops. Crucially, the $\alpha$ slider does \emph{not} restart playback: when a sample begins playing, all seven of its pre-rendered loops are wired in parallel through a Web Audio gain-crossfade graph (one gain node per $\alpha$ stop, all summing into a master gain that drives the destination). Dragging the slider executes a sample-accurate gain crossfade with a $5$~ms time-constant between the active and target $\alpha$ stops, so the concept transition is heard \emph{continuously} rather than as a sequence of discrete switches -- which is what makes the \emph{Smoothness} question perceptually meaningful in a listening setting. Only one sample plays at a time: pressing play on another row auto-pauses the currently-active one. To reduce wait time during navigation, the audio buffers for the next trial are prefetched with $6$-way parallelism while the current trial is being rated.

\paragraph{Likert questions.}
Beneath the player are three Likert ($1$--$5$) scales:
\begin{itemize}
    \item \textbf{Seamless Edit.} ``How successfully did the audio adopt the target concept (as you increased $\alpha$) without sounding unnatural or losing its original musical identity?'' Anchors: $1$~=~\textit{Failed / ruined the original context}, $5$~=~\textit{Perfectly seamless}.
    \item \textbf{General Audio Quality.} ``Rate the overall acoustic quality and realism of the generated audio.'' Anchors: $1$~=~\textit{Poor / highly distorted}, $5$~=~\textit{Excellent / studio quality}.
    \item \textbf{Smoothness.} ``How natural and gradual is the transition of the target concept when changing the steering strength?'' Anchors: $1$~=~\textit{Abrupt / glitchy}, $5$~=~\textit{Very smooth / natural}.
\end{itemize}
\newpage
A trial cannot be submitted until all three scales have been answered for every one of its seven samples. The landing page additionally asks the listener to self-rate their musical-skill level on a $1$--$5$ scale ($1$~=~\textit{trouble distinguishing instruments in a recording}, through $5$~=~\textit{professional musician}); this rating is logged but is not used for filtering or for any analysis in the main paper.

\paragraph{Logging, resumption, and analysis filter.}
For every (sample, trial) pair we log the session id (a UUIDv4 minted on the landing page), the trial's presentation index, the sample index within the trial, the trial concept and prompt, all $\alpha$ values offered (\texttt{alphas}), the $\alpha$ stops that were active at any point \emph{while} the sample was playing (\texttt{alphas\_visited}; values selected with the slider while paused are not logged), whether the play button was pressed at least once (\texttt{listened}), and the seven shuffled methods chosen for that trial (\texttt{selected\_methods}).
Submission is trial-by-trial: every \emph{Next} click POSTs the
which appends seven scores to the final \texttt{results.csv}.

In our analyses, we keep only rows with \texttt{listened}~=~$1$, i.e.\ samples for which the listener actually pressed play. This filter (a) removes a small number of rows where a listener clicked through Likerts without auditioning the sample and (b) ensures that the engagement statistics reported below describe rows that contributed to the published numbers.

\paragraph{Participants and engagement.}
The study collected $1279$ rated samples across $32$ sessions, with $7$ listeners completing all $15$ trials.

Self-rated musical skill skewed towards casual listeners ($2$ sessions at level~$1$, $16$ at level~$2$, $9$ at level~$3$, $4$ at level~$4$, $1$ at level~$5$; mean $2.56$). Engagement with the $\alpha$ slider was high: in $98.4\%$ of rated samples the listener visited at least three distinct $\alpha$ stops while listening, and the mean number of $\alpha$ stops visited per sample was $6.2$ out of $7$ -- consistent with active sweeping rather than rating the default position.

\paragraph{Question correlations.}
Although the three Likert scales received broadly similar mean ratings (\emph{Seamless Edit}: $2.92$, \emph{Audio Quality}: $3.00$, \emph{Smoothness}: $2.92$), they are far from redundant. The Pearson correlations between the per-rating scores are $r{=}0.61$ between Audio Quality and Smoothness, $r{=}0.52$ between Seamless Edit and Smoothness, and $r{=}0.44$ between Seamless Edit and Audio Quality. \emph{Seamless Edit}, the dimension that asks specifically whether the steering succeeded, is therefore the most informationally independent of the three, supporting its central role in the per-method and pairwise analyses above.

\paragraph{Skill vs. method interaction.}
Self-reported musical skill modulates which paradigm a listener prefers (\cref{tab:user_study_skill_interaction}). Splitting the $32$ sessions into \emph{casual listeners} (skill $\leq 2$, $497$ rated samples) and \emph{trained musicians} (skill $\geq 3$, $782$ rated samples), the localized activation-space methods $\text{CAA}$ and $\text{AUSteer}$ receive \emph{higher} mean Seamless-Edit ratings from trained musicians than from casual listeners (CAA: $3.40$ vs.\ $3.23$; AUSteer: $3.42$ vs.\ $3.10$), while $\text{FreeSliders}$, PCI, Text Emb, and Token Emb move in the opposite direction (e.g.\ FreeSliders: $2.71$ vs.\ $3.38$, a gap of $0.67$ points). One natural reading is that musically trained listeners are more sensitive to the artifacts introduced by score- and prompt-level edits, while activation-space edits remain perceptually clean enough to score even higher with critical ears. We caveat that only one session was rated at the highest skill level ($n{=}105$ ratings), so we group skill levels $3$--$5$ together rather than reporting a finer breakdown.

\begin{table}[h]
\centering
\small
\caption{\textbf{Mean Seamless-Edit rating per method (Localized variant), split by self-reported musical skill.} Casual listener = skill $\leq 2$; trained musician = skill $\geq 3$. Activation-space methods are preferred more strongly by trained listeners; score- and prompt-level methods less so.}
\label{tab:user_study_skill_interaction}
\begin{tabular}{lccc}
\toprule
\textbf{Method (loc.)} & \textbf{Casual listeners ($\leq 2$)} & \textbf{Trained musicians ($\geq 3$)} & $\boldsymbol{\Delta}$ \\
\midrule
$\text{PCI}$         & $3.20$ & $2.80$ & $-0.40$ \\
$\text{Text Embeddings}$    & $2.69$ & $2.32$ & $-0.37$ \\
$\text{Token Embeddins}$   & $2.64$ & $2.58$ & $-0.06$ \\
$\text{FreeSliders}$ & $3.38$ & $2.71$ & $-0.67$ \\
$\text{CAA}$         & $3.23$ & $3.40$ & $+0.17$ \\
$\text{AUSteer}$     & $3.10$ & $3.42$ & $+0.32$ \\
$\text{SAE}$         & $3.27$ & $3.17$ & $-0.10$ \\
\bottomrule
\end{tabular}
\end{table}

\section{Studied audio diffusion architectures} \label{app:audio_arch}

In our experiments, we employ audio generative models with diverse architectures, namely \textit{\ldm} \citep{liu2024audioldm}, \textit{\stableaudio} \citep{evans2024stable}, and \textit{\ace} \citep{gong2025ace0step0}. While \ldm and \stableaudio are trained with diffusion loss \citep{ho2020denoising}, \ace are trained to learn a velocity field within the flow matching paradigm \citep{flowmatching}. \ldm leverages a U-Net \citep{ronneberger2015unet} architecture as a backbone and employs FLAN-T5 \citep{chung2024scaling} as text encoder. Conversely, \stableaudio is a Diffusion Transformer (DiT, \citet{Peebles_2023_ICCV}) architecture with a T5 \citep{t5model} text embedding model. Finally, \ace uses mT5 \citep{chung2023unimax} for text encoding, an autoregressive model \citep{liu2025songgen} for lyric encoding, and employs the Linear DiT \citep{xie2025sana} backbone for the denoising process.

\clearpage

\section{Additional results for localized activation steering}\label{app:loc_steering_tables}
\subsection{Localized activation steering}
\begin{figure}[ht]
    \centering
    \includegraphics[width=1.0\linewidth]{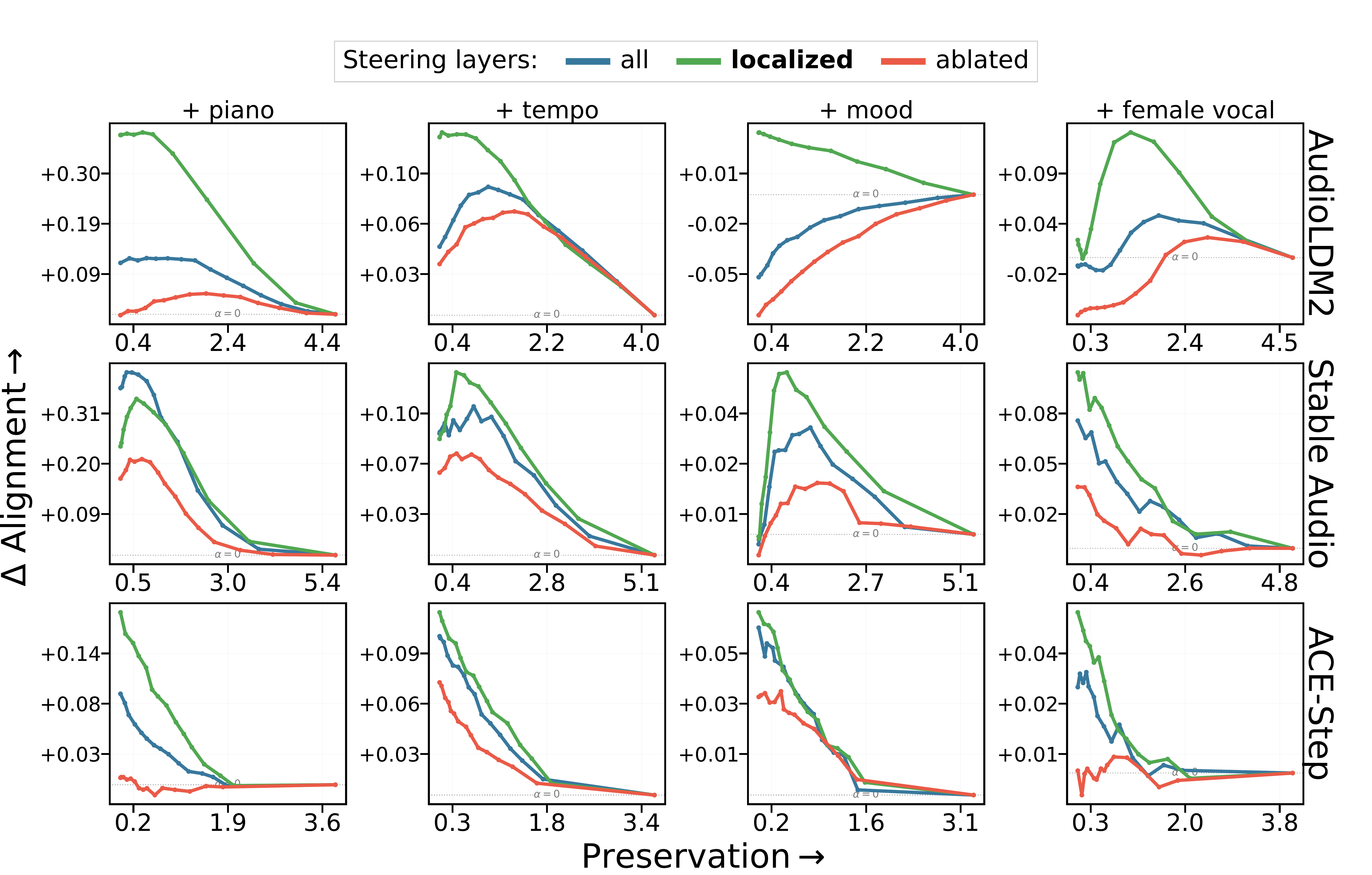}
    \caption{\textbf{Layer selection in activation steering is crucial for audio concept modulation.} Steering the activations in localized layers outperforms steering all of the layers, while steering all layers except the localized ones leads to drastic degradation.}
    \label{fig:loc_steering_full}
\end{figure}

In \cref{fig:loc_steering_full}, we present a generalization of our localized activation steering with CAA across $4$ musical concepts and $3$ audio diffusion models. Notably, for some concepts (mood in AudioLDM2 and Stable Audio Open, and female vocal in ACE-Step), steering without our layers results in concept degradation. In \cref{tab:audioldm2_auc_csm,tab:stableaudio_auc_csm,tab:acestep_auc_csm} we present quantitative results supporting this.

\begin{table}[h]
\centering
\caption{\textbf{Per-concept steering metrics for AudioLDM2 with CAA.} Higher is better for AUC, lower for Smoothness. The best per row is \textbf{bolded}.}
\label{tab:audioldm2_auc_csm}

    \resizebox{1.0\linewidth}{!}{
\begin{tabular}{llcccc}
\toprule
Concept & Layers & AUC (MuQ) $\uparrow$ & AUC (CLAP) $\uparrow$ & Smoothness (MuQ) $\downarrow$ & Smoothness (CLAP) $\downarrow$ \\
\midrule
\multirow{3}{*}{Piano} & all & 0.331 & 0.175 & \textbf{0.081} & 0.098 \\
& localized & \textbf{0.930} & \textbf{0.390} & 0.111 & \textbf{0.064} \\
& ablated & 0.099 & 0.099 & 0.176 & 0.171 \\
\midrule
\multirow{3}{*}{Tempo} & all & 0.236 & 0.008 & 0.125 & 0.222 \\
& localized & \textbf{0.300} & \textbf{0.067} & \textbf{0.065} & \textbf{0.073} \\
& ablated & 0.201 & -0.007 & 0.135 & 0.605 \\
\midrule
\multirow{3}{*}{Mood} & all & -0.066 & -0.005 & -- & 0.361 \\
& localized & \textbf{0.089} & \textbf{0.152} & \textbf{0.069} & \textbf{0.076} \\
& ablated & -0.130 & -0.073 & -- & 0.827 \\
\midrule
\multirow{3}{*}{Female vocal} & all & 0.095 & 0.107 & 0.226 & 0.233 \\
& localized & \textbf{0.304} & \textbf{0.337} & \textbf{0.193} & \textbf{0.128} \\
& ablated & -0.045 & -0.061 & 0.437 & 0.472 \\
\midrule
\multirow{3}{*}{\textbf{Average}} & all & 0.149 & 0.071 & 0.144 & 0.229 \\
& localized & \textbf{0.406} & \textbf{0.236} & \textbf{0.109} & \textbf{0.086} \\
& ablated & 0.031 & -0.011 & 0.249 & 0.519 \\
\bottomrule
\end{tabular}}
\end{table}

\begin{table}[h]
\centering
\caption{\textbf{Per-concept steering metrics for Stable Audio Open with CAA.} Higher is better for AUC, lower for Smoothness. The best per row is \textbf{bolded}.}
\label{tab:stableaudio_auc_csm}

    \resizebox{1.0\linewidth}{!}{
\begin{tabular}{llcccc}
\toprule
Concept & Layers & AUC (MuQ) $\uparrow$ & AUC (CLAP) $\uparrow$ & Smoothness (MuQ) $\downarrow$ & Smoothness (CLAP) $\downarrow$ \\
\midrule
\multirow{3}{*}{Piano} & all & \textbf{0.792} & \textbf{0.228} & 0.093 & \textbf{0.105} \\
& localized & 0.742 & 0.220 & 0.128 & 0.154 \\
& ablated & 0.383 & 0.094 & \textbf{0.086} & 0.129 \\
\midrule
\multirow{3}{*}{Tempo} & all & 0.268 & 0.108 & 0.110 & 0.100 \\
& localized & \textbf{0.323} & \textbf{0.216} & 0.113 & \textbf{0.085} \\
& ablated & 0.183 & 0.031 & \textbf{0.096} & 0.202 \\
\midrule
\multirow{3}{*}{Mood} & all & 0.074 & 0.241 & 0.199 & 0.099 \\
& localized & \textbf{0.119} & \textbf{0.246} & \textbf{0.183} & \textbf{0.099} \\
& ablated & 0.034 & 0.163 & 0.240 & 0.099 \\
\midrule
\multirow{3}{*}{Female vocal} & all & 0.106 & \textbf{0.122} & \textbf{0.089} & \textbf{0.041} \\
& localized & \textbf{0.153} & 0.097 & 0.110 & 0.062 \\
& ablated & 0.030 & 0.087 & 0.148 & 0.066 \\
\midrule
\multirow{3}{*}{\textbf{Average}} & all & 0.310 & 0.175 & \textbf{0.123} & \textbf{0.086} \\
& localized & \textbf{0.334} & \textbf{0.195} & 0.134 & 0.100 \\
& ablated & 0.158 & 0.094 & 0.142 & 0.124 \\
\bottomrule
\end{tabular}}
\end{table}

\begin{table}[h]
\centering
\caption{\textbf{Per-concept steering metrics for ACE-Step with CAA.} Higher is better for AUC, lower for Smoothness. The best per row is \textbf{bolded}.}
\label{tab:acestep_auc_csm}

    \resizebox{1.0\linewidth}{!}{
\begin{tabular}{llcccc}
\toprule
Concept & Layers & AUC (MuQ) $\uparrow$ & AUC (CLAP) $\uparrow$ & Smoothness (MuQ) $\downarrow$ & Smoothness (CLAP) $\downarrow$ \\
\midrule
\multirow{3}{*}{Piano} & all & 0.060 & 0.026 & 0.038 & 0.063 \\
& localized & \textbf{0.141} & \textbf{0.061} & \textbf{0.033} & \textbf{0.032} \\
& ablated & -0.008 & -0.010 & 0.488 & -- \\
\midrule
\multirow{3}{*}{Tempo} & all & 0.095 & 0.022 & 0.032 & 0.075 \\
& localized & \textbf{0.111} & \textbf{0.023} & \textbf{0.029} & \textbf{0.042} \\
& ablated & 0.059 & 0.018 & 0.034 & 0.076 \\
\midrule
\multirow{3}{*}{Mood} & all & 0.043 & \textbf{0.049} & 0.069 & 0.038 \\
& localized & \textbf{0.049} & 0.046 & \textbf{0.042} & \textbf{0.035} \\
& ablated & 0.036 & 0.044 & 0.087 & 0.064 \\
\midrule
\multirow{3}{*}{Female vocal} & all & 0.020 & 0.104 & 0.138 & 0.034 \\
& localized & \textbf{0.029} & \textbf{0.116} & \textbf{0.068} & \textbf{0.034} \\
& ablated & -0.002 & 0.058 & 0.736 & 0.082 \\
\midrule
\multirow{3}{*}{\textbf{Average}} & all & 0.055 & 0.051 & 0.069 & 0.053 \\
& localized & \textbf{0.083} & \textbf{0.061} & \textbf{0.043} & \textbf{0.036} \\
& ablated & 0.021 & 0.027 & 0.336 & 0.074 \\
\bottomrule
\end{tabular}}
\end{table}

\clearpage

\subsection{Quantitative methods' configurations comparison}
In \cref{tab:steering_auc_localization}, we provide a full table comparing global (all layers steered) and localized (functional layers steered) configurations for all steering methods. 

\begin{table}[h]
    \centering
    \caption{\textbf{Effect of layer localization on each steering method.} For every method, we compare its standard variant (steering all diffusion model layers) against our localized variant (applying the intervention only at layers \{7,8\}). Mean $\pm$ standard error pooled across 9 musical concepts and both steering directions (see \cref{tab:steering_auc_piano,tab:steering_auc_violin,tab:steering_auc_guitar,tab:steering_auc_tempo,tab:steering_auc_mood,tab:steering_auc_vocal_gender,tab:steering_auc_vocal_style,tab:steering_auc_popjazz,tab:steering_auc_classical_electronic} in Appendix). The $\Delta$ row reports the relative gain from localization, expressed as a \% change of the Standard mean (sign chosen so that ``$+$'' always denotes improvement): \textcolor{gainpos}{\textbf{green}} when localization helps and \textcolor{gainneg}{\textbf{red}} when it hurts, faded when the absolute gain is below the standard error of the difference.}
    \label{tab:steering_auc_localization}
    \vspace{0.5em}
    \resizebox{1.0\linewidth}{!}{
        \begin{tabular}{lccccc}
            \toprule
            \multirow{2}{*}{\textbf{Method}}
            & \multicolumn{2}{c}{\textbf{AUC} ($\uparrow$)}
            & \multicolumn{2}{c}{\textbf{Smoothness} ($\downarrow$)}
            & \multirow{2}{*}{\shortstack{\textbf{ Audio}\\\textbf{Quality} ($\uparrow$)}} \\
            \cmidrule(lr){2-3}\cmidrule(lr){4-5}
            & MuQ & CLAP & MuQ & CLAP & \\
            \midrule
            PCI~\textcolor{gray}{(\citet{gorgun2026temporal})} & $0.084_{\pm.019}$ & $0.049_{\pm.010}$ & $0.185_{\pm.064}$ & $0.223_{\pm.066}$ & $6.803_{\pm.016}$ \\
            \quad + \textbf{localized} & $0.068_{\pm.016}$ & $0.038_{\pm.008}$ & $0.277_{\pm.143}$ & $0.122_{\pm.017}$ & $6.789_{\pm.014}$ \\
            \quad $\Delta$ & \dneg{-19\%} & \dneg{-22\%} & \dneg{-50\%} & \dpos{+21\%} & \dneg{-0.2\%} \\
            \midrule
            Text Embeddings~\textcolor{gray}{(\citet{ezra2025freesliders0})} & $0.014_{\pm.007}$ & $0.011_{\pm.004}$ & $0.439_{\pm.224}$ & $0.681_{\pm.277}$ & $6.690_{\pm.027}$ \\
            \quad + \textbf{localized} & $0.023_{\pm.006}$ & $0.014_{\pm.004}$ & $0.334_{\pm.123}$ & $0.366_{\pm.140}$ & $6.780_{\pm.014}$ \\
            \quad $\Delta$ & \dpos{+62\%} & \dpos{+31\%} & \dpos{+25\%} & \dposf{+48\%} & \dpos{+1.4\%} \\
            \midrule
            Token Embeddings~\textcolor{gray}{(\citet{baumann2025continuous})} & $0.035_{\pm.011}$ & $0.024_{\pm.006}$ & $0.267_{\pm.063}$ & $0.211_{\pm.058}$ & $6.776_{\pm.023}$ \\
            \quad + \textbf{localized} & $0.032_{\pm.009}$ & $0.020_{\pm.006}$ & $0.208_{\pm.073}$ & $0.282_{\pm.098}$ & $6.794_{\pm.013}$ \\
            \quad $\Delta$ & \dnegf{-8\%} & \dneg{-20\%} & \dposf{+18\%} & \dneg{-33\%} & \dpos{+0.3\%} \\
            \midrule
            FreeSliders~\textcolor{gray}{(\citet{ezra2025freesliders0})} & $0.079_{\pm.012}$ & $0.049_{\pm.009}$ & $0.055_{\pm.007}$ & $0.066_{\pm.012}$ & $6.790_{\pm.017}$ \\
            \quad + \textbf{localized} & $0.080_{\pm.012}$ & $0.049_{\pm.009}$ & $0.065_{\pm.014}$ & $0.074_{\pm.011}$ & $6.790_{\pm.017}$ \\
            \quad $\Delta$ & \dposf{+1\%} & \dnegf{-1\%} & \dneg{-18\%} & \dneg{-13\%} & \dnegf{-0.0\%} \\
            \midrule
            Concept Sliders~\textcolor{gray}{(\citet{gandikota2024concept})} & $0.086_{\pm.010}$ & $0.067_{\pm.011}$ & $0.056_{\pm.006}$ & $0.045_{\pm.006}$ & $6.762_{\pm.019}$ \\
            \quad + \textbf{localized} & $0.068_{\pm.010}$ & $0.050_{\pm.009}$ & $0.099_{\pm.037}$ & $0.060_{\pm.009}$ & $6.776_{\pm.019}$ \\
            \quad $\Delta$ & \dneg{-21\%} & \dneg{-25\%} & \dneg{-77\%} & \dneg{-34\%} & \dpos{+0.2\%} \\
            \midrule
            AUSteer~\textcolor{gray}{(\citet{feng2026finegrained})} & $0.081_{\pm.020}$ & $0.058_{\pm.013}$ & $0.212_{\pm.069}$ & $0.065_{\pm.011}$ & $6.758_{\pm.014}$ \\
            \quad + \textbf{localized} & $0.096_{\pm.016}$ & $0.063_{\pm.012}$ & $0.054_{\pm.006}$ & $0.050_{\pm.005}$ & $6.775_{\pm.018}$ \\
            \quad $\Delta$ & \dpos{+19\%} & \dposf{+8\%} & \dpos{+74\%} & \dpos{+22\%} & \dpos{+0.3\%} \\
            \midrule
            CAA~\textcolor{gray}{(\citet{rimsky-caa})} & $0.075_{\pm.016}$ & $0.046_{\pm.010}$ & $0.092_{\pm.034}$ & $0.071_{\pm.015}$ & $6.787_{\pm.016}$ \\
            \quad + \textbf{localized} & $0.104_{\pm.019}$ & $0.065_{\pm.011}$ & $0.063_{\pm.018}$ & $0.052_{\pm.010}$ & $6.792_{\pm.017}$ \\
            \quad $\Delta$ & \dpos{+39\%} & \dpos{+40\%} & \dpos{+32\%} & \dpos{+26\%} & \dpos{+0.1\%} \\
            \bottomrule
        \end{tabular}
    }
\end{table}

\clearpage

\section{Audio concept steering benchmark}\label{app:steering_benchmark}

\subsection{Methods comparison across concepts}\label{app:subsection_methods_comparison}
We compare all steering methods across $9$ musical features, in both positive and negative directions. \cref{tab:steering_auc_piano,tab:steering_auc_mood,tab:steering_auc_tempo,tab:steering_auc_vocal_gender,tab:steering_auc_vocal_style,tab:steering_auc_violin,tab:steering_auc_classical_electronic,tab:steering_auc_guitar,tab:steering_auc_popjazz} present evaluation of methods across musical concepts.
\setcounter{topnumber}{4}
\setcounter{totalnumber}{4}
\renewcommand{\topfraction}{1.0}
\renewcommand{\textfraction}{0.0}
\setlength{\textfloatsep}{6pt plus 2pt minus 2pt}
\setlength{\floatsep}{6pt plus 2pt minus 2pt}
\setlength{\intextsep}{6pt plus 2pt minus 2pt}
\vspace{-0em}
\begin{table}[h]
    \centering
    \caption{\textbf{Evaluating steering \textit{piano} concept with \ace.} We describe metrics in \cref{sec:metrics}. Symbols \textit{+} and \textit{-} denote the quality of steering towards positive (\textit{add piano}) and negative (\textit{remove piano}) directions, \textit{avg} = (pos + neg) / 2. \textbf{Best} and \uline{second} results are highlighted.}
    \label{tab:steering_auc_piano}
    \vspace{0.5em}
    \resizebox{1.0\linewidth}{!}{
        \begin{tabular}{lccccccccccccc}
            \toprule
            \multirow{2}{*}{\shortstack[l]{\textbf{Method}\\(\textit{piano})}}
                & \multicolumn{3}{c}{\textbf{AUC MuQ} ($\uparrow$)}
                & \multicolumn{3}{c}{\textbf{AUC CLAP} ($\uparrow$)}
                & \multicolumn{3}{c}{\textbf{Smoothness MuQ} ($\downarrow$)}
                & \multicolumn{3}{c}{\textbf{Smoothness CLAP} ($\downarrow$)}
                & \multirow{2}{*}{\shortstack{\textbf{Avg}\\\textbf{Quality} ($\uparrow$)}} \\
            \cmidrule(lr){2-4}\cmidrule(lr){5-7}\cmidrule(lr){8-10}\cmidrule(lr){11-13}
                & + & - & avg & + & - & avg & + & - & avg & + & - & avg & \\
            \midrule
  PCI & $0.104$ & $0.004$ & $0.054$ & $0.044$ & $0.001$ & $0.022$ & $0.066$ & $0.418$ & $0.242$ & $0.077$ & $0.501$ & $0.289$ & $6.814$ \\
  PCI \textbf{(loc.)} & $0.081$ & $0.007$ & $0.044$ & $0.027$ & $0.010$ & $0.018$ & $0.080$ & $0.547$ & $0.313$ & $0.094$ & $0.270$ & $0.182$ & $6.809$ \\
  Text Embeddings & $0.056$ & $0.005$ & $0.030$ & $0.027$ & $0.022$ & $0.024$ & $0.065$ & $0.415$ & $0.240$ & $0.073$ & $0.051$ & $0.062$ & $6.803$ \\
  Text Embeddings \textbf{(loc.)} & $0.060$ & $0.015$ & $0.038$ & $0.020$ & $0.019$ & $0.019$ & $0.069$ & $0.385$ & $0.227$ & $0.075$ & $0.151$ & $0.113$ & $6.796$ \\
  Token Embeddings & $0.127$ & $0.040$ & $0.084$ & $0.049$ & $0.021$ & $0.035$ & $0.084$ & $0.085$ & $0.084$ & $0.078$ & $0.082$ & $0.080$ & $6.793$ \\
  Token Embeddings \textbf{(loc.)} & $0.102$ & $0.040$ & $0.071$ & $0.031$ & $0.026$ & $0.028$ & $0.081$ & $0.084$ & $0.083$ & $0.089$ & $0.099$ & $0.094$ & $6.780$ \\
  FreeSliders & $0.085$ & $0.049$ & $0.067$ & $0.033$ & $0.026$ & $0.030$ & $0.049$ & $0.038$ & $0.044$ & $0.052$ & $0.069$ & $0.061$ & $6.787$ \\
  FreeSliders \textbf{(loc.)} & $0.093$ & $0.056$ & $0.075$ & $0.036$ & $0.028$ & $0.032$ & $0.038$ & $0.031$ & \textbf{0.035} & $0.038$ & $0.066$ & $0.052$ & $6.786$ \\
  Concept Sliders & $0.098$ & $0.059$ & $0.078$ & $0.056$ & $0.032$ & $0.044$ & $0.041$ & $0.029$ & \uline{0.035} & $0.031$ & $0.057$ & $0.044$ & $6.793$ \\
  Concept Sliders \textbf{(loc.)} & $0.071$ & $0.050$ & $0.060$ & $0.039$ & $0.026$ & $0.033$ & $0.056$ & $0.041$ & $0.048$ & $0.050$ & $0.070$ & $0.060$ & $6.795$ \\
  AUSteer & $0.097$ & $0.000$ & $0.049$ & $0.012$ & $0.029$ & $0.020$ & $0.039$ & $0.653$ & $0.346$ & $0.133$ & $0.097$ & $0.115$ & $6.762$ \\
  AUSteer \textbf{(loc.)} & $0.153$ & $0.052$ & $0.102$ & $0.040$ & $0.034$ & $0.037$ & $0.022$ & $0.053$ & $0.037$ & $0.030$ & $0.049$ & $0.040$ & $6.784$ \\
  CAA & $0.060$ & $0.044$ & $0.052$ & $0.026$ & $0.024$ & $0.025$ & $0.038$ & $0.066$ & $0.052$ & $0.063$ & $0.095$ & $0.079$ & $6.785$ \\
  CAA \textbf{(loc.)} & $0.141$ & $0.073$ & \uline{0.107} & $0.061$ & $0.044$ & \uline{0.053} & $0.033$ & $0.051$ & $0.042$ & $0.032$ & $0.044$ & \textbf{0.038} & $6.788$ \\
  SAE \textbf{(loc.)} & $0.196$ & $0.084$ & \textbf{0.140} & $0.080$ & $0.034$ & \textbf{0.057} & $0.025$ & $0.055$ & $0.040$ & $0.028$ & $0.049$ & \uline{0.038} & $6.791$ \\
            \bottomrule
        \end{tabular}
    }
\end{table}

\vspace{-0em}
\begin{table}[h]
    \centering
    \caption{\textbf{Evaluating steering \textit{tempo} concept with \ace.} We describe metrics in \cref{sec:metrics}. Symbols \textit{+} and \textit{-} denote the quality of steering towards positive (\textit{fast song}) and negative (\textit{slow song}) directions, \textit{avg} = (pos + neg) / 2. \textbf{Best} and \uline{second} results are highlighted.}
    \label{tab:steering_auc_tempo}
    \vspace{0.5em}
    \resizebox{1.0\linewidth}{!}{
        \begin{tabular}{lccccccccccccc}
            \toprule
            \multirow{2}{*}{\shortstack[l]{\textbf{Method}\\(\textit{tempo})}}
                & \multicolumn{3}{c}{\textbf{AUC MuQ} ($\uparrow$)}
                & \multicolumn{3}{c}{\textbf{AUC CLAP} ($\uparrow$)}
                & \multicolumn{3}{c}{\textbf{Smoothness MuQ} ($\downarrow$)}
                & \multicolumn{3}{c}{\textbf{Smoothness CLAP} ($\downarrow$)}
                & \multirow{2}{*}{\shortstack{\textbf{Avg}\\\textbf{Quality} ($\uparrow$)}} \\
            \cmidrule(lr){2-4}\cmidrule(lr){5-7}\cmidrule(lr){8-10}\cmidrule(lr){11-13}
                & + & - & avg & + & - & avg & + & - & avg & + & - & avg & \\
            \midrule
  PCI & $0.052$ & $0.075$ & $0.063$ & $0.008$ & $0.027$ & $0.017$ & $0.081$ & $0.055$ & $0.068$ & $0.190$ & $0.076$ & $0.133$ & $6.746$ \\
  PCI \textbf{(loc.)} & $0.047$ & $0.054$ & $0.051$ & $0.007$ & $0.019$ & $0.013$ & $0.081$ & $0.059$ & $0.070$ & $0.204$ & $0.070$ & $0.137$ & $6.742$ \\
  Text Embeddings & $0.015$ & $0.011$ & $0.013$ & $-0.016$ & $0.021$ & $0.002$ & $0.234$ & $0.115$ & $0.175$ & $2.731$ & $0.072$ & $1.402$ & $6.678$ \\
  Text Embeddings \textbf{(loc.)} & $0.016$ & $0.015$ & $0.016$ & $-0.001$ & $-0.001$ & $-0.001$ & $0.208$ & $0.119$ & $0.163$ & $0.404$ & $0.684$ & $0.544$ & $6.782$ \\
  Token Embeddings & $0.012$ & $0.025$ & $0.018$ & $0.011$ & $0.020$ & $0.015$ & $0.166$ & $0.150$ & $0.158$ & $0.082$ & $0.088$ & $0.085$ & $6.741$ \\
  Token Embeddings \textbf{(loc.)} & $0.025$ & $0.012$ & $0.019$ & $0.013$ & $0.016$ & $0.014$ & $0.097$ & $0.118$ & $0.107$ & $0.077$ & $0.094$ & $0.086$ & $6.778$ \\
  FreeSliders & $0.071$ & $0.076$ & $0.073$ & $0.014$ & $0.031$ & $0.023$ & $0.029$ & $0.033$ & \textbf{0.031} & $0.078$ & $0.046$ & $0.062$ & $6.759$ \\
  FreeSliders \textbf{(loc.)} & $0.080$ & $0.075$ & $0.078$ & $0.015$ & $0.029$ & $0.022$ & $0.034$ & $0.028$ & \uline{0.031} & $0.080$ & $0.073$ & $0.077$ & $6.762$ \\
  Concept Sliders & $0.065$ & $0.059$ & $0.062$ & $0.014$ & $0.019$ & $0.017$ & $0.050$ & $0.030$ & $0.040$ & $0.119$ & $0.054$ & $0.087$ & $6.760$ \\
  Concept Sliders \textbf{(loc.)} & $0.063$ & $0.061$ & $0.062$ & $0.016$ & $0.017$ & $0.017$ & $0.049$ & $0.029$ & $0.039$ & $0.132$ & $0.060$ & $0.096$ & $6.760$ \\
  AUSteer & $0.095$ & $0.090$ & $0.092$ & $0.020$ & $0.038$ & \uline{0.029} & $0.038$ & $0.042$ & $0.040$ & $0.070$ & $0.046$ & $0.058$ & $6.769$ \\
  AUSteer \textbf{(loc.)} & $0.077$ & $0.082$ & $0.080$ & $0.019$ & $0.036$ & $0.028$ & $0.060$ & $0.038$ & $0.049$ & $0.064$ & $0.045$ & $0.054$ & $6.769$ \\
  CAA & $0.095$ & $0.073$ & $0.084$ & $0.022$ & $0.036$ & \textbf{0.029} & $0.032$ & $0.039$ & $0.036$ & $0.075$ & $0.032$ & \uline{0.054} & $6.781$ \\
  CAA \textbf{(loc.)} & $0.111$ & $0.099$ & \uline{0.105} & $0.023$ & $0.028$ & $0.025$ & $0.029$ & $0.037$ & $0.033$ & $0.042$ & $0.040$ & \textbf{0.041} & $6.786$ \\
  SAE \textbf{(loc.)} & $0.121$ & $0.100$ & \textbf{0.110} & $0.021$ & $0.014$ & $0.018$ & $0.048$ & $0.032$ & $0.040$ & $0.054$ & $0.072$ & $0.063$ & $6.761$ \\
            \bottomrule
        \end{tabular}
    }
\end{table}

\vspace{-0em}
\begin{table}[h]
    \centering
    \caption{\textbf{Evaluating steering \textit{mood} concept with \ace.} We describe metrics in \cref{sec:metrics}. Symbols \textit{+} and \textit{-} denote the quality of steering towards positive (\textit{happy song}) and negative (\textit{sad song}) directions, \textit{avg} = (pos + neg) / 2. \textbf{Best} and \uline{second} results are highlighted.}
    \label{tab:steering_auc_mood}
    \vspace{0.5em}
    \resizebox{1.0\linewidth}{!}{
        \begin{tabular}{lccccccccccccc}
            \toprule
            \multirow{2}{*}{\shortstack[l]{\textbf{Method}\\(\textit{mood})}}
                & \multicolumn{3}{c}{\textbf{AUC MuQ} ($\uparrow$)}
                & \multicolumn{3}{c}{\textbf{AUC CLAP} ($\uparrow$)}
                & \multicolumn{3}{c}{\textbf{Smoothness MuQ} ($\downarrow$)}
                & \multicolumn{3}{c}{\textbf{Smoothness CLAP} ($\downarrow$)}
                & \multirow{2}{*}{\shortstack{\textbf{Avg}\\\textbf{Quality} ($\uparrow$)}} \\
            \cmidrule(lr){2-4}\cmidrule(lr){5-7}\cmidrule(lr){8-10}\cmidrule(lr){11-13}
                & + & - & avg & + & - & avg & + & - & avg & + & - & avg & \\
            \midrule
  PCI & $0.028$ & $0.027$ & $0.027$ & $0.020$ & $0.049$ & $0.034$ & $0.102$ & $0.097$ & $0.099$ & $0.103$ & $0.104$ & $0.103$ & $6.766$ \\
  PCI \textbf{(loc.)} & $0.008$ & $0.031$ & $0.020$ & $0.008$ & $0.044$ & $0.026$ & $0.178$ & $0.077$ & $0.127$ & $0.103$ & $0.069$ & $0.086$ & $6.748$ \\
  Text Embeddings & $0.010$ & $-0.001$ & $0.004$ & $-0.002$ & $0.019$ & $0.008$ & $0.281$ & $0.181$ & $0.231$ & $0.573$ & $0.062$ & $0.318$ & $6.693$ \\
  Text Embeddings \textbf{(loc.)} & $0.007$ & $0.009$ & $0.008$ & $0.000$ & $0.027$ & $0.014$ & $0.178$ & $0.184$ & $0.181$ & $0.407$ & $0.084$ & $0.246$ & $6.764$ \\
  Token Embeddings & $0.002$ & $0.001$ & $0.001$ & $-0.005$ & $0.026$ & $0.010$ & $0.462$ & $0.265$ & $0.364$ & $0.938$ & $0.116$ & $0.527$ & $6.710$ \\
  Token Embeddings \textbf{(loc.)} & $-0.005$ & $0.001$ & $-0.002$ & $-0.009$ & $0.017$ & $0.004$ & $-$ & $0.291$ & $-$ & $1.839$ & $0.102$ & $0.971$ & $6.755$ \\
  FreeSliders & $0.034$ & $0.026$ & $0.030$ & $0.024$ & $0.043$ & $0.033$ & $0.052$ & $0.058$ & \uline{0.055} & $0.050$ & $0.028$ & $0.039$ & $6.769$ \\
  FreeSliders \textbf{(loc.)} & $0.028$ & $0.025$ & $0.027$ & $0.015$ & $0.043$ & $0.029$ & $0.054$ & $0.064$ & $0.059$ & $0.129$ & $0.051$ & $0.090$ & $6.766$ \\
  Concept Sliders & $0.036$ & $0.065$ & \uline{0.051} & $0.040$ & $0.099$ & $0.070$ & $0.066$ & $0.051$ & $0.058$ & $0.044$ & $0.024$ & \uline{0.034} & $6.716$ \\
  Concept Sliders \textbf{(loc.)} & $0.024$ & $0.043$ & $0.033$ & $0.032$ & $0.075$ & $0.054$ & $0.103$ & $0.046$ & $0.074$ & $0.052$ & $0.025$ & $0.039$ & $6.745$ \\
  AUSteer & $0.007$ & $0.044$ & $0.026$ & $0.031$ & $0.097$ & $0.064$ & $0.459$ & $0.058$ & $0.259$ & $0.093$ & $0.038$ & $0.066$ & $6.723$ \\
  AUSteer \textbf{(loc.)} & $0.037$ & $0.069$ & \textbf{0.053} & $0.045$ & $0.105$ & \textbf{0.075} & $0.091$ & $0.040$ & $0.066$ & $0.053$ & $0.025$ & $0.039$ & $6.732$ \\
  CAA & $0.043$ & $0.048$ & $0.046$ & $0.049$ & $0.088$ & $0.069$ & $0.069$ & $0.059$ & $0.064$ & $0.038$ & $0.034$ & $0.036$ & $6.739$ \\
  CAA \textbf{(loc.)} & $0.049$ & $0.048$ & $0.049$ & $0.046$ & $0.095$ & \uline{0.071} & $0.042$ & $0.058$ & \textbf{0.050} & $0.035$ & $0.019$ & \textbf{0.027} & $6.746$ \\
  SAE \textbf{(loc.)} & $0.023$ & $0.070$ & $0.046$ & $0.036$ & $0.091$ & $0.063$ & $0.090$ & $0.067$ & $0.078$ & $0.059$ & $0.075$ & $0.067$ & $6.735$ \\
            \bottomrule
        \end{tabular}
    }
\end{table}

\begin{table}[h]
    \centering
    \caption{\textbf{Evaluating steering \textit{vocal gender} concept with \ace.} We describe metrics in \cref{sec:metrics}. Symbols \textit{+} and \textit{-} denote the quality of steering towards positive (\textit{female vocal}) and negative (\textit{male vocal}) directions, \textit{avg} = (pos + neg) / 2. \textbf{Best} and \uline{second} results are highlighted.}
    \label{tab:steering_auc_vocal_gender}
    \vspace{0.5em}
    \resizebox{1.0\linewidth}{!}{
        \begin{tabular}{lccccccccccccc}
            \toprule
            \multirow{2}{*}{\shortstack[l]{\textbf{Method}\\(\textit{vocal gender})}}
                & \multicolumn{3}{c}{\textbf{AUC MuQ} ($\uparrow$)}
                & \multicolumn{3}{c}{\textbf{AUC CLAP} ($\uparrow$)}
                & \multicolumn{3}{c}{\textbf{Smoothness MuQ} ($\downarrow$)}
                & \multicolumn{3}{c}{\textbf{Smoothness CLAP} ($\downarrow$)}
                & \multirow{2}{*}{\shortstack{\textbf{Avg}\\\textbf{Quality} ($\uparrow$)}} \\
            \cmidrule(lr){2-4}\cmidrule(lr){5-7}\cmidrule(lr){8-10}\cmidrule(lr){11-13}
                & + & - & avg & + & - & avg & + & - & avg & + & - & avg & \\
            \midrule
  PCI & $0.045$ & $0.010$ & $0.027$ & $0.132$ & $0.058$ & $0.095$ & $0.088$ & $0.249$ & $0.168$ & $0.053$ & $0.109$ & $0.081$ & $6.882$ \\
  PCI \textbf{(loc.)} & $0.021$ & $0.015$ & $0.018$ & $0.101$ & $0.023$ & $0.062$ & $0.123$ & $0.316$ & $0.219$ & $0.046$ & $0.105$ & $0.076$ & $6.845$ \\
  Text Embeddings & $-0.075$ & $0.041$ & $-0.017$ & $-0.017$ & $0.028$ & $0.006$ & $-$ & $0.096$ & $-$ & $4.537$ & $0.074$ & $2.306$ & $6.693$ \\
  Text Embeddings \textbf{(loc.)} & $-0.008$ & $0.001$ & $-0.004$ & $0.025$ & $0.007$ & $0.016$ & $0.187$ & $0.563$ & $0.375$ & $0.088$ & $0.119$ & $0.103$ & $6.821$ \\
  Token Embeddings & $-0.005$ & $-0.008$ & $-0.007$ & $0.004$ & $0.010$ & $0.007$ & $0.597$ & $0.454$ & $0.525$ & $0.160$ & $0.241$ & $0.201$ & $6.882$ \\
  Token Embeddings \textbf{(loc.)} & $-0.008$ & $0.005$ & $-0.002$ & $0.025$ & $0.001$ & $0.013$ & $1.021$ & $0.374$ & $0.698$ & $0.143$ & $0.474$ & $0.308$ & $6.850$ \\
  FreeSliders & $0.044$ & $0.025$ & $0.035$ & $0.106$ & $0.068$ & $0.087$ & $0.064$ & $0.136$ & $0.100$ & $0.043$ & $0.075$ & $0.059$ & $6.864$ \\
  FreeSliders \textbf{(loc.)} & $0.047$ & $0.027$ & $0.037$ & $0.106$ & $0.072$ & $0.089$ & $0.084$ & $0.217$ & $0.151$ & $0.043$ & $0.087$ & $0.065$ & $6.862$ \\
  Concept Sliders & $0.072$ & $0.081$ & \textbf{0.076} & $0.112$ & $0.060$ & $0.086$ & $0.051$ & $0.027$ & \textbf{0.039} & $0.024$ & $0.038$ & \textbf{0.031} & $6.842$ \\
  Concept Sliders \textbf{(loc.)} & $0.045$ & $0.043$ & \uline{0.044} & $0.084$ & $0.034$ & $0.059$ & $0.051$ & $0.100$ & \uline{0.075} & $0.030$ & $0.068$ & \uline{0.049} & $6.865$ \\
  AUSteer & $-0.002$ & $0.040$ & $0.019$ & $0.135$ & $0.101$ & \textbf{0.118} & $0.853$ & $0.065$ & $0.459$ & $0.049$ & $0.079$ & $0.064$ & $6.794$ \\
  AUSteer \textbf{(loc.)} & $0.018$ & $0.044$ & $0.031$ & $0.108$ & $0.078$ & $0.093$ & $0.141$ & $0.044$ & $0.092$ & $0.042$ & $0.079$ & $0.060$ & $6.845$ \\
  CAA & $0.020$ & $0.005$ & $0.012$ & $0.104$ & $0.070$ & $0.087$ & $0.138$ & $0.574$ & $0.356$ & $0.034$ & $0.082$ & $0.058$ & $6.848$ \\
  CAA \textbf{(loc.)} & $0.029$ & $0.007$ & $0.018$ & $0.116$ & $0.085$ & \uline{0.101} & $0.068$ & $0.334$ & $0.201$ & $0.034$ & $0.081$ & $0.057$ & $6.864$ \\
  SAE \textbf{(loc.)} & $0.044$ & $0.018$ & $0.031$ & $0.122$ & $0.068$ & $0.095$ & $0.066$ & $0.094$ & $0.080$ & $0.038$ & $0.072$ & $0.055$ & $6.852$ \\
            \bottomrule
        \end{tabular}
    }
\end{table}

\vspace{-2em}
\begin{table}[h]
    \centering
    \caption{\textbf{Evaluating steering \textit{vocal style} concept with \ace.} We describe metrics in \cref{sec:metrics}. Symbols \textit{+} and \textit{-} denote the quality of steering towards positive (\textit{rap vocal}) and negative (\textit{sing vocal}) directions, \textit{avg} = (pos + neg) / 2. \textbf{Best} and \uline{second} results are highlighted.}
    \label{tab:steering_auc_vocal_style}
    \vspace{0.5em}
    \resizebox{1.0\linewidth}{!}{
        \begin{tabular}{lccccccccccccc}
            \toprule
            \multirow{2}{*}{\shortstack[l]{\textbf{Method}\\(\textit{vocal style})}}
                & \multicolumn{3}{c}{\textbf{AUC MuQ} ($\uparrow$)}
                & \multicolumn{3}{c}{\textbf{AUC CLAP} ($\uparrow$)}
                & \multicolumn{3}{c}{\textbf{Smoothness MuQ} ($\downarrow$)}
                & \multicolumn{3}{c}{\textbf{Smoothness CLAP} ($\downarrow$)}
                & \multirow{2}{*}{\shortstack{\textbf{Avg}\\\textbf{Quality} ($\uparrow$)}} \\
            \cmidrule(lr){2-4}\cmidrule(lr){5-7}\cmidrule(lr){8-10}\cmidrule(lr){11-13}
                & + & - & avg & + & - & avg & + & - & avg & + & - & avg & \\
            \midrule
  PCI & $0.155$ & $-0.010$ & $0.073$ & $0.056$ & $-0.001$ & $0.028$ & $0.068$ & $1.277$ & $0.672$ & $0.074$ & $1.250$ & $0.662$ & $6.823$ \\
  PCI \textbf{(loc.)} & $0.136$ & $-0.010$ & $0.063$ & $0.053$ & $-0.011$ & $0.021$ & $0.069$ & $2.739$ & $1.404$ & $0.091$ & $-$ & $-$ & $6.800$ \\
  Text Embeddings & $-0.005$ & $-0.014$ & $-0.010$ & $-0.026$ & $0.006$ & $-0.010$ & $1.029$ & $2.983$ & $2.006$ & $0.630$ & $0.122$ & $0.376$ & $6.674$ \\
  Text Embeddings \textbf{(loc.)} & $0.036$ & $-0.002$ & $0.017$ & $0.024$ & $-0.010$ & $0.007$ & $0.077$ & $2.508$ & $1.293$ & $0.246$ & $2.549$ & $1.398$ & $6.811$ \\
  Token Embeddings & $0.045$ & $0.007$ & $0.026$ & $0.032$ & $0.004$ & $0.018$ & $0.089$ & $0.190$ & $0.139$ & $0.081$ & $0.197$ & $0.139$ & $6.877$ \\
  Token Embeddings \textbf{(loc.)} & $0.043$ & $0.005$ & $0.024$ & $0.033$ & $-0.005$ & $0.014$ & $0.103$ & $0.365$ & $0.234$ & $0.085$ & $0.324$ & $0.204$ & $6.842$ \\
  FreeSliders & $0.118$ & $0.019$ & $0.068$ & $0.051$ & $0.011$ & $0.031$ & $0.033$ & $0.093$ & $0.063$ & $0.049$ & $0.254$ & $0.152$ & $6.842$ \\
  FreeSliders \textbf{(loc.)} & $0.121$ & $0.019$ & $0.070$ & $0.052$ & $0.011$ & $0.031$ & $0.031$ & $0.087$ & $0.059$ & $0.034$ & $0.216$ & $0.125$ & $6.843$ \\
  Concept Sliders & $0.148$ & $0.031$ & $0.089$ & $0.099$ & $0.019$ & \textbf{0.059} & $0.047$ & $0.055$ & $0.051$ & $0.046$ & $0.043$ & \uline{0.044} & $6.822$ \\
  Concept Sliders \textbf{(loc.)} & $0.107$ & $0.016$ & $0.061$ & $0.049$ & $0.005$ & $0.027$ & $0.043$ & $0.126$ & $0.084$ & $0.057$ & $0.156$ & $0.106$ & $6.824$ \\
  AUSteer & $0.094$ & $0.044$ & $0.069$ & $0.058$ & $0.042$ & \uline{0.050} & $0.045$ & $0.033$ & \uline{0.039} & $0.046$ & $0.035$ & \textbf{0.041} & $6.813$ \\
  AUSteer \textbf{(loc.)} & $0.106$ & $0.037$ & $0.072$ & $0.030$ & $0.023$ & $0.027$ & $0.026$ & $0.069$ & $0.048$ & $0.070$ & $0.046$ & $0.058$ & $6.815$ \\
  CAA & $0.122$ & $0.023$ & $0.073$ & $0.046$ & $0.020$ & $0.033$ & $0.045$ & $0.073$ & $0.059$ & $0.050$ & $0.073$ & $0.061$ & $6.822$ \\
  CAA \textbf{(loc.)} & $0.163$ & $0.020$ & \uline{0.091} & $0.055$ & $0.012$ & $0.034$ & $0.034$ & $0.081$ & $0.057$ & $0.053$ & $0.103$ & $0.078$ & $6.831$ \\
  SAE \textbf{(loc.)} & $0.179$ & $0.042$ & \textbf{0.110} & $0.053$ & $0.021$ & $0.037$ & $0.021$ & $0.037$ & \textbf{0.029} & $0.045$ & $0.085$ & $0.065$ & $6.832$ \\
            \bottomrule
        \end{tabular}
    }
\end{table}

\vspace{-2em}
\begin{table}[h]
    \centering
    \caption{\textbf{Evaluating steering \textit{violin} concept with \ace.} We describe metrics in \cref{sec:metrics}. Symbols \textit{+} and \textit{-} denote the quality of steering towards positive (\textit{add violin}) and negative (\textit{remove violin}) directions, \textit{avg} = (pos + neg) / 2. \textbf{Best} and \uline{second} results are highlighted.}
    \label{tab:steering_auc_violin}
    \vspace{0.5em}
    \resizebox{1.0\linewidth}{!}{
        \begin{tabular}{lccccccccccccc}
            \toprule
            \multirow{2}{*}{\shortstack[l]{\textbf{Method}\\(\textit{violin})}}
                & \multicolumn{3}{c}{\textbf{AUC MuQ} ($\uparrow$)}
                & \multicolumn{3}{c}{\textbf{AUC CLAP} ($\uparrow$)}
                & \multicolumn{3}{c}{\textbf{Smoothness MuQ} ($\downarrow$)}
                & \multicolumn{3}{c}{\textbf{Smoothness CLAP} ($\downarrow$)}
                & \multirow{2}{*}{\shortstack{\textbf{Avg}\\\textbf{Quality} ($\uparrow$)}} \\
            \cmidrule(lr){2-4}\cmidrule(lr){5-7}\cmidrule(lr){8-10}\cmidrule(lr){11-13}
                & + & - & avg & + & - & avg & + & - & avg & + & - & avg & \\
            \midrule
  PCI & $0.148$ & $0.018$ & $0.083$ & $0.065$ & $0.011$ & $0.038$ & $0.076$ & $0.239$ & $0.158$ & $0.075$ & $0.186$ & $0.130$ & $6.828$ \\
  PCI \textbf{(loc.)} & $0.127$ & $0.019$ & $0.073$ & $0.051$ & $0.014$ & $0.033$ & $0.064$ & $0.146$ & $0.105$ & $0.076$ & $0.165$ & $0.120$ & $6.833$ \\
  Text Embeddings & $0.088$ & $-0.006$ & $0.041$ & $0.045$ & $0.007$ & $0.026$ & $0.055$ & $0.472$ & $0.263$ & $0.045$ & $0.105$ & $0.075$ & $6.824$ \\
  Text Embeddings \textbf{(loc.)} & $0.089$ & $0.004$ & $0.047$ & $0.033$ & $0.016$ & $0.025$ & $0.051$ & $0.473$ & $0.262$ & $0.057$ & $0.126$ & $0.092$ & $6.822$ \\
  Token Embeddings & $0.146$ & $0.027$ & $0.087$ & $0.062$ & $0.018$ & $0.040$ & $0.037$ & $0.175$ & $0.106$ & $0.041$ & $0.104$ & $0.072$ & $6.817$ \\
  Token Embeddings \textbf{(loc.)} & $0.113$ & $0.024$ & $0.069$ & $0.049$ & $0.023$ & $0.036$ & $0.044$ & $0.165$ & $0.105$ & $0.048$ & $0.137$ & $0.092$ & $6.814$ \\
  FreeSliders & $0.100$ & $0.053$ & $0.077$ & $0.050$ & $0.027$ & $0.038$ & $0.040$ & $0.074$ & $0.057$ & $0.036$ & $0.051$ & $0.043$ & $6.807$ \\
  FreeSliders \textbf{(loc.)} & $0.114$ & $0.058$ & $0.086$ & $0.054$ & $0.027$ & $0.040$ & $0.025$ & $0.056$ & $0.041$ & $0.033$ & $0.069$ & $0.051$ & $6.811$ \\
  Concept Sliders & $0.160$ & $0.052$ & $0.106$ & $0.060$ & $0.046$ & \textbf{0.053} & $0.026$ & $0.084$ & $0.055$ & $0.039$ & $0.045$ & \uline{0.042} & $6.791$ \\
  Concept Sliders \textbf{(loc.)} & $0.137$ & $0.048$ & $0.092$ & $0.052$ & $0.033$ & $0.043$ & $0.030$ & $0.062$ & $0.046$ & $0.062$ & $0.062$ & $0.062$ & $6.810$ \\
  AUSteer & $0.070$ & $-0.006$ & $0.032$ & $0.006$ & $0.020$ & $0.013$ & $0.049$ & $1.097$ & $0.573$ & $0.150$ & $0.085$ & $0.118$ & $6.785$ \\
  AUSteer \textbf{(loc.)} & $0.151$ & $0.056$ & $0.104$ & $0.053$ & $0.041$ & $0.047$ & $0.018$ & $0.081$ & $0.050$ & $0.034$ & $0.056$ & $0.045$ & $6.804$ \\
  CAA & $0.071$ & $0.025$ & $0.048$ & $-0.007$ & $0.013$ & $0.003$ & $0.043$ & $0.062$ & $0.053$ & $0.204$ & $0.107$ & $0.155$ & $6.798$ \\
  CAA \textbf{(loc.)} & $0.147$ & $0.068$ & \uline{0.108} & $0.054$ & $0.050$ & \uline{0.052} & $0.020$ & $0.046$ & \textbf{0.033} & $0.028$ & $0.043$ & \textbf{0.036} & $6.811$ \\
  SAE \textbf{(loc.)} & $0.181$ & $0.078$ & \textbf{0.130} & $0.049$ & $0.037$ & $0.043$ & $0.024$ & $0.042$ & \uline{0.033} & $0.052$ & $0.033$ & $0.042$ & $6.778$ \\
            \bottomrule
        \end{tabular}
    }
\end{table}

\vspace{-2em}
\begin{table}[h]
    \centering
    \caption{\textbf{Evaluating steering \textit{guitar type} concept with \ace.} We describe metrics in \cref{sec:metrics}. Symbols \textit{+} and \textit{-} denote the quality of steering towards positive (\textit{acoustic guitar}) and negative (\textit{electric guitar}) directions, \textit{avg} = (pos + neg) / 2. \textbf{Best} and \uline{second} results are highlighted.}
    \label{tab:steering_auc_guitar}
    \vspace{0.5em}
    \resizebox{1.0\linewidth}{!}{
        \begin{tabular}{lccccccccccccc}
            \toprule
            \multirow{2}{*}{\shortstack[l]{\textbf{Method}\\(\textit{guitar type})}}
                & \multicolumn{3}{c}{\textbf{AUC MuQ} ($\uparrow$)}
                & \multicolumn{3}{c}{\textbf{AUC CLAP} ($\uparrow$)}
                & \multicolumn{3}{c}{\textbf{Smoothness MuQ} ($\downarrow$)}
                & \multicolumn{3}{c}{\textbf{Smoothness CLAP} ($\downarrow$)}
                & \multirow{2}{*}{\shortstack{\textbf{Avg}\\\textbf{Quality} ($\uparrow$)}} \\
            \cmidrule(lr){2-4}\cmidrule(lr){5-7}\cmidrule(lr){8-10}\cmidrule(lr){11-13}
                & + & - & avg & + & - & avg & + & - & avg & + & - & avg & \\
            \midrule
  PCI & $0.146$ & $0.012$ & $0.079$ & $0.073$ & $0.001$ & $0.037$ & $0.059$ & $0.139$ & $0.099$ & $0.059$ & $0.762$ & $0.410$ & $6.730$ \\
  PCI \textbf{(loc.)} & $0.117$ & $0.013$ & $0.065$ & $0.049$ & $0.015$ & $0.032$ & $0.056$ & $0.139$ & $0.098$ & $0.067$ & $0.329$ & $0.198$ & $6.723$ \\
  Text Embeddings & $0.010$ & $0.007$ & $0.009$ & $-0.029$ & $0.026$ & $-0.002$ & $0.139$ & $0.305$ & $0.222$ & $-$ & $0.068$ & $-$ & $6.569$ \\
  Text Embeddings \textbf{(loc.)} & $0.005$ & $0.005$ & $0.005$ & $-0.020$ & $0.019$ & $-0.001$ & $0.193$ & $0.300$ & $0.247$ & $0.680$ & $0.235$ & $0.457$ & $6.683$ \\
  Token Embeddings & $0.084$ & $0.004$ & $0.044$ & $0.038$ & $-0.007$ & $0.015$ & $0.062$ & $0.646$ & $0.354$ & $0.058$ & $0.679$ & $0.369$ & $6.730$ \\
  Token Embeddings \textbf{(loc.)} & $0.061$ & $0.011$ & $0.036$ & $0.011$ & $0.006$ & $0.009$ & $0.059$ & $0.242$ & $0.150$ & $0.126$ & $0.375$ & $0.250$ & $6.728$ \\
  FreeSliders & $0.156$ & $0.019$ & $0.087$ & $0.065$ & $0.018$ & $0.042$ & $0.026$ & $0.114$ & $0.070$ & $0.033$ & $0.148$ & $0.090$ & $6.694$ \\
  FreeSliders \textbf{(loc.)} & $0.152$ & $0.018$ & $0.085$ & $0.064$ & $0.017$ & $0.040$ & $0.023$ & $0.218$ & $0.121$ & $0.046$ & $0.193$ & $0.120$ & $6.692$ \\
  Concept Sliders & $0.148$ & $0.030$ & $0.089$ & $0.074$ & $0.042$ & \uline{0.058} & $0.030$ & $0.137$ & $0.084$ & $0.028$ & $0.063$ & \uline{0.046} & $6.653$ \\
  Concept Sliders \textbf{(loc.)} & $0.139$ & $0.028$ & $0.084$ & $0.060$ & $0.034$ & $0.047$ & $0.036$ & $0.152$ & $0.094$ & $0.037$ & $0.084$ & $0.060$ & $6.660$ \\
  AUSteer & $0.158$ & $0.040$ & \uline{0.099} & $0.045$ & $0.026$ & $0.036$ & $0.027$ & $0.151$ & $0.089$ & $0.032$ & $0.075$ & $0.053$ & $6.670$ \\
  AUSteer \textbf{(loc.)} & $0.110$ & $0.044$ & $0.077$ & $0.039$ & $0.020$ & $0.029$ & $0.025$ & $0.090$ & \uline{0.057} & $0.036$ & $0.126$ & $0.081$ & $6.657$ \\
  CAA & $0.107$ & $0.014$ & $0.060$ & $0.027$ & $0.010$ & $0.019$ & $0.031$ & $0.216$ & $0.124$ & $0.046$ & $0.210$ & $0.128$ & $6.690$ \\
  CAA \textbf{(loc.)} & $0.153$ & $0.027$ & $0.090$ & $0.055$ & $0.016$ & $0.036$ & $0.025$ & $0.119$ & $0.072$ & $0.050$ & $0.196$ & $0.123$ & $6.686$ \\
  SAE \textbf{(loc.)} & $0.205$ & $0.069$ & \textbf{0.137} & $0.077$ & $0.046$ & \textbf{0.061} & $0.016$ & $0.099$ & \textbf{0.057} & $0.020$ & $0.052$ & \textbf{0.036} & $6.649$ \\
            \bottomrule
        \end{tabular}
    }
\end{table}

\vspace{-2em}
\begin{table}[h]
    \centering
    \caption{\textbf{Evaluating steering  \textit{jazz genre} concept with \ace.} We describe metrics in \cref{sec:metrics}. Symbols \textit{+} and \textit{-} denote the quality of steering towards positive (\textit{jazz song}) and negative (\textit{rock song}) directions, \textit{avg} = (pos + neg) / 2. \textbf{Best} and \uline{second} results are highlighted.}
    \label{tab:steering_auc_popjazz}
    \vspace{0.5em}
    \resizebox{1.0\linewidth}{!}{
        \begin{tabular}{lccccccccccccc}
            \toprule
            \multirow{2}{*}{\shortstack[l]{\textbf{Method}\\(\textit{jazz genre})}}
                & \multicolumn{3}{c}{\textbf{AUC MuQ} ($\uparrow$)}
                & \multicolumn{3}{c}{\textbf{AUC CLAP} ($\uparrow$)}
                & \multicolumn{3}{c}{\textbf{Smoothness MuQ} ($\downarrow$)}
                & \multicolumn{3}{c}{\textbf{Smoothness CLAP} ($\downarrow$)}
                & \multirow{2}{*}{\shortstack{\textbf{Avg}\\\textbf{Quality} ($\uparrow$)}} \\
            \cmidrule(lr){2-4}\cmidrule(lr){5-7}\cmidrule(lr){8-10}\cmidrule(lr){11-13}
                & + & - & avg & + & - & avg & + & - & avg & + & - & avg & \\
            \midrule
  PCI & $0.363$ & $0.049$ & $0.206$ & $0.176$ & $0.010$ & $0.093$ & $0.032$ & $0.140$ & $0.086$ & $0.042$ & $0.231$ & $0.137$ & $6.822$ \\
  PCI \textbf{(loc.)} & $0.297$ & $0.038$ & $0.168$ & $0.146$ & $0.012$ & $0.079$ & $0.034$ & $0.114$ & $0.074$ & $0.040$ & $0.160$ & $0.100$ & $6.801$ \\
  Text Embeddings & $0.014$ & $0.028$ & $0.021$ & $-0.025$ & $0.067$ & $0.021$ & $0.352$ & $0.092$ & $0.222$ & $1.184$ & $0.065$ & $0.624$ & $6.633$ \\
  Text Embeddings \textbf{(loc.)} & $0.058$ & $0.031$ & $0.045$ & $0.005$ & $0.042$ & $0.024$ & $0.107$ & $0.077$ & $0.092$ & $0.245$ & $0.058$ & $0.151$ & $6.760$ \\
  Token Embeddings & $-0.011$ & $0.031$ & $0.010$ & $-0.006$ & $0.043$ & $0.019$ & $0.965$ & $0.170$ & $0.568$ & $0.697$ & $0.076$ & $0.386$ & $6.706$ \\
  Token Embeddings \textbf{(loc.)} & $0.006$ & $0.038$ & $0.022$ & $-0.006$ & $0.015$ & $0.004$ & $0.318$ & $0.091$ & $0.205$ & $0.830$ & $0.135$ & $0.482$ & $6.794$ \\
  FreeSliders & $0.145$ & $0.115$ & $0.130$ & $0.086$ & $0.068$ & $0.077$ & $0.035$ & $0.046$ & \textbf{0.041} & $0.035$ & $0.064$ & $0.049$ & $6.783$ \\
  FreeSliders \textbf{(loc.)} & $0.138$ & $0.108$ & $0.123$ & $0.078$ & $0.067$ & $0.072$ & $0.033$ & $0.055$ & $0.044$ & $0.035$ & $0.073$ & $0.054$ & $6.782$ \\
  Concept Sliders & $0.189$ & $0.114$ & $0.152$ & $0.157$ & $0.106$ & \uline{0.132} & $0.041$ & $0.074$ & $0.057$ & $0.021$ & $0.052$ & $0.037$ & $6.727$ \\
  Concept Sliders \textbf{(loc.)} & $0.170$ & $0.098$ & $0.134$ & $0.119$ & $0.081$ & $0.100$ & $0.026$ & $0.055$ & \uline{0.041} & $0.020$ & $0.041$ & \textbf{0.030} & $6.753$ \\
  AUSteer & $0.240$ & $0.134$ & $0.187$ & $0.112$ & $0.079$ & $0.095$ & $0.032$ & $0.073$ & $0.053$ & $0.027$ & $0.039$ & \uline{0.033} & $6.761$ \\
  AUSteer \textbf{(loc.)} & $0.250$ & $0.120$ & $0.185$ & $0.132$ & $0.078$ & $0.105$ & $0.034$ & $0.077$ & $0.056$ & $0.025$ & $0.043$ & $0.034$ & $6.786$ \\
  CAA & $0.216$ & $0.120$ & $0.168$ & $0.091$ & $0.056$ & $0.073$ & $0.039$ & $0.063$ & $0.051$ & $0.030$ & $0.040$ & $0.035$ & $6.810$ \\
  CAA \textbf{(loc.)} & $0.270$ & $0.142$ & \uline{0.206} & $0.146$ & $0.083$ & $0.115$ & $0.026$ & $0.060$ & $0.043$ & $0.019$ & $0.051$ & $0.035$ & $6.811$ \\
  SAE \textbf{(loc.)} & $0.318$ & $0.126$ & \textbf{0.222} & $0.170$ & $0.103$ & \textbf{0.137} & $0.051$ & $0.120$ & $0.085$ & $0.041$ & $0.078$ & $0.060$ & $6.776$ \\
            \bottomrule
        \end{tabular}
    }
\end{table}

\vspace{-2em}
\begin{table}[h]
    \centering
    \caption{\textbf{Evaluating steering \textit{classical genre} concept with \ace.} We describe metrics in \cref{sec:metrics}. Symbols \textit{+} and \textit{-} denote the quality of steering towards positive (\textit{classical song}) and negative (\textit{electronic song}) directions, \textit{avg} = (pos + neg) / 2. \textbf{Best} and \uline{second} results are highlighted.}
    \label{tab:steering_auc_classical_electronic}
    \vspace{0.5em}
    \resizebox{1.0\linewidth}{!}{
        \begin{tabular}{lccccccccccccc}
            \toprule
            \multirow{2}{*}{\shortstack[l]{\textbf{Method}\\(\textit{classical genre})}}
                & \multicolumn{3}{c}{\textbf{AUC MuQ} ($\uparrow$)}
                & \multicolumn{3}{c}{\textbf{AUC CLAP} ($\uparrow$)}
                & \multicolumn{3}{c}{\textbf{Smoothness MuQ} ($\downarrow$)}
                & \multicolumn{3}{c}{\textbf{Smoothness CLAP} ($\downarrow$)}
                & \multirow{2}{*}{\shortstack{\textbf{Avg}\\\textbf{Quality} ($\uparrow$)}} \\
            \cmidrule(lr){2-4}\cmidrule(lr){5-7}\cmidrule(lr){8-10}\cmidrule(lr){11-13}
                & + & - & avg & + & - & avg & + & - & avg & + & - & avg & \\
            \midrule
  PCI & $0.239$ & $0.049$ & $0.144$ & $0.117$ & $0.028$ & $0.072$ & $0.036$ & $0.108$ & $0.072$ & $0.038$ & $0.088$ & $0.063$ & $6.813$ \\
  PCI \textbf{(loc.)} & $0.206$ & $0.019$ & $0.113$ & $0.102$ & $0.013$ & $0.057$ & $0.034$ & $0.133$ & $0.084$ & $0.043$ & $0.104$ & $0.073$ & $6.798$ \\
  Text Embeddings & $0.075$ & $-0.006$ & $0.034$ & $-0.009$ & $0.055$ & $0.023$ & $0.068$ & $0.242$ & $0.155$ & $0.515$ & $0.051$ & $0.283$ & $6.642$ \\
  Text Embeddings \textbf{(loc.)} & $0.067$ & $-0.000$ & $0.033$ & $0.010$ & $0.044$ & $0.027$ & $0.114$ & $0.215$ & $0.164$ & $0.319$ & $0.068$ & $0.194$ & $6.783$ \\
  Token Embeddings & $0.079$ & $0.022$ & $0.050$ & $0.055$ & $0.064$ & $0.060$ & $0.043$ & $0.163$ & $0.103$ & $0.035$ & $0.052$ & $0.044$ & $6.732$ \\
  Token Embeddings \textbf{(loc.)} & $0.074$ & $0.025$ & $0.050$ & $0.056$ & $0.051$ & $0.053$ & $0.046$ & $0.118$ & $0.082$ & $0.037$ & $0.053$ & $0.045$ & $6.805$ \\
  FreeSliders & $0.176$ & $0.106$ & $0.141$ & $0.088$ & $0.081$ & $0.084$ & $0.029$ & $0.041$ & $0.035$ & $0.039$ & $0.033$ & $0.036$ & $6.810$ \\
  FreeSliders \textbf{(loc.)} & $0.171$ & $0.100$ & $0.136$ & $0.088$ & $0.080$ & $0.084$ & $0.043$ & $0.041$ & $0.042$ & $0.039$ & $0.031$ & $0.035$ & $6.809$ \\
  Concept Sliders & $0.111$ & $0.035$ & $0.073$ & $0.085$ & $0.092$ & $0.089$ & $0.053$ & $0.118$ & $0.086$ & $0.048$ & $0.027$ & $0.037$ & $6.756$ \\
  Concept Sliders \textbf{(loc.)} & $0.088$ & $-0.000$ & $0.044$ & $0.075$ & $0.075$ & $0.075$ & $0.047$ & $0.739$ & $0.393$ & $0.051$ & $0.020$ & $0.035$ & $6.776$ \\
  AUSteer & $0.205$ & $0.106$ & $0.156$ & $0.095$ & $0.098$ & $0.097$ & $0.033$ & $0.061$ & $0.047$ & $0.041$ & $0.027$ & \uline{0.034} & $6.748$ \\
  AUSteer \textbf{(loc.)} & $0.204$ & $0.115$ & \uline{0.159} & $0.133$ & $0.112$ & \textbf{0.122} & $0.026$ & $0.042$ & \uline{0.034} & $0.047$ & $0.034$ & $0.041$ & $6.788$ \\
  CAA & $0.177$ & $0.091$ & $0.134$ & $0.075$ & $0.085$ & $0.080$ & $0.022$ & $0.050$ & $0.036$ & $0.025$ & $0.032$ & \textbf{0.029} & $6.811$ \\
  CAA \textbf{(loc.)} & $0.211$ & $0.119$ & \textbf{0.165} & $0.109$ & $0.091$ & $0.100$ & $0.030$ & $0.037$ & \textbf{0.033} & $0.034$ & $0.038$ & $0.036$ & $6.806$ \\
  SAE \textbf{(loc.)} & $0.179$ & $0.095$ & $0.137$ & $0.112$ & $0.125$ & \uline{0.118} & $0.043$ & $0.100$ & $0.072$ & $0.073$ & $0.064$ & $0.068$ & $6.752$ \\
            \bottomrule
        \end{tabular}
    }
\end{table}

\clearpage

\subsection{Alignment--preservation curves}\label{app:auc_curves}

\begin{figure}[h]
    \centering
    \begin{subfigure}{0.49\linewidth}
        \includegraphics[width=\linewidth]{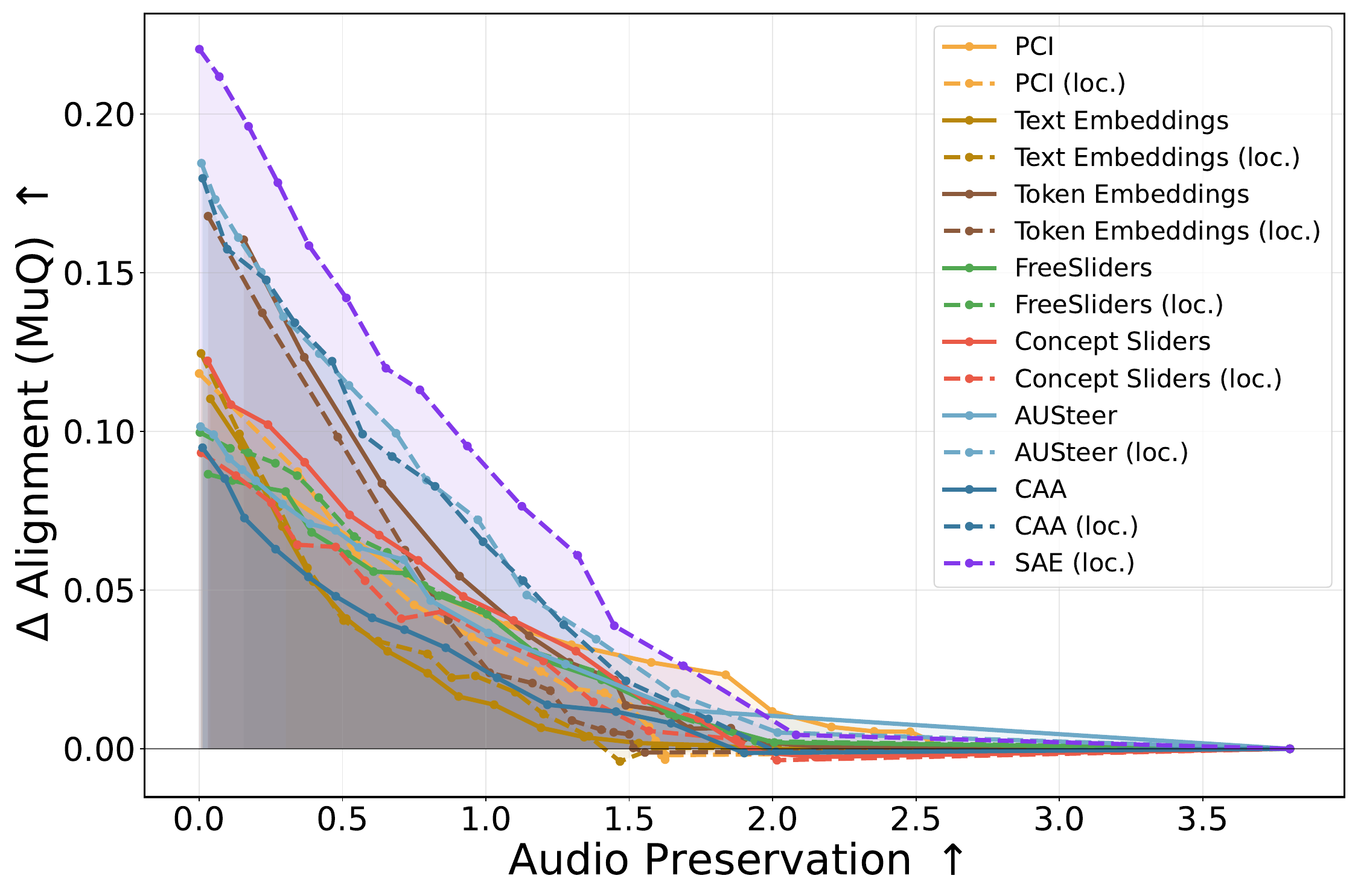}
        \caption{add piano (+)}
    \end{subfigure}
    \hfill
    \begin{subfigure}{0.49\linewidth}
        \includegraphics[width=\linewidth]{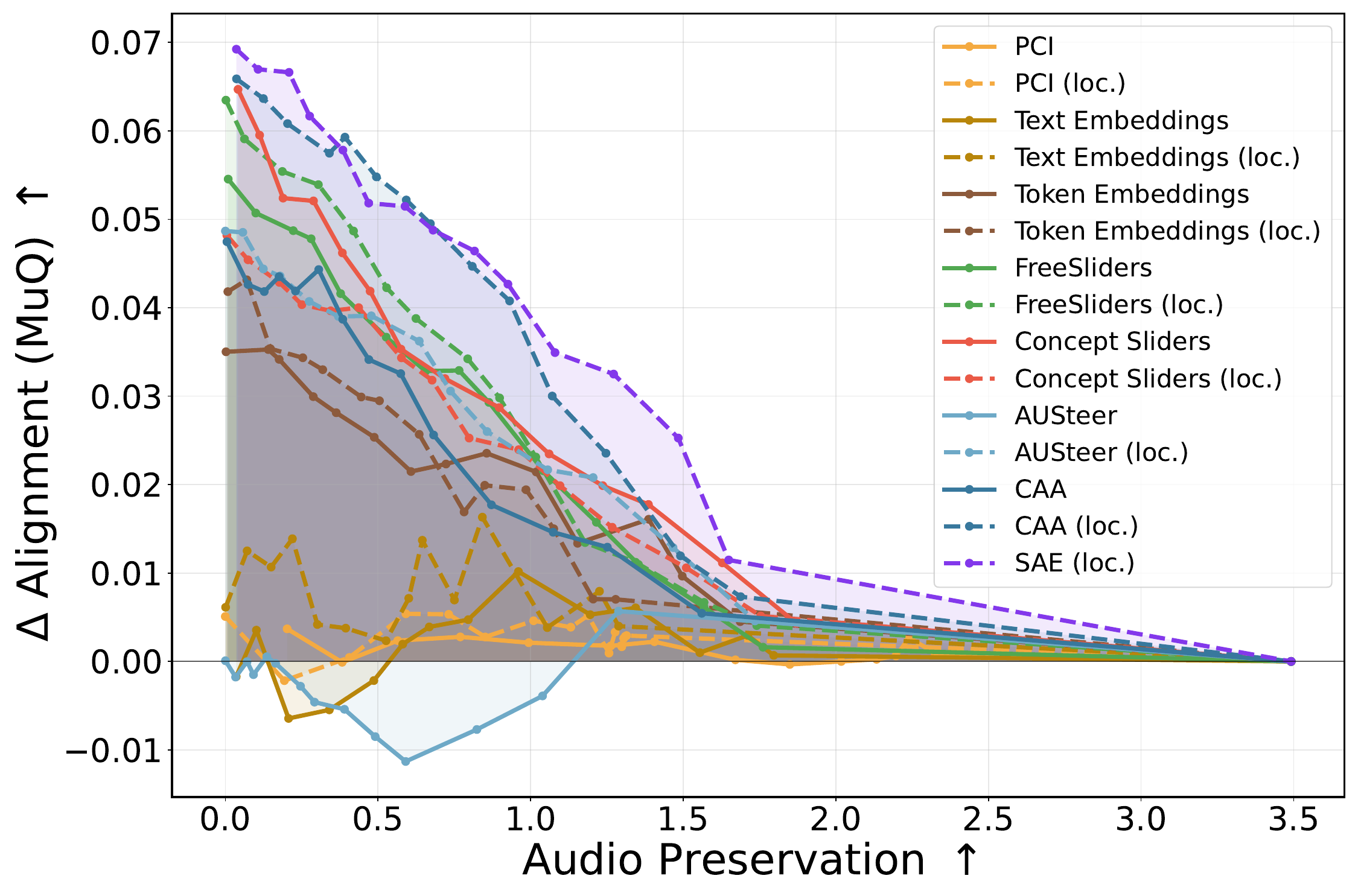}
        \caption{remove piano (--)}
    \end{subfigure}
    \caption{\textbf{Alignment--preservation curves for \textit{piano} steering.} Used for computing AUC values in Tab.~\ref{tab:steering_auc_piano}.}
    \label{fig:auc_curves_piano}
\end{figure}

\begin{figure}[h]
    \centering
    \begin{subfigure}{0.49\linewidth}
        \includegraphics[width=\linewidth]{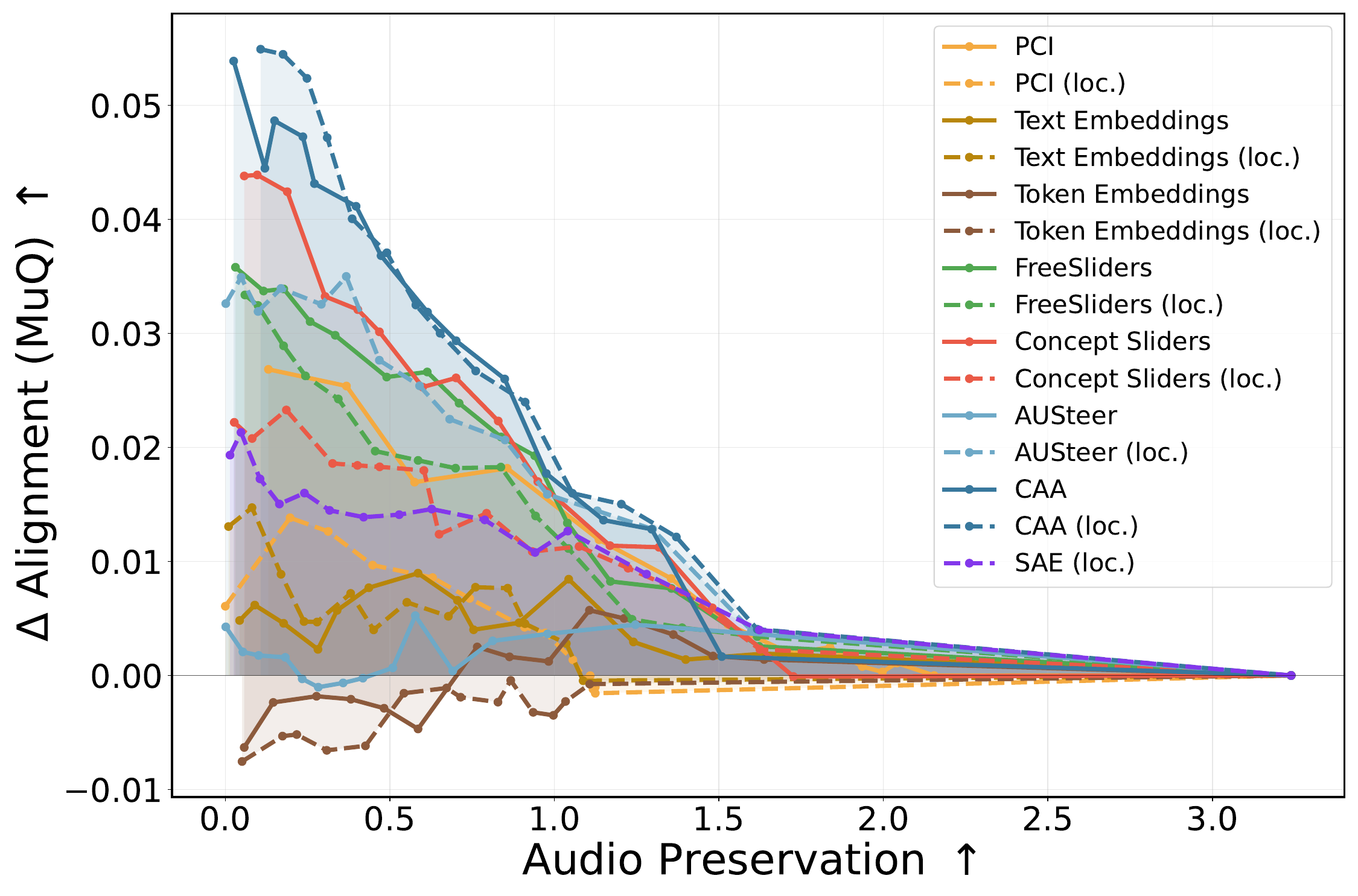}
        \caption{brighter mood (+)}
    \end{subfigure}
    \hfill
    \begin{subfigure}{0.49\linewidth}
        \includegraphics[width=\linewidth]{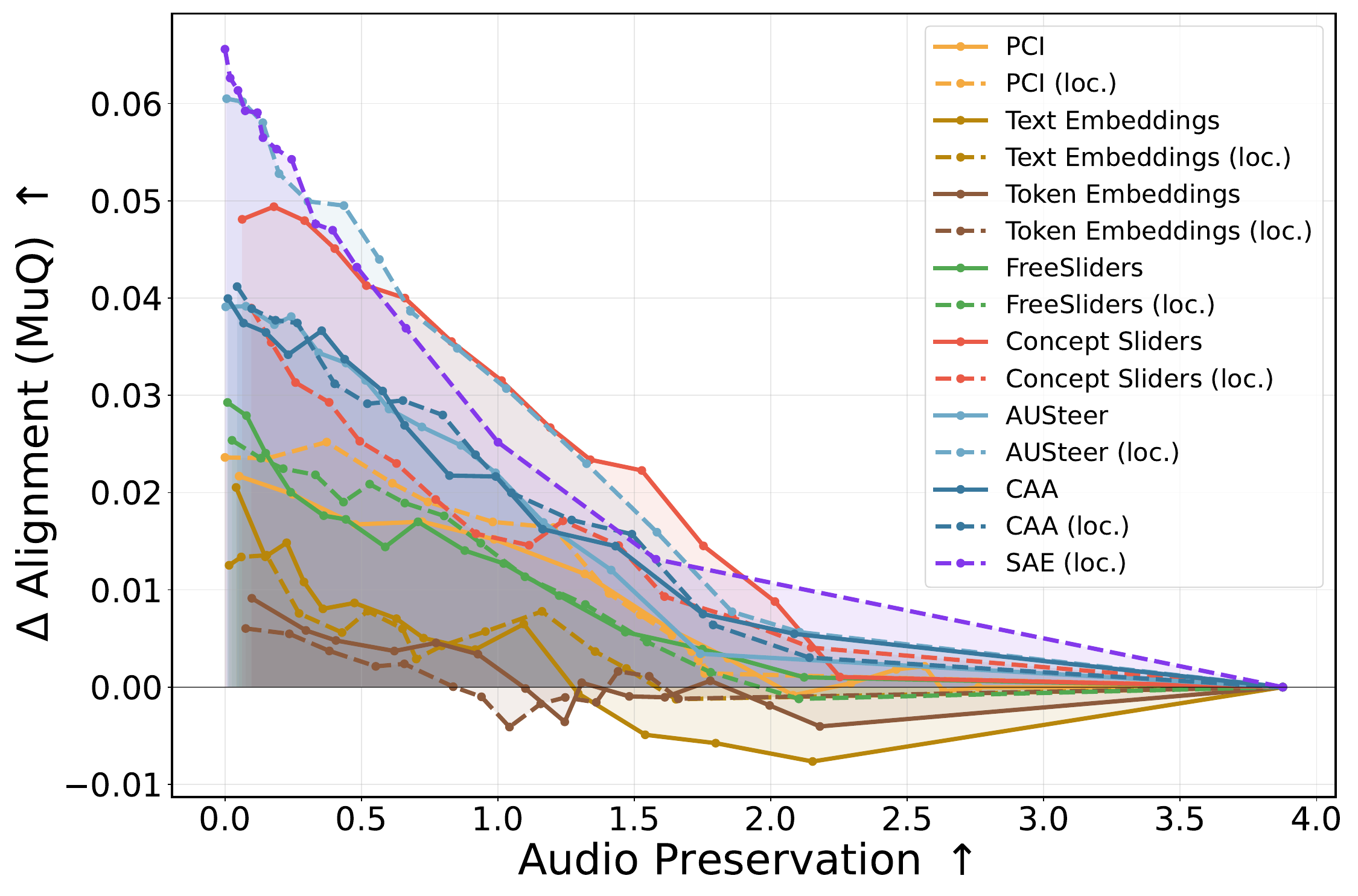}
        \caption{darker mood (--)}
    \end{subfigure}
    \caption{\textbf{Alignment--preservation curves for \textit{mood} steering.} Used for computing AUC values in Tab.~\ref{tab:steering_auc_mood}.}
    \label{fig:auc_curves_mood}
\end{figure}

\begin{figure}[h]
    \centering
    \begin{subfigure}{0.49\linewidth}
        \includegraphics[width=\linewidth]{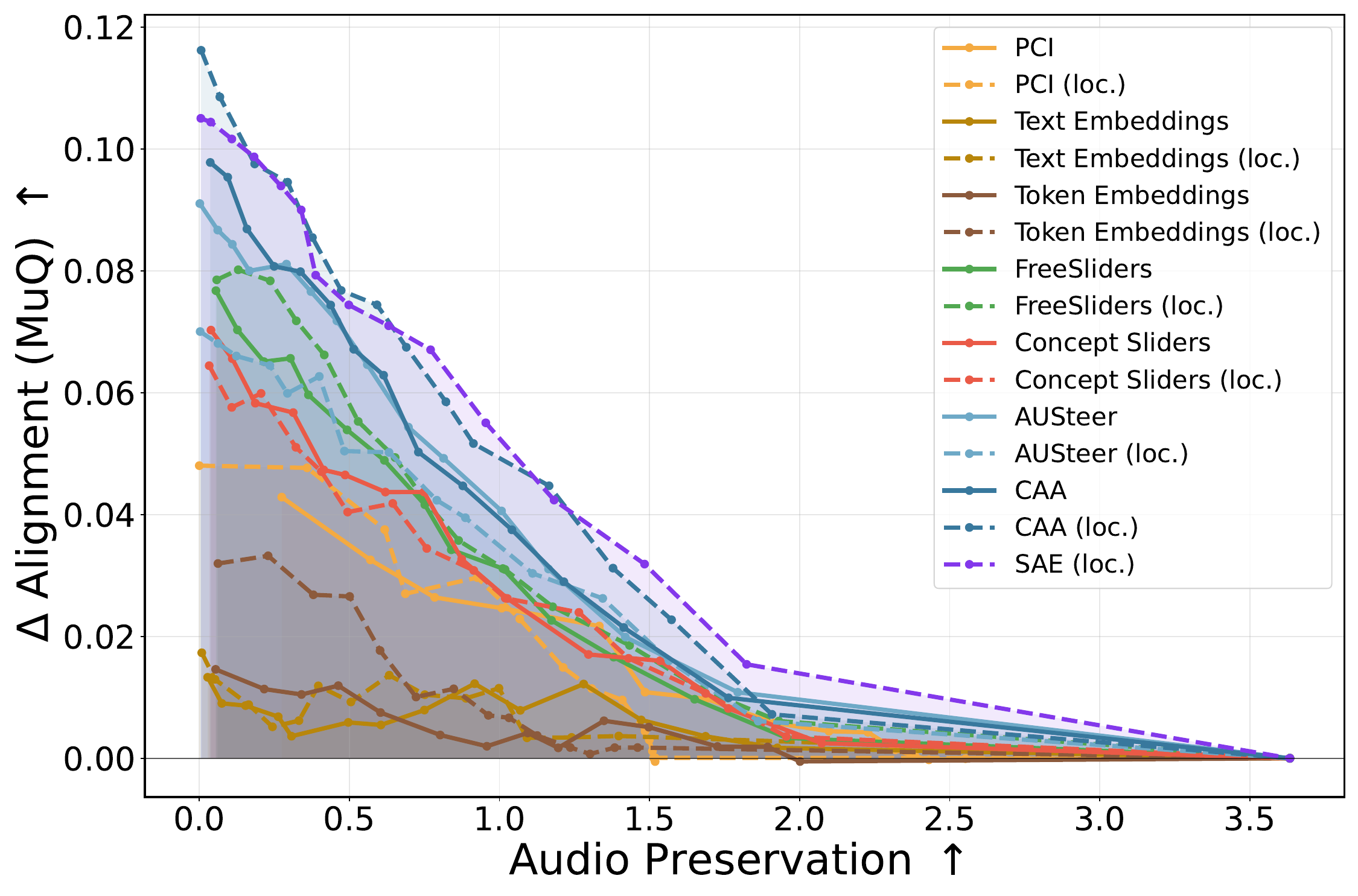}
        \caption{faster tempo (+)}
    \end{subfigure}
    \hfill
    \begin{subfigure}{0.49\linewidth}
        \includegraphics[width=\linewidth]{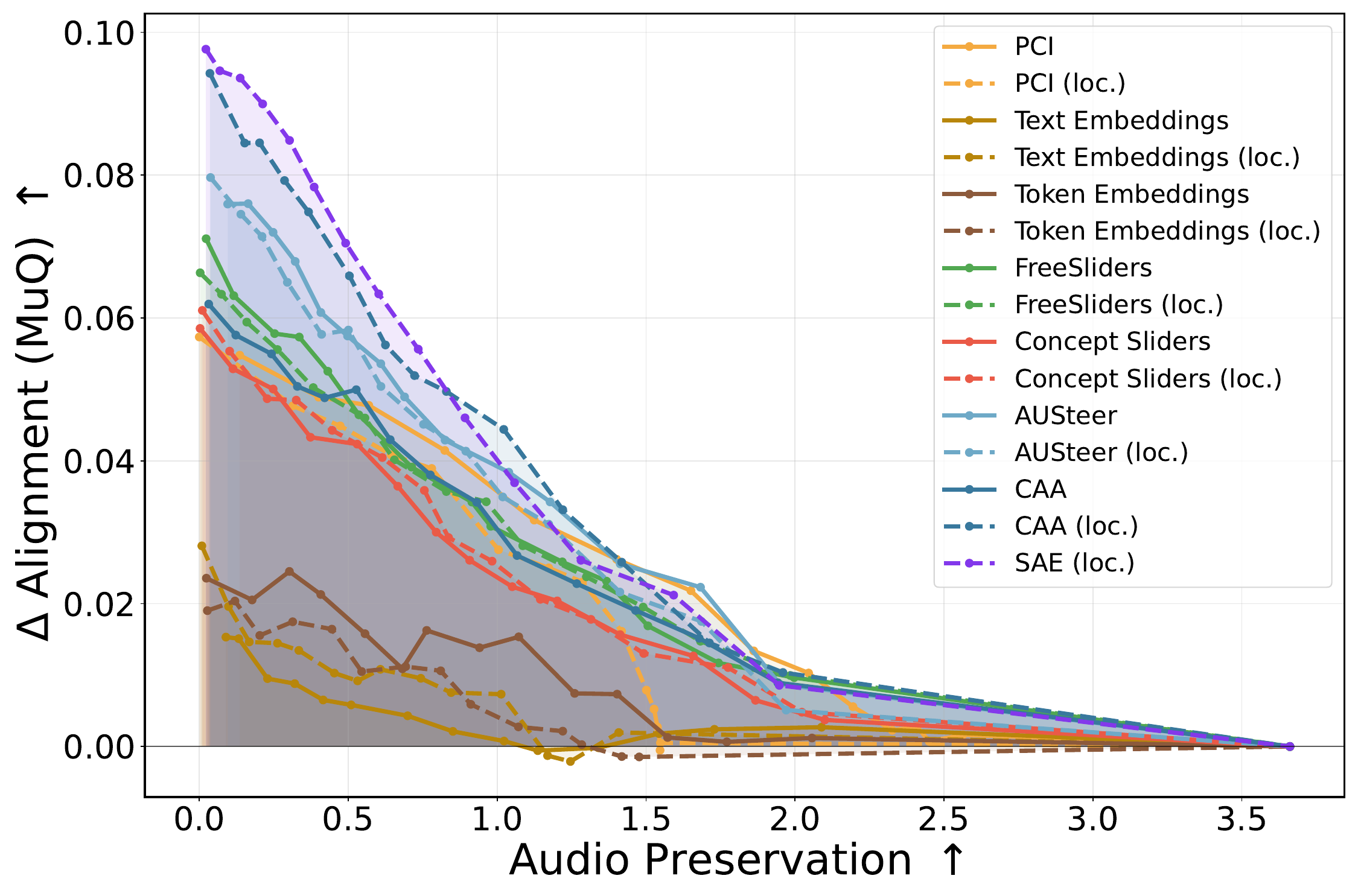}
        \caption{slower tempo (--)}
    \end{subfigure}
    \caption{\textbf{Alignment--preservation curves for \textit{tempo} steering.} Used for computing AUC values in Tab.~\ref{tab:steering_auc_tempo}.}
    \label{fig:auc_curves_tempo}
\end{figure}

\begin{figure}[h]
    \centering
    \begin{subfigure}{0.49\linewidth}
        \includegraphics[width=\linewidth]{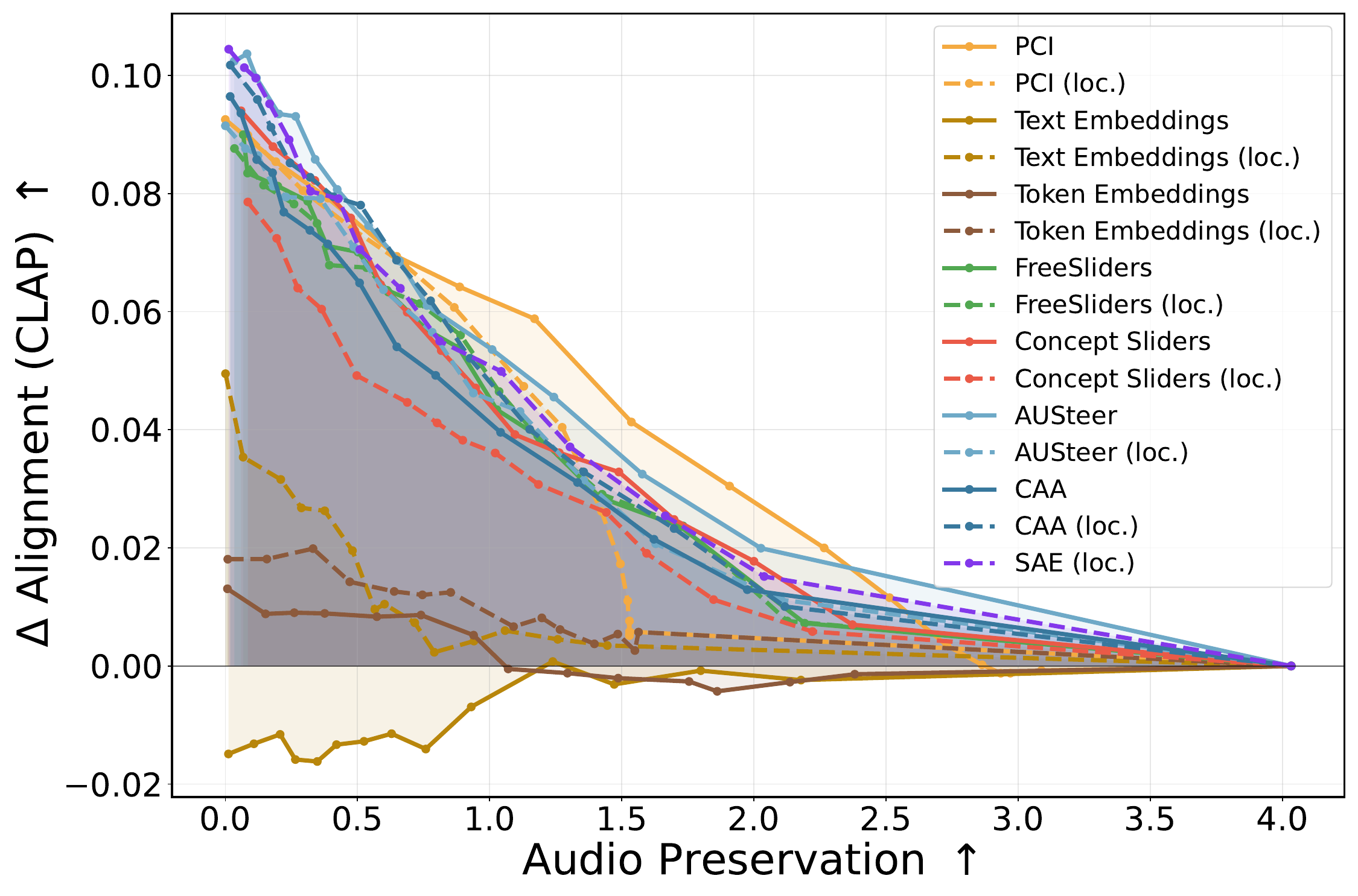}
        \caption{female vocals (+)}
    \end{subfigure}
    \hfill
    \begin{subfigure}{0.49\linewidth}
        \includegraphics[width=\linewidth]{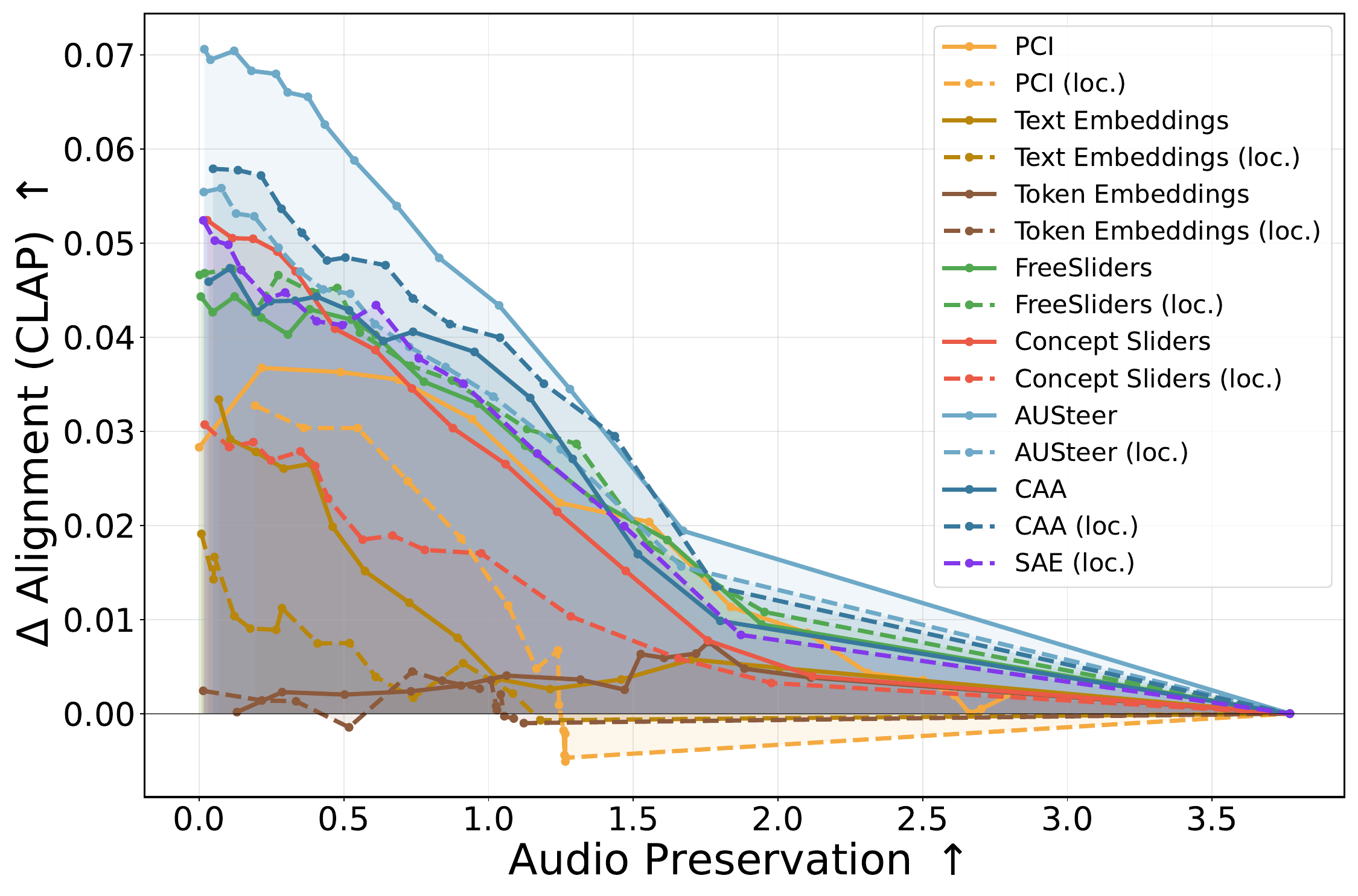}
        \caption{male vocals (--)}
    \end{subfigure}
    \caption{\textbf{Alignment--preservation curves for \textit{vocal gender} steering.} Used for computing AUC values in Tab.~\ref{tab:steering_auc_vocal_gender}.}
    \label{fig:auc_curves_vocals}
\end{figure}

\begin{figure}[h]
    \centering
    \begin{subfigure}{0.49\linewidth}
        \includegraphics[width=\linewidth]{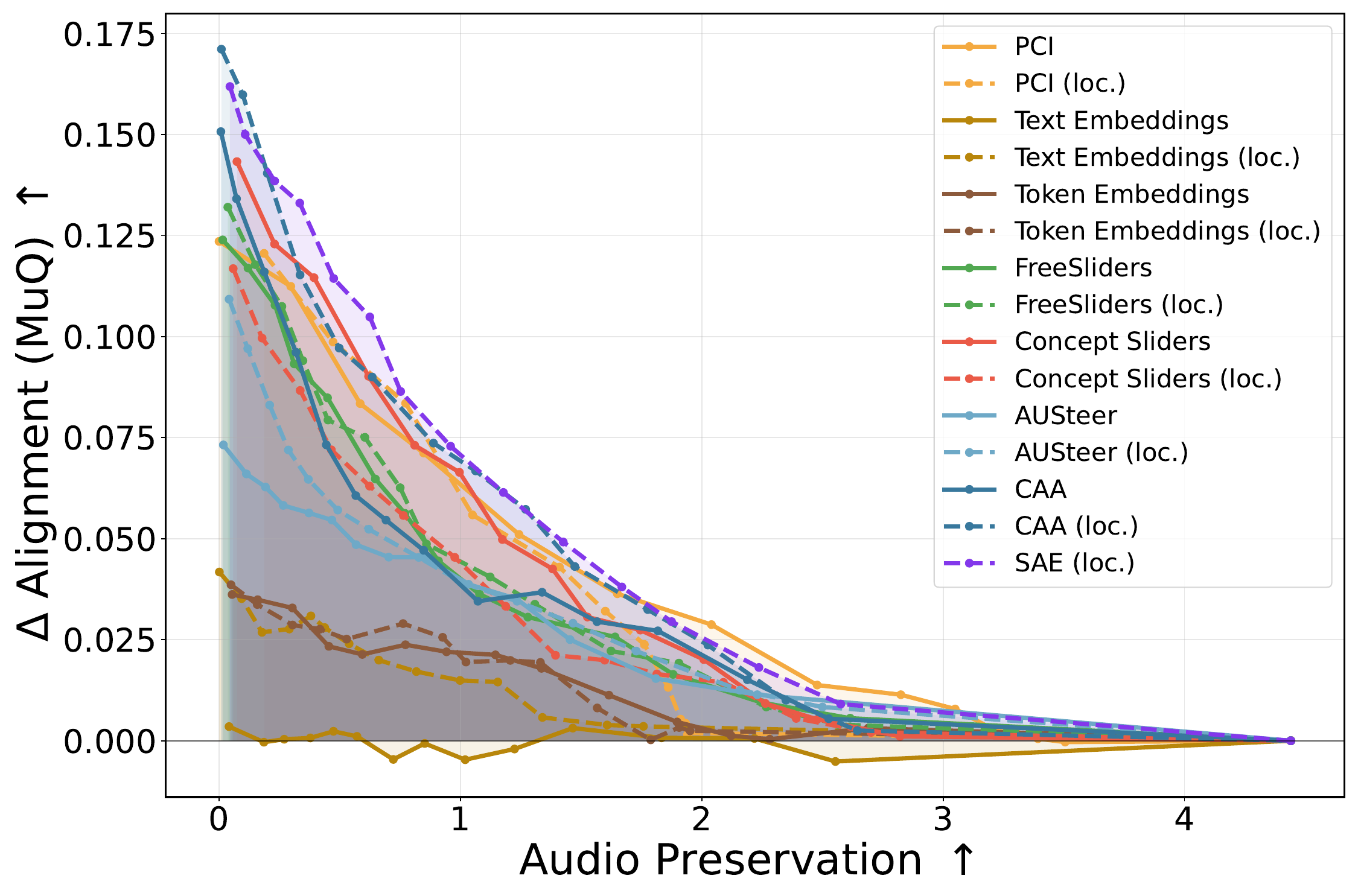}
        \caption{rap vocal style (+)}
    \end{subfigure}
    \hfill
    \begin{subfigure}{0.49\linewidth}
        \includegraphics[width=\linewidth]{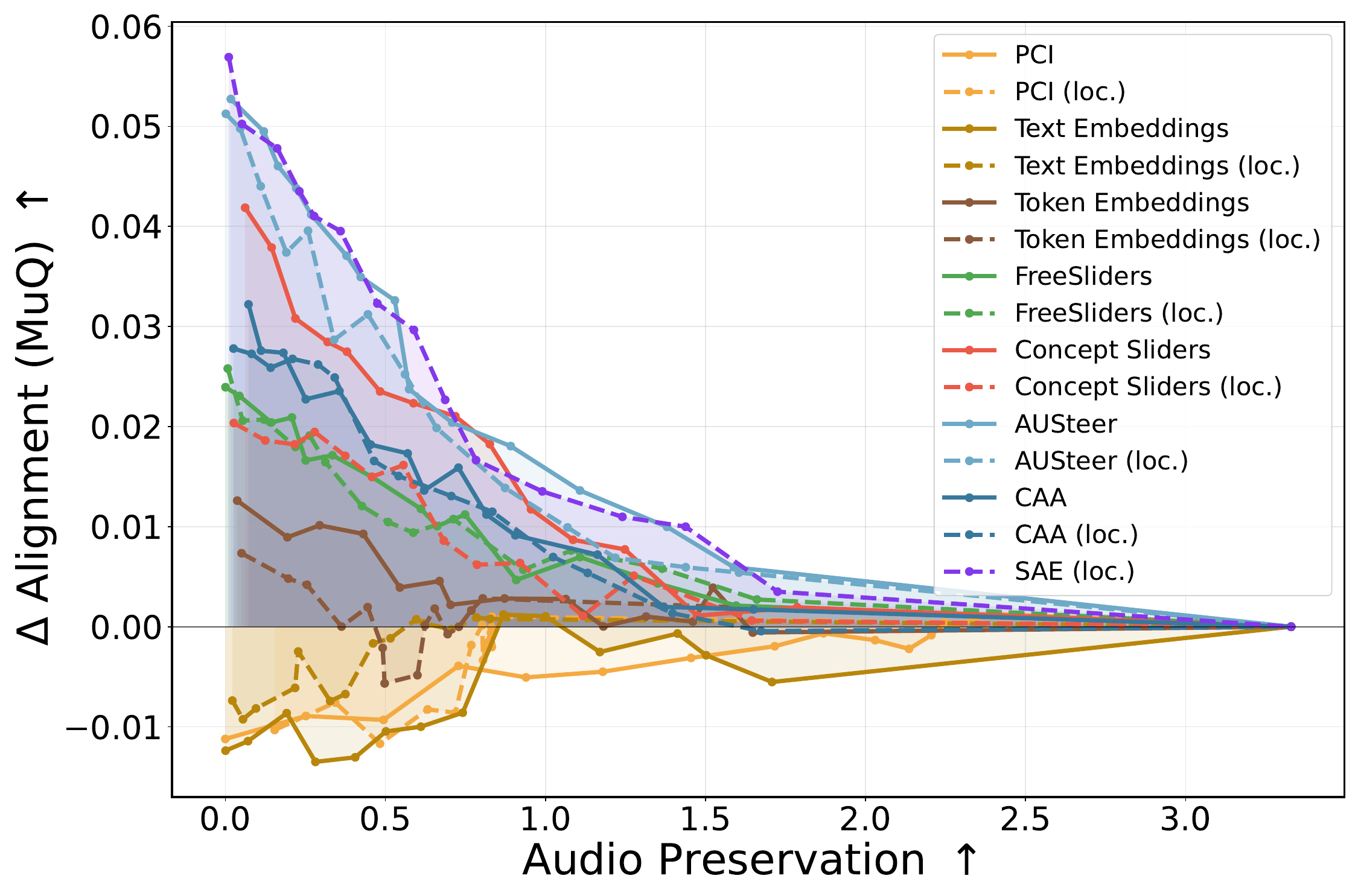}
        \caption{singing vocal style (--)}
    \end{subfigure}
    \caption{\textbf{Alignment--preservation curves for \textit{vocal style} steering.} Used for computing AUC values in Tab.~\ref{tab:steering_auc_vocal_style}.}
    \label{fig:auc_curves_vocal_style}
\end{figure}

\begin{figure}[h]
    \centering
    \begin{subfigure}{0.49\linewidth}
        \includegraphics[width=\linewidth]{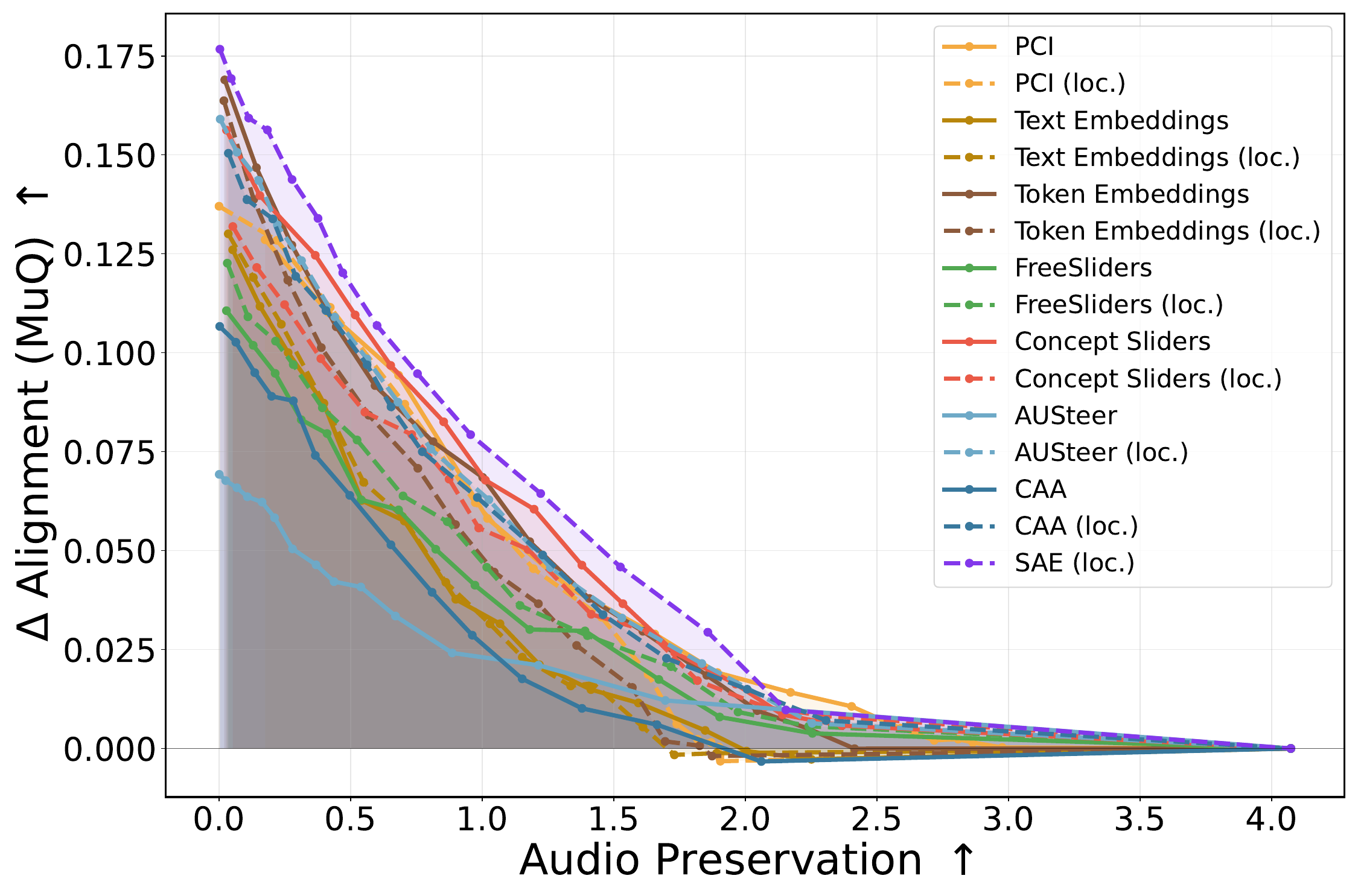}
        \caption{add violin (+)}
    \end{subfigure}
    \hfill
    \begin{subfigure}{0.49\linewidth}
        \includegraphics[width=\linewidth]{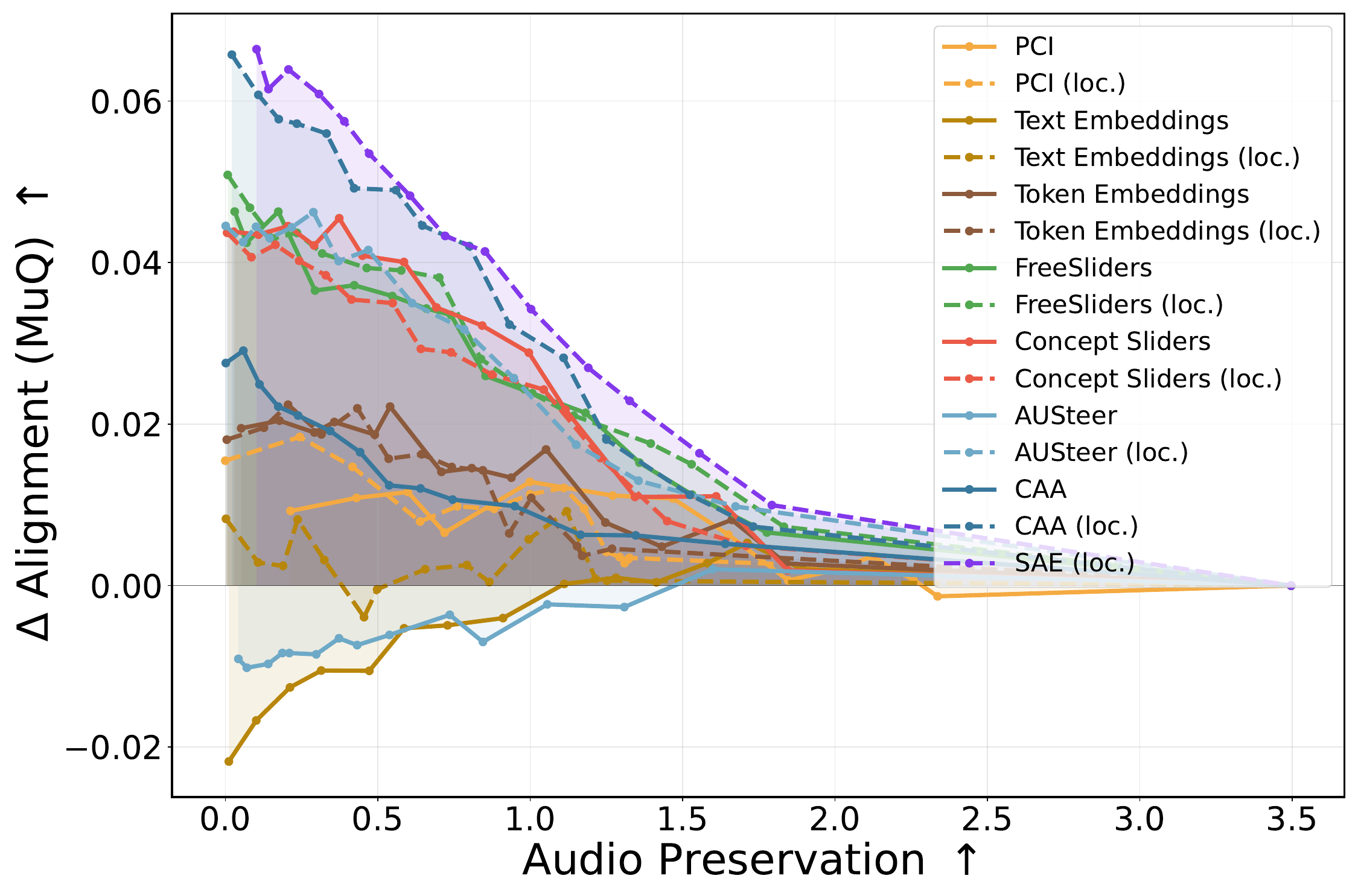}
        \caption{remove violin (--)}
    \end{subfigure}
    \caption{\textbf{Alignment--preservation curves for \textit{violin} steering.} Used for computing AUC values in Tab.~\ref{tab:steering_auc_violin}.}
    \label{fig:auc_curves_violin}
\end{figure}

\begin{figure}[h]
    \centering
    \begin{subfigure}{0.49\linewidth}
        \includegraphics[width=\linewidth]{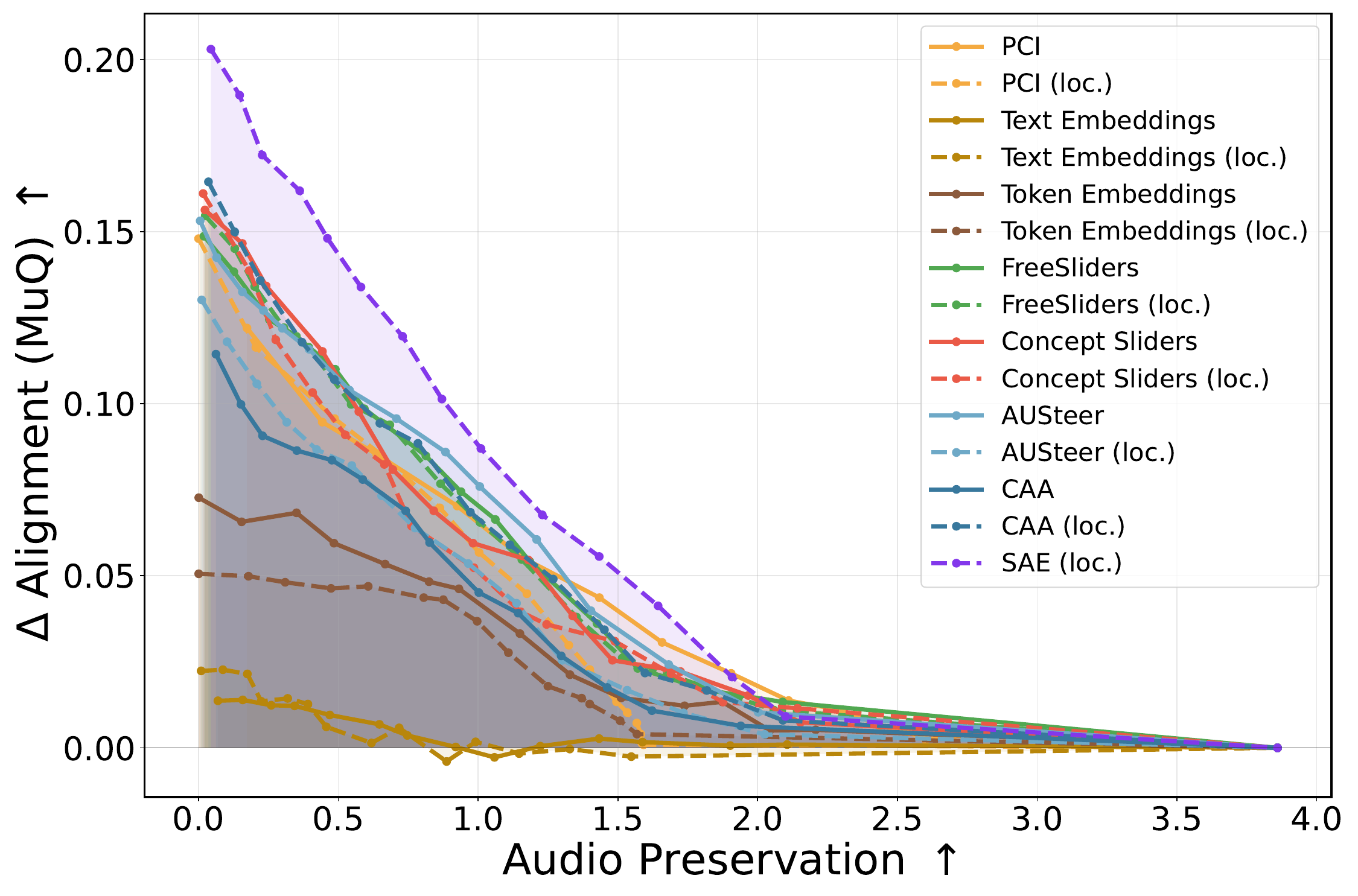}
        \caption{acoustic guitar (+)}
    \end{subfigure}
    \hfill
    \begin{subfigure}{0.49\linewidth}
        \includegraphics[width=\linewidth]{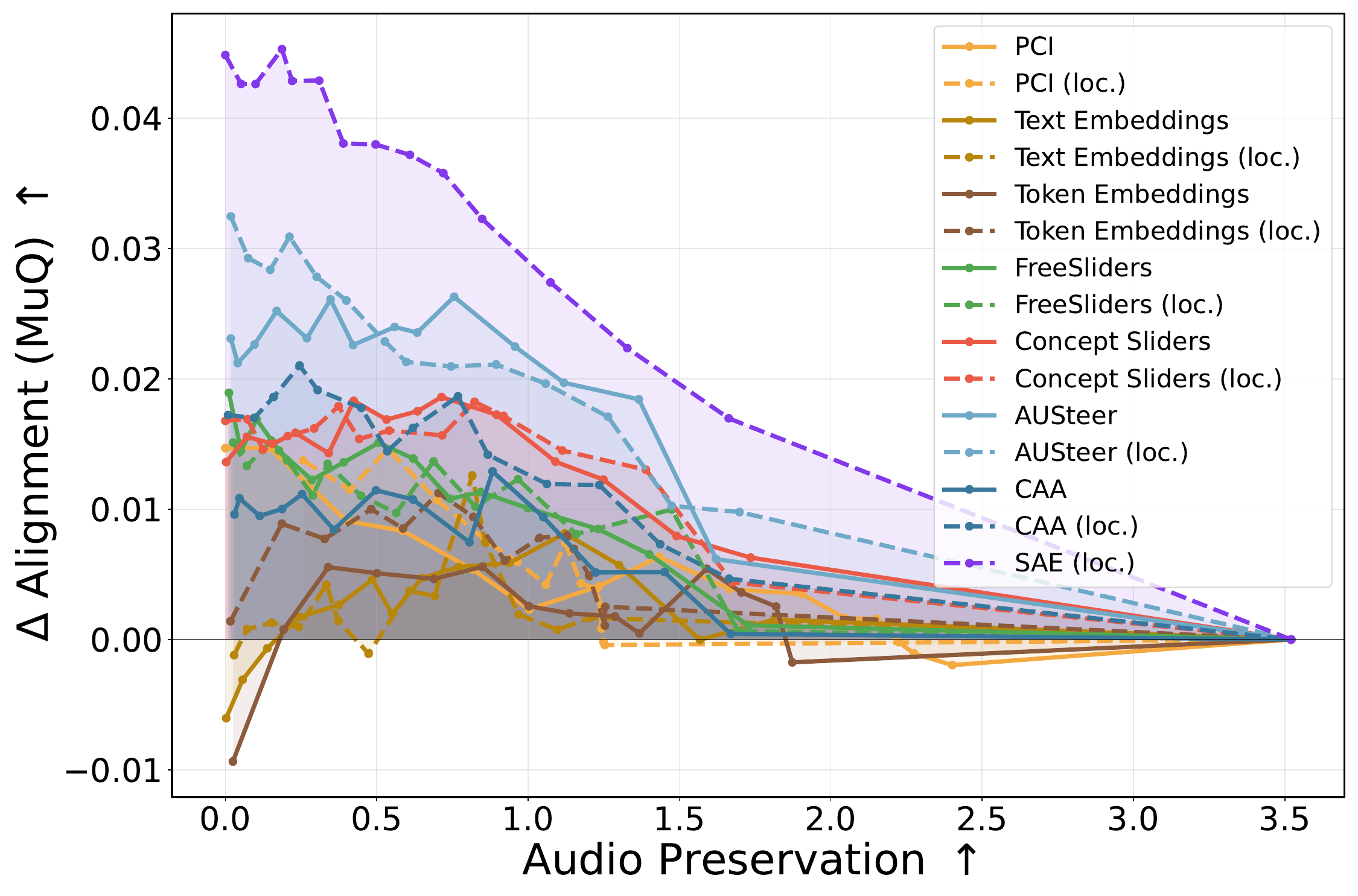}
        \caption{electric guitar (--)}
    \end{subfigure}
    \caption{\textbf{Alignment--preservation curves for \textit{guitar type} steering.} Used for computing AUC values in Tab.~\ref{tab:steering_auc_guitar}.}
    \label{fig:auc_curves_guitar}
\end{figure}

\begin{figure}[h]
    \centering
    \begin{subfigure}{0.49\linewidth}
        \includegraphics[width=\linewidth]{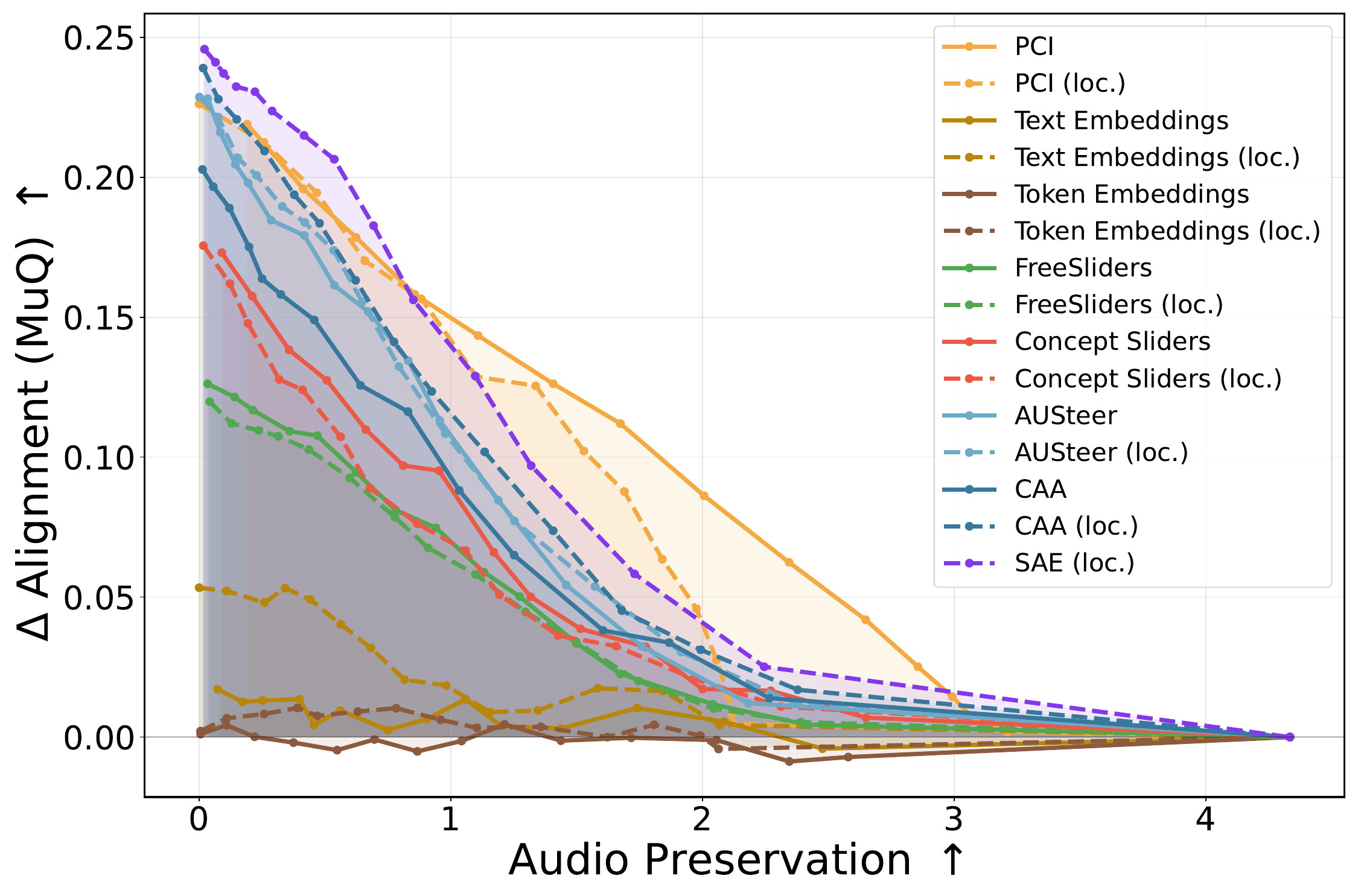}
        \caption{jazz genre (+)}
    \end{subfigure}
    \hfill
    \begin{subfigure}{0.49\linewidth}
        \includegraphics[width=\linewidth]{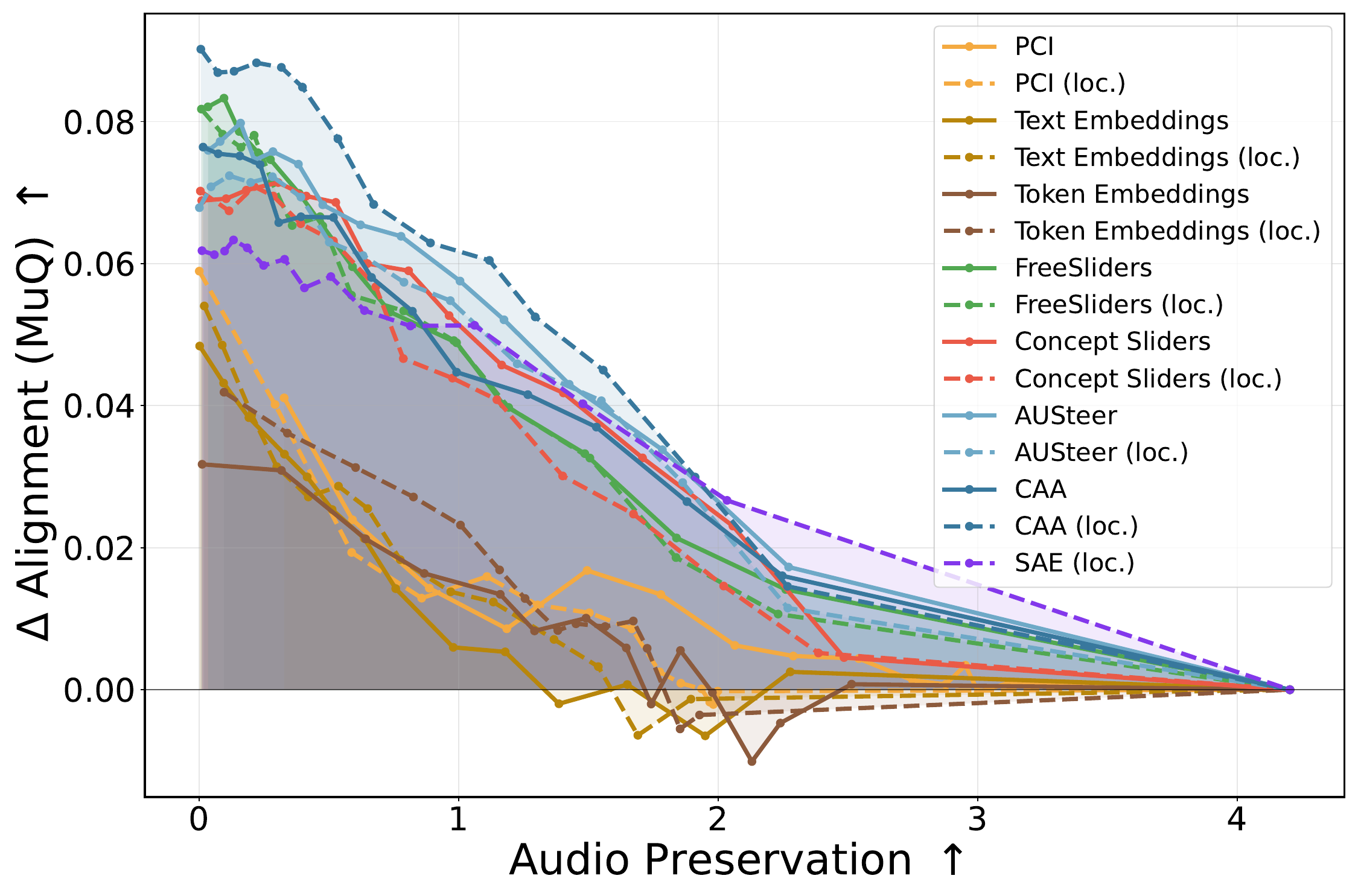}
        \caption{rock genre (--)}
    \end{subfigure}
    \caption{\textbf{Alignment--preservation curves for \textit{jazz genre} steering.} Used for computing AUC values in Tab.~\ref{tab:steering_auc_popjazz}.}
    \label{fig:auc_curves_jazz}
\end{figure}

\begin{figure}[h]
    \centering
    \begin{subfigure}{0.49\linewidth}
        \includegraphics[width=\linewidth]{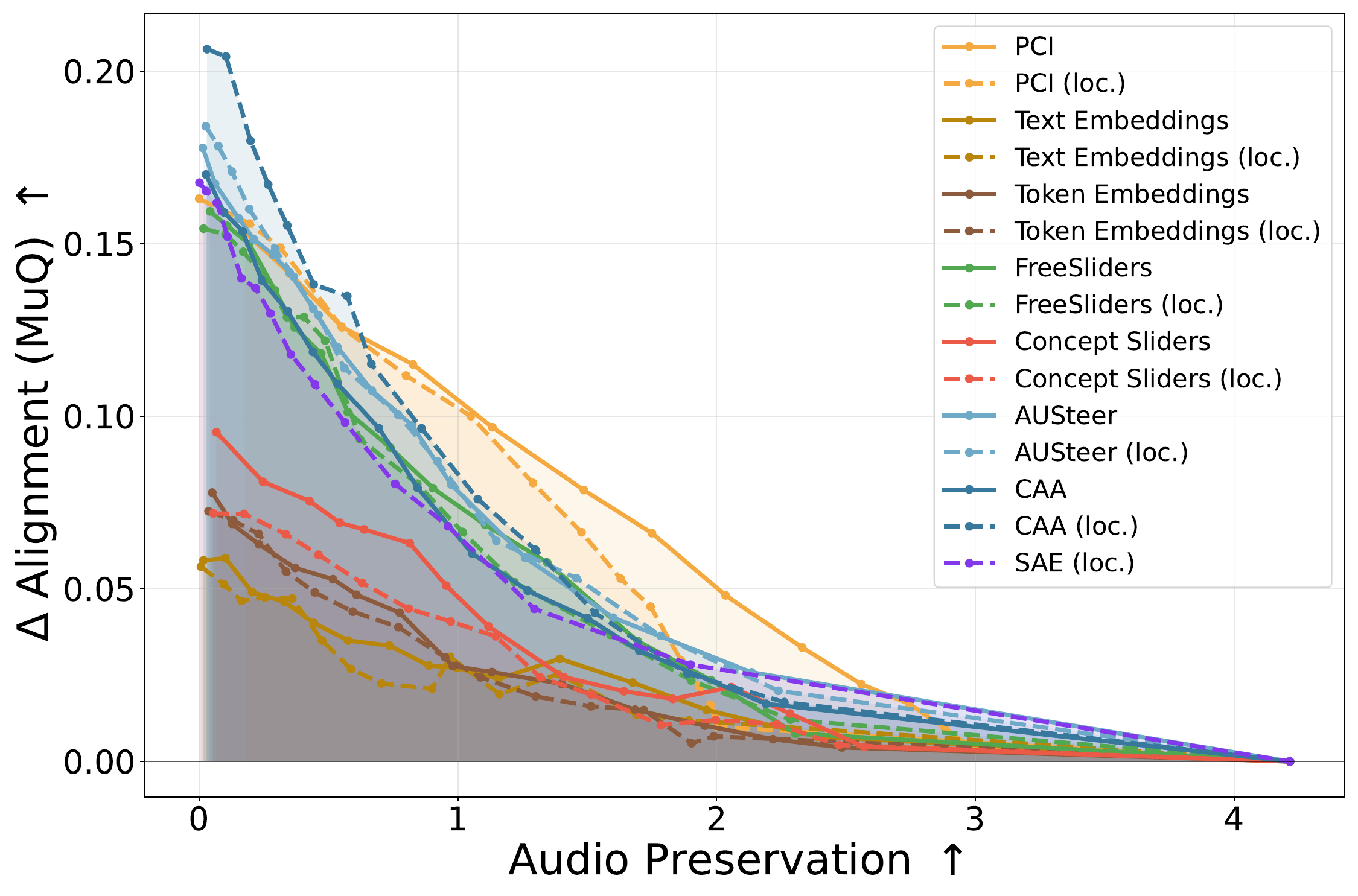}
        \caption{classical genre (+)}
    \end{subfigure}
    \hfill
    \begin{subfigure}{0.49\linewidth}
        \includegraphics[width=\linewidth]{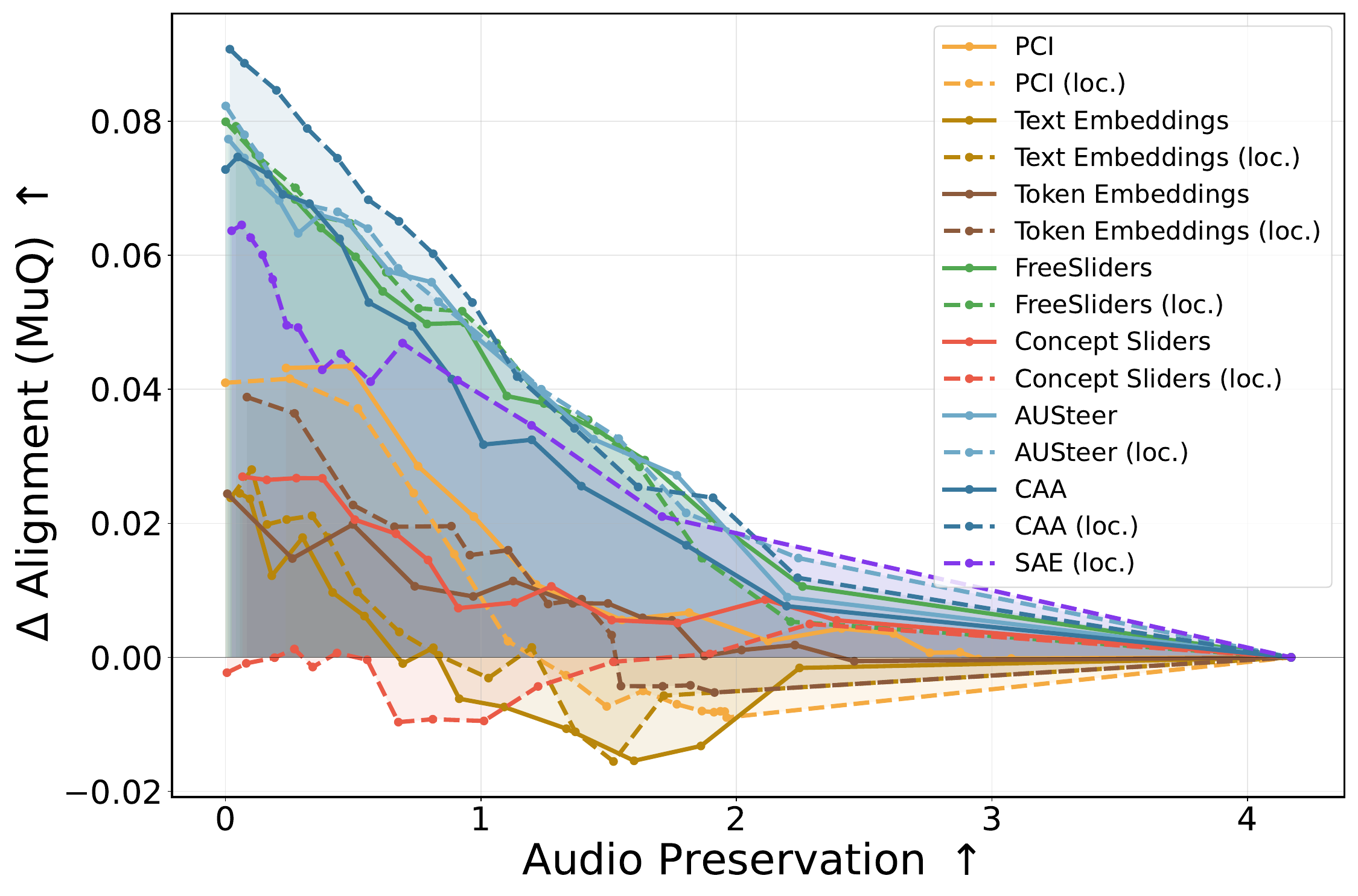}
        \caption{electronic genre (--)}
    \end{subfigure}
    \caption{\textbf{Alignment--preservation curves for \textit{classical genre} steering.} Used for computing AUC values in Tab.~\ref{tab:steering_auc_classical_electronic}.}
    \label{fig:auc_curves_classical}
\end{figure}

\Cref{fig:auc_curves_piano,fig:auc_curves_mood,fig:auc_curves_tempo,fig:auc_curves_vocals,fig:auc_curves_vocal_style,fig:auc_curves_violin,fig:auc_curves_guitar,fig:auc_curves_jazz,fig:auc_curves_classical} visualize the alignment--preservation curves whose areas yield the AUC scores in \cref{tab:steering_auc_piano,tab:steering_auc_mood,tab:steering_auc_tempo,tab:steering_auc_vocal_gender,tab:steering_auc_vocal_style,tab:steering_auc_violin,tab:steering_auc_classical_electronic,tab:steering_auc_guitar,tab:steering_auc_popjazz}. For all nine benchmark concepts and both steering directions, each panel sweeps the steering strength $\alpha$ for every evaluated method, plotting the sign-corrected alignment delta $\Delta a(\alpha)$ against the preservation score $p(\alpha)$ (\cref{sec:metrics}). The AUC metric value reported in the tables is the area under the corresponding curve, integrated up to the shared cutoff $P_{\max}$. Curves closer to the upper-right corner represent better gain in alignment at matched preservation.

\clearpage
\section{Spectrograms}

\begin{figure}[h]
      \centering
      \begin{subfigure}{\linewidth}
          \includegraphics[width=\linewidth]{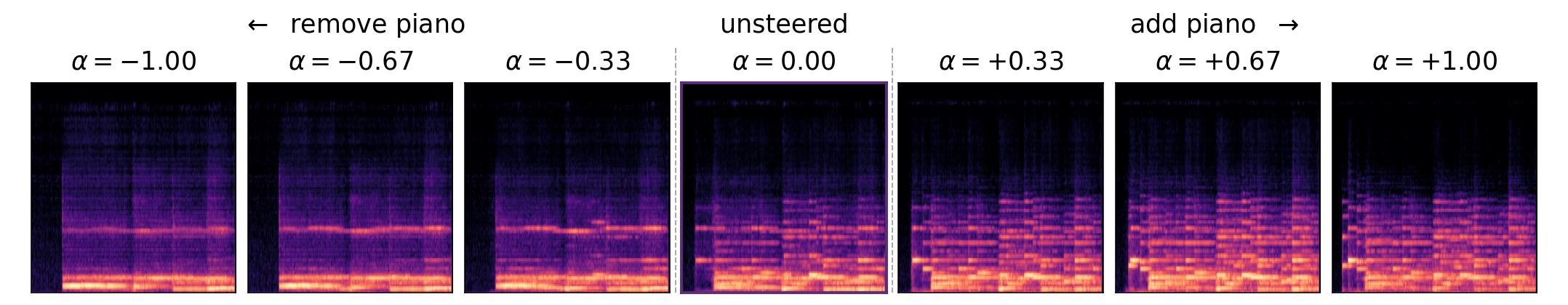}
          \caption{Piano: ``calm coldplay style song with slight piano''}
      \end{subfigure}

      \begin{subfigure}{\linewidth}
          \includegraphics[width=\linewidth]{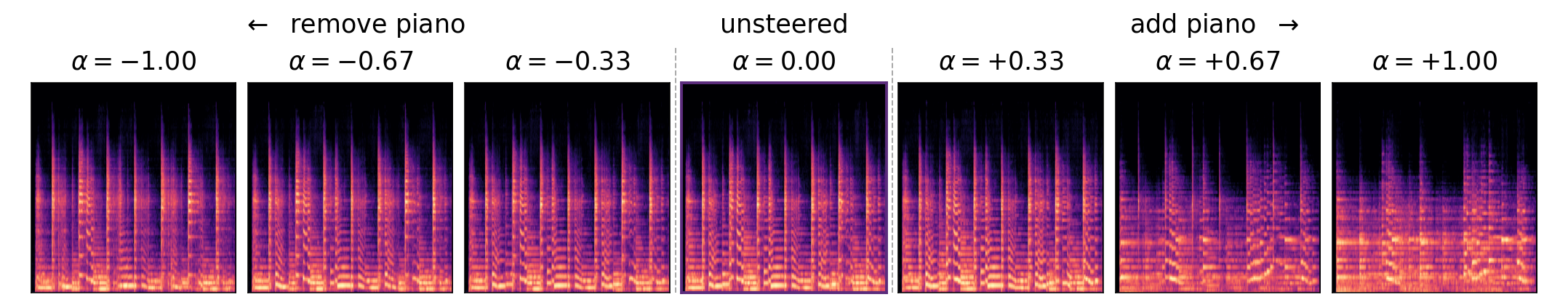}
          \caption{Piano: ``Latin Spanish Music with subtle guitar''}
      \end{subfigure}


      \begin{subfigure}{\linewidth}
          \includegraphics[width=\linewidth]{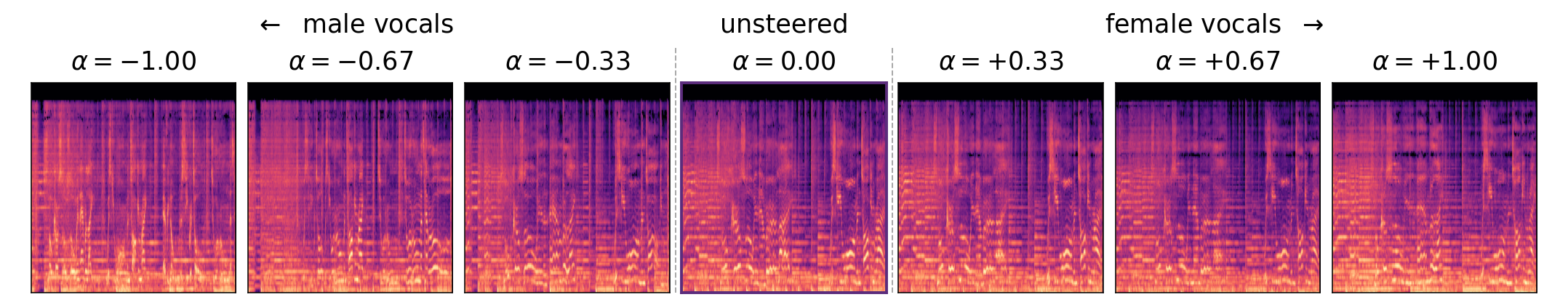}
          \caption{Vocal: ``jazz trio with brushed drums, walking upright bass, and smoky piano''}
      \end{subfigure}

      \begin{subfigure}{\linewidth}
          \includegraphics[width=\linewidth]{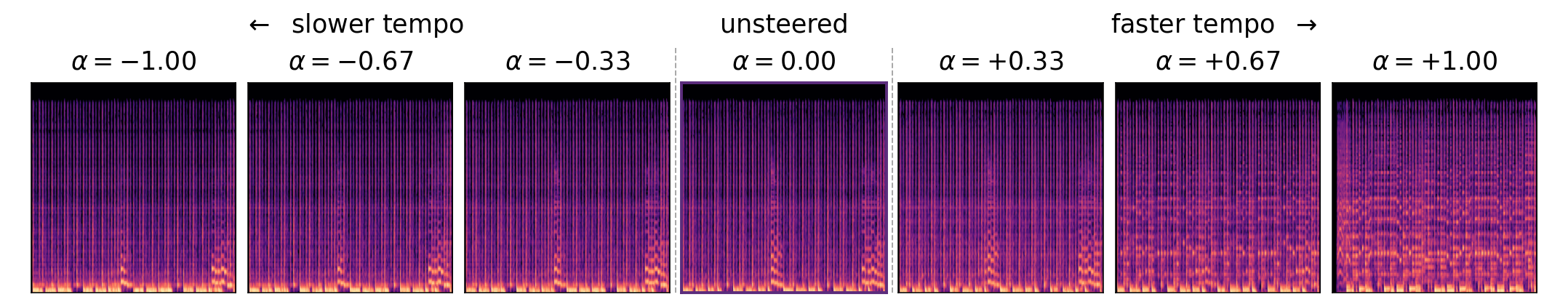}
          \caption{Tempo: ``Latin music with percussion and guitar''}
      \end{subfigure}

      \begin{subfigure}{\linewidth}
          \includegraphics[width=\linewidth]{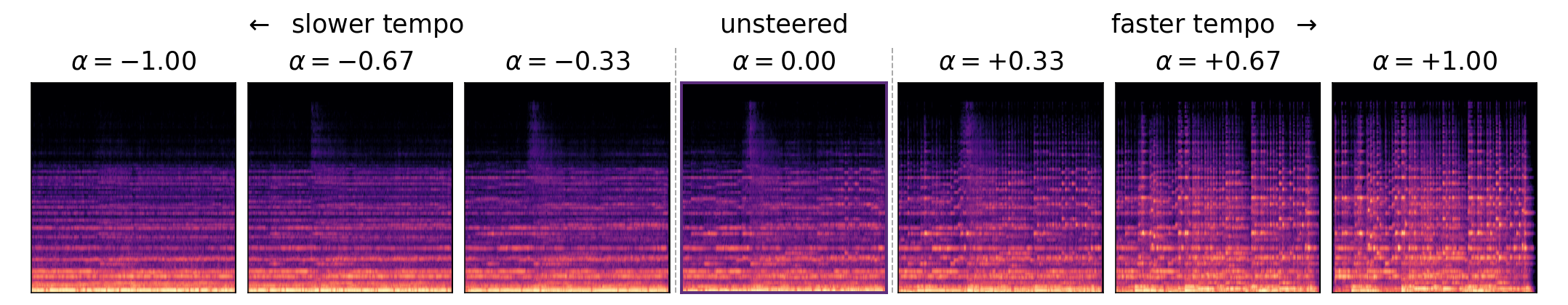}
          \caption{Tempo: ``dramatic opera, lush orchestral strings, and swelling brass''}
      \end{subfigure}

      \caption{\textbf{Mel spectrograms for audio concept modulation.} Each row presents the result of the steering audio generation process with SAEs. The central column ($\alpha=0$) is the
  unsteered baseline.}
      \label{fig:sae_spectrograms}
  \end{figure}

We complement our results with qualitative examples in \Cref{fig:sae_spectrograms}, showing mel-spectrograms of generated audio modulated via SAE steering. For each prompt, we sweep the steering coefficient $\alpha$ over seven evenly spaced values, with $\alpha=0$ corresponding to the unsteered track. 

The spectrograms make it visually apparent how SAE interventions controllably edit the time--frequency content of the generation as $\alpha$ is modulated.

\clearpage
\section{Details on multi-concept steering}\label{app:multi_concept_steering}

  We evaluate activation steering methods on multi-concept steering with nine combinations, including
  \begin{enumerate}
      \item five pairs: \textit{piano with violin} (\textbf{P}+\textbf{V}),
            \textit{piano with female vocals} (\textbf{P}+\textbf{FV}),
            \textit{acoustic guitar with female vocals} (\textbf{A}+\textbf{FV}),
            \textit{piano with male vocals} (\textbf{P}+$\overline{\textbf{FV}}$), and
            \textit{jazz with slower tempo} (\textbf{J}+$\overline{\textbf{T}}$);
      \item four triples: \textit{piano with violin and jazz}
                (\textbf{P}+\textbf{V}+\textbf{J}),
            \textit{acoustic guitar with female vocals and brighter mood}
                (\textbf{A}+\textbf{FV}+\textbf{M}),
            \textit{piano with violin and slower tempo}
                (\textbf{P}+\textbf{V}+$\overline{\textbf{T}}$), and
            \textit{acoustic guitar with male vocals and darker mood}
                (\textbf{A}+$\overline{\textbf{FV}}$+$\overline{\textbf{M}}$).
  \end{enumerate}

For CAA, AUSteer, and SAE, we construct the multi-concept steering vector as the unit-weight summation of the single-concept vectors. Negative direction vectors (e.g., male vocal or darker mood) are constructed by a sign flip of the corresponding positive direction vectors.

All generations share the main benchmark settings (\ace model with $T=30$ inference steps and CFG: $\omega=5$). The configuration consists of 100 tracks, each $30$ seconds long. We sweep $16$ alphas linearly in one direction. For each concept combination evaluation, we calculate the alignment separately for each constituent concept with
CLAP or MUQ-T, and report the mean across concepts. Steering strengths are calibrated to match the average preservation cutoff for the main benchmark.

Results in \cref{tab:multi-concept-auc} compare activation steering methods in non-localized and localized variants across diverse concept combinations for both (MUQ and CLAP) text-audio alignment models. Localized steering variants nearly always offer better steering performance: 
CAA$_{\text{loc}}$ improves on CAA$_{\text{all}}$ on all nine combinations in both
metrics, and AUSteer$_{\text{loc}}$ improves on AUSteer$_{\text{all}}$ on eight of nine under MUQ-T. In $15$ out of $18$ cases, localized steering offers the best performance.

\begin{table}[h]
\centering
\small
\caption{\textbf{Multi-concept steering performance measured with average Area Under LPAPS-MUQ and LPAPS-CLAP Curves.} Best result per combination of concepts is \textbf{bolded}. Single column denotes steering on combinations of concepts, where \textbf{P} = Piano, \textbf{V} = Violin, \textbf{FV} = Female Vocal, \textbf{A} = Acoustic Guitar, \textbf{J} = Jazz Music, \textbf{M} = Brighter Mood, $\overline{\text{\textbf{FV}}}$ = Male Vocal, $\overline{\text{\textbf{T}}}$ = Slower Tempo, and $\overline{\text{\textbf{M}}}$ = Darker Mood.}
\label{tab:multi-concept-auc}
\resizebox{\linewidth}{!}{%
\begin{tabular}{lccccccccc}
\toprule
\multirow{2}{*}{\textbf{Method}} & \multicolumn{5}{c}{Steering Two Concepts} & \multicolumn{4}{c}{Steering Three Concepts} \\
\cmidrule(lr){2-6}\cmidrule(lr){7-10}
& \textbf{P\,+\,V} & \textbf{P\,+\,FV} & \textbf{A\,+\,FV} & \textbf{P\,+\,$\overline{\text{FV}}$} & \textbf{J\,+\,$\overline{\text{T}}$} & \textbf{P\,+\,V\,+\,J} & \textbf{A\,+\,FV\,+\,M} & \textbf{P\,+\,V\,+\,$\overline{\text{T}}$} & \textbf{A\,+\,$\overline{\text{FV}}$\,+\,$\overline{\text{M}}$} \\
\midrule
\multicolumn{10}{l}{\textit{AUC MuQ}} \\
\midrule
AUSteer$_{\text{all}}$ & +0.057 & +0.001 & +0.022 & +0.013 & +0.090 & +0.062 & +0.005 & +0.048 & +0.045 \\
CAA$_{\text{all}}$ & +0.048 & +0.011 & +0.028 & +0.008 & +0.068 & +0.047 & +0.026 & +0.029 & +0.032 \\
\midrule
AUSteer$_{\text{loc}}$ & +0.099 & +0.018 & +0.029 & \textbf{+0.039} & +0.086 & +0.088 & +0.021 & +0.069 & +0.051 \\
CAA$_{\text{loc}}$ & +0.094 & \textbf{+0.034} & \textbf{+0.052} & +0.027 & +0.103 & +0.081 & \textbf{+0.035} & +0.061 & +0.039 \\
SAE$_{\text{loc}}$ & \textbf{+0.114} & +0.027 & +0.041 & +0.034 & \textbf{+0.110} & \textbf{+0.100} & +0.018 & \textbf{+0.081} & \textbf{+0.063} \\
\midrule
\multicolumn{10}{l}{\textit{AUC CLAP}} \\
\midrule
AUSteer$_{\text{all}}$ & -0.002 & +0.047 & \textbf{+0.042} & \textbf{+0.051} & +0.035 & +0.012 & +0.028 & +0.001 & \textbf{+0.054} \\
CAA$_{\text{all}}$ & -0.003 & +0.037 & +0.019 & +0.036 & +0.036 & -0.002 & +0.036 & -0.003 & +0.044 \\
\midrule
AUSteer$_{\text{loc}}$ & +0.022 & +0.046 & +0.026 & +0.037 & +0.050 & +0.021 & +0.036 & +0.010 & +0.050 \\
CAA$_{\text{loc}}$ & \textbf{+0.027} & +0.048 & +0.032 & +0.049 & +0.050 & \textbf{+0.029} & \textbf{+0.047} & +0.013 & +0.047 \\
SAE$_{\text{loc}}$ & +0.018 & \textbf{+0.049} & +0.041 & +0.045 & \textbf{+0.056} & +0.029 & +0.027 & \textbf{+0.014} & +0.050 \\
\bottomrule
\end{tabular}
}
\end{table}



\end{document}